\documentclass{article}
\usepackage{graphicx}
\usepackage{amsmath}
\usepackage{amsfonts}
\usepackage{braket}
\usepackage{mathrsfs}
\usepackage{booktabs}
\usepackage{comment}
\usepackage{tikz}
\usepackage{ytableau}
\usepackage[a4paper, margin=1.25in]{geometry}

\usepackage{graphicx}
\usepackage{latexsym,amsmath,amsfonts,amssymb}
\usepackage{bbm}
\usepackage{tikz}
\usepackage[makeroom]{cancel}
\usetikzlibrary{decorations.pathmorphing,cd}
\usepackage[latin1]{inputenc}
\usepackage[american]{babel}
\usepackage{bbm}
\usepackage{cite}
\usepackage[colorlinks=true, citecolor=blue, linkcolor=blue, linktocpage=true]{hyperref}
\usepackage{adjustbox}
\usepackage{array}
\usepackage{float}
\usepackage{tabu}
\usepackage{comment}

\usepackage{amsfonts}
\usepackage{amscd}
\usepackage{amssymb}
\usepackage{amsmath,bbm}
\usepackage{graphicx}
\usepackage{epsfig}
\usepackage{latexsym}
\usepackage{mathtools}
\usepackage{hyperref}
\usepackage{tikz-cd}
\usepackage[vcentermath]{youngtab}
    
\def \be  {\begin{equation}}
\def \ee  {\end{equation}}
\def \bea {\begin{equation}\begin{aligned}}
\def \eea {\end{aligned}\end{equation}}
\def \ba  {\begin{eqnarray}}
\def \ea  {\end{eqnarray}}
\def \bb  {\begin{thebibliography}}
\def \eb  {\end{thebibliography}}
\def \lab #1 {\label{#1}}

\newcommand\ep{\epsilon}

\newcommand\cB{\mathcal{B}}
\newcommand\cC{\mathcal{C}}

\newcommand\cF{\mathcal{F}}

\newcommand\cH{\mathcal{H}}
\newcommand\cI{\mathcal{I}}

\newcommand\cL{\mathcal{L}}

\newcommand\cN{\mathcal{N}}
\newcommand\cO{\mathcal{O}}

\newcommand\cS{\mathcal{S}}

\newcommand\cW{\mathcal{W}}

\newcommand\bC{\mathbb{C }}
\newcommand\bP{\mathbb{P}}

\newcommand\bR{\mathbb{R}}

\newcommand\bZ{\mathbb{Z}}

\newcommand\fm{\mathfrak{m}}

\newcommand\fn{\mathfrak{n}}

\renewcommand{\d}{\mathrm{d}}

\newcommand{\wt}{\widetilde}

\numberwithin{equation}{section}

\def\repa{\raise4pt\hbox{$\square$}\mkern-14mu\raise-4pt\hbox{$\square$}}
\def\repab{\overline{\raise4pt\hbox{$\square$}\mkern-14mu\raise-4pt\hbox{$\square$}\mkern-1mu}}

\def\smileface{\ensuremath{\hbox{\large$\bigcirc$}\mkern-15mu\raise-1pt\hbox{\scriptsize$\smallsmile$}%
		\mkern-10mu\raise4pt\hbox{..}\mkern4mu}}
\def\frownface{\ensuremath{\hbox{\large$\bigcirc$}\mkern-15mu\raise-1pt\hbox{\scriptsize$\smallfrown$}%
		\mkern-10mu\raise4pt\hbox{..}\mkern4mu}}

\DeclareMathOperator{\sgn}{sgn}

\DeclareMathOperator{\re}{\mathbb{R}e}

\DeclareMathOperator{\Li}{Li}

\definecolor{cardinal}{rgb}{0.6,0,0}
\definecolor{darkgreen}{rgb}{0,0.5,0}
\definecolor{golden}{rgb}{0.92, 0.7, 0}
\definecolor{midnight}{rgb}{0, 0, 0.5}
\definecolor{darkblue}{rgb}{0.2, 0, 0.8}

\usepackage{titlesec}
\titleformat*{\section}{\Large\bfseries}
\titleformat*{\subsection}{\large\bfseries}
\titleformat*{\subsubsection}{\normalsize\bfseries}
\titleformat*{\paragraph}{\normalsize\bfseries}

\usepackage{authblk}

\newtheorem{conjecture}{Conjecture}

\begin{document}

\title{\textbf{$\text{AdS}_4$ Holography and the Hilbert Scheme}}

\author[1,4]{Samuel Crew \thanks{samuel.crew24@imperial.ac.uk}}
\author[2,3,4]{Daniel Zhang \thanks{
daniel.zhang@sjc.ox.ac.uk}}
\author[5]{Ziruo Zhang \thanks{zhangziruo@gmail.com}}

\affil[1]{\small Department of Mathematics, Imperial College London, London SW7 1NE, U.K.}
\affil[2]{\small St John's College, University of Oxford, St Giles', Oxford, U.K.}
\affil[3]{\small Mathematical Institute, University of Oxford, Woodstock Road, Oxford, U.K.}
\affil[4]{\small Theoretical Sciences Visiting Program (TSVP), Okinawa Institute of Science and Technology Graduate University, Onna, 904-0495, Japan}
\affil[5]{\small Kavli Institute for Theoretical Sciences (KITS), University of the Chinese Academy of
Sciences, Beijing 100190, China}

\maketitle

\setcounter{page}{1}

\begin{abstract}
We elucidate a holographic relationship between the enumerative geometry of the Hilbert scheme of $N$ points in the plane $\mathbb{C}^2$, with $N$ large, and the entropy of certain magnetically charged black holes with $\text{AdS}_4$ asymptotics. Specifically, we demonstrate how the entropy functional arises from the asymptotics of 't Hooft and Wilson line operators in a 3d $\mathcal{N}= 4$ gauge theory. The gauge-Bethe correspondence allows us to interpret this calculation in terms of the enumerative geometry of the Hilbert scheme and thereby conjecture that the entropy is saturated by expectation values of certain natural operators in the quantum $K$-theory ring acting on the localised $K$-theory of the Hilbert scheme. 
We give numerical evidence that the large $N$ limit is saturated by contributions from a certain vacuum/fixed point on the Hilbert scheme, associated to a particular triangular-shaped Young diagram, by evolving solutions to the Bethe equations numerically at finite (but large) $N$ towards the classical limit. We thus conjecture a concrete geometric holographic dual of the so-called gravitational/Cardy block.
\end{abstract}

\tableofcontents

\section{Introduction}

There is by now a rich interplay and convergence between the subjects of symplectic singularities and supersymmetric 3d gauge theories. Some examples of this include vortex counting and quasimaps \cite{bullimore2016vortices, bullimore2019twisted, dimofte2011vortex,Inglese:2023tyc, Haouzi:2023doo, Kimura:2024xpr}; symplectic duality and 3d mirror symmetry \cite{bullimore2016boundaries}; 3d-3d correspondences \cite{cheng20243, dimofte2014gauge, dimofte20133,Hayashi:2024jof} and correspondences between boundary conditions and generalised cohomology \cite{Bullimore:2021rnr, Dedushenko:2021mds, Dedushenko:2023qjq, Zhang:2022wwy, Hayashi:2024jof,Panerai:2020boq}. In this work, we are interested in extending this interplay and investigating geometric aspects of the holographic relationship between 3d gauge theory and $\text{AdS}_4$ gravity. 

Physically, it is well-known that the Bekenstein-Hawking entropy of certain static BPS magnetically charged black holes in M-theory on $\text{AdS}_4\times S^7$ can be accounted for by the large $N$ asymptotics of the topologically twisted index of the holographically dual 3d $\mathcal{N}=4$ ADHM theory with one flavour and $U(N)$ gauge group \cite{benini2016black,Hosseini:2016tor, Hosseini:2016ume, Choi:2019dfu, Choi:2019zpz}.\footnote{More precisely, the black holes are dual to BPS states in topologically twisted ABJM theory with $k = 1$ and at large $N$ via the standard AdS/CFT duality in \cite{Aharony:2008ug}. The $k = 1$ ABJM theory is in turn dual to the low energy limit of ADHM theory with one flavour \cite{Aharony:2008ug,Kapustin:2010xq}.} As we will detail later, the theory has flavour symmetry group $T$, whose maximal torus is generated by the charges $(F_1,F_2,F_C)$. $F_C$ is the 3d topological symmetry (the Coulomb branch flavour symmetry) under which monopole operators are charged. The index is a function of fluxes $(\fn_{1},\fn_{2},\fn_C)$ along $(F_1,F_2,F_C)$, and fugacities $(t_1,t_2,\zeta)$ conjugate to $(F_1,F_2,F_C)$. One of the main goals of our work is to provide a concrete geometric interpretation of the black hole microstate counting from the perspective of enumerative geometry and integrability.

Geometrically, we study the Hilbert scheme of $N$ points in the plane, denoted $X_N = \text{Hilb}^N(\mathbb{C}^2)$. This variety arises as the moduli space of Higgs branch vacua of the aforementioned gauge theory; described as a quiver in figure \ref{fig:adhm_quiver}. Enumerative counts of quasimaps in $X_N$ \cite{smirnov2020characters} are related to physical hemisphere partition functions \cite{Crew:2020psc} and the $K$-theory of the Hilbert scheme realises an integrable system associated to the quantum toroidal algebra $\mathfrak{gl}_1$ \cite{feigin2013representations, hernandez2009quantum}. The quantum $K$-theory ring $QK_T(X_N)$ naturally acts on this Hilbert space---physically as line operators acting on states generated by boundary conditions---and the expectation values of operators in this ring are controlled by the Bethe equations associated to this integrable system. We then realise the known entropy functional 
\begin{equation}\label{eq: entropy function}
    \cS =N^\frac{3}{2}\frac{\sqrt{2}}{3}\sqrt{\wt\Delta_1\wt\Delta_2\wt\Delta_3\wt\Delta_4}\sum_{i=1}^4\frac{\wt\fn_i}{\wt\Delta_i} \,,\qquad \sum_{i=1}^4\wt\Delta_i = 2\pi\,,
\end{equation}
as large $N$ expectation values of operators in the quantum $K$-theory ring of the Hilbert scheme of points. $\wt\Delta_i$ are a redundant parametrisation of the chemical potentials $(\log t_1,$ $\log t_2,$ $\log\zeta)$ refining the index, and $\wt\fn_i$ satisfying $\sum_{i=1}^4\wt\fn_i=-2$ are a redundant parametrisation of the fluxes $(\fn_{1},$ $\fn_{2},$ $\fn_C)$. 

Along the way to recovering the entropy functional from the enumerative geometry of the Hilbert scheme, we identify a dominant Higgs branch vacuum providing a geometric dual to the gravitational/Cardy block \cite{hosseini2019gluing, Choi:2019dfu, Choi:2019zpz}. We develop new numerical techniques to give significant evidence for a novel conjecture that this vacuum corresponds to a fixed point on the Hilbert scheme described by a particular triangular-shaped Young diagram. We give a precise statement of this conjecture in section $\ref{sec:BHpartition}$.

\subsection*{Summary}
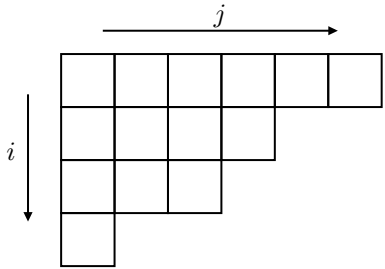
\begin{figure}
    \centering

\tikzset{every picture/.style={line width=0.75pt}} 

\begin{tikzpicture}[x=0.4pt,y=0.4pt,yscale=-1,xscale=1]

\draw   (171,62) -- (221,62) -- (221,112) -- (171,112) -- cycle ;
\draw   (221,62) -- (271,62) -- (271,112) -- (221,112) -- cycle ;
\draw   (271,62) -- (321,62) -- (321,112) -- (271,112) -- cycle ;
\draw   (321,62) -- (371,62) -- (371,112) -- (321,112) -- cycle ;
\draw   (371,62) -- (421,62) -- (421,112) -- (371,112) -- cycle ;
\draw   (171,112) -- (221,112) -- (221,162) -- (171,162) -- cycle ;
\draw   (171,162) -- (221,162) -- (221,212) -- (171,212) -- cycle ;
\draw   (171,212) -- (221,212) -- (221,262) -- (171,262) -- cycle ;
\draw   (221,162) -- (271,162) -- (271,212) -- (221,212) -- cycle ;
\draw   (271,162) -- (321,162) -- (321,212) -- (271,212) -- cycle ;
\draw   (221,112) -- (271,112) -- (271,162) -- (221,162) -- cycle ;
\draw   (271,112) -- (321,112) -- (321,162) -- (271,162) -- cycle ;
\draw   (321,112) -- (371,112) -- (371,162) -- (321,162) -- cycle ;
\draw   (421,62) -- (471,62) -- (471,112) -- (421,112) -- cycle ;
\draw    (140,100) -- (140,217) ;
\draw [shift={(140,220)}, rotate = 270] [fill={rgb, 255:red, 0; green, 0; blue, 0 }  ][line width=0.08]  [draw opacity=0] (8.93,-4.29) -- (0,0) -- (8.93,4.29) -- cycle    ;
\draw    (210,40) -- (427,40) ;
\draw [shift={(430,40)}, rotate = 180] [fill={rgb, 255:red, 0; green, 0; blue, 0 }  ][line width=0.08]  [draw opacity=0] (8.93,-4.29) -- (0,0) -- (8.93,4.29) -- cycle    ;

\draw (117,142.4) node [anchor=north west][inner sep=0.75pt]    {$i$};
\draw (311,12.4) node [anchor=north west][inner sep=0.75pt]    {$j$};

\end{tikzpicture}
\caption{\small Partitions are specified by their parts $\lambda=(\lambda_1,\lambda_2,\ldots)$. The transpose partition is denoted $\lambda^{\vee}$. Partitions can be written as Young diagrams in $\mathbb{Z}^2$ with boxes labelled by $s=(i,j) \in \lambda$, where $(i,j)$ run over the rows and columns respectively. The arm and leg lengths of $s \in \lambda$ are defined as 
$a_{\lambda}(s) = \lambda_{i} - j$, $l_{\lambda}(s) = \lambda^\vee_j-i$. The hook and the content of a box $s$ are $h_{\lambda}(s) = a_{\lambda}(s) + l_{\lambda}(s) +1$, $c_{\lambda}(s) = j-i$,}\label{fig:YTeg}
\end{figure}

The main object of study is the (equivariant) quantum $K$-theory ring $QK_T(X_N)$ of the Hilbert scheme of $N$ points in the plane $X_N = \text{Hilb}^N(\mathbb{C}^2)$. The physical construction is a $B$-twisted 3d $\mathcal{N}=4$ gauge theory on $S^2 \times_B S^1$ where the ring is realised as the ring of BPS Wilson line operators wrapping $S^1$. We write
\begin{equation}
    QK_T(X_N) = \mathbb{Z}[s_1^{\pm1},\ldots,s_N^{\pm N}, t_1^{\pm1},t_2^{\pm1}] / \mathcal{I} \,,
\end{equation}
where the equivariant parameters $\{s_a\}_{a=1}^N$ and $t_1$ and $t_2$ are realised physically as gauge and flavour fugacities (complexified Wilson lines) for the field theory, the latter corresponding to the group of flavour symmetries $T$. The ideal $\mathcal{I}$ is generated by the Bethe equations which correspond to the vacuum equations of the 3d $\mathcal{N}=4$ gauge theory
\begin{equation}\label{eq:outlineBEs}
    e^{s_a \partial_{s_a} \mathcal{W}} = 1.
\end{equation}
Here $\mathcal{W}$ is the effective twisted superpotential of the theory considered as a 2d theory after a $S^1$-compactification. The Bethe equations are a set of rational equations for $\{s_a\}$ dependent on the equivariant parameters $t_1$, $t_2$. We write them explicitly in \eqref{eq:bethe_eqns}. As we shall explain, the ring of operators $QK_T(X_N)$ acts naturally on the Hilbert space of localised $K$-theory denoted $\mathcal{H}_N := K_T^{\text{loc.}}(X_N)$, which corresponds physically to the space of supersymmetric ground states of the theory on a torus with complex structure $\tau$, in the limit where $q=e^{2\pi i \tau}\rightarrow 1$. The supersymmetric ground states correspond to \textit{classical} solutions to the Bethe equations \eqref{eq:outlineBEs}, meaning solutions in the limit $\zeta \rightarrow 0$. This Hilbert space\footnote{In fact \cite{smirnov2016rationality}, the Hilbert space may be identified with the space of Macdonald polynomials and the inner product structure arises as the Macdonald inner product.} has a linear basis of torus fixed points, corresponding to massive vacua, which for $X_N$ correspond to partitions \textit{i.e.} Young diagrams $|\lambda\rangle$ with $N$ boxes (see figure \ref{fig:YTeg} for our Young diagram conventions). 

Via the gauge-Bethe correspondence \cite{nekrasov2009supersymmetric, nekrasov2009quantum, okounkov2015lectures, feigin2013representations, hernandez2009quantum},  $\mathcal{W}$ may be alternatively viewed as the Yang-Yang potential of the quantum integrable system associated to the Hilbert scheme and the representation theory of quantum toroidal $\mathfrak{gl}_1$. The space $\bigoplus_N \mathcal{H}_N$ may then be viewed as a Fock module for this quantum group \cite{smirnov2016rationality}.

\begin{figure}
    \centering
    \hspace{-15mm}
    \includegraphics[scale=0.2]{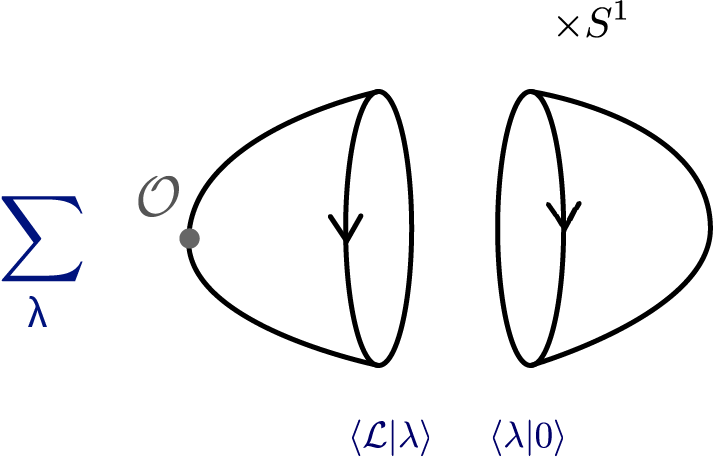}
    \caption{The factorisation setup. The path integral on $S^2 \times S^1$ in the presence of a line operator is sliced along a $S^1 \times S^1$ boundary with a complete set of states $|\lambda \rangle \langle \lambda |$ inserted.}
    \label{fig:factorisation}
\end{figure}

In this work we are interested in vacuum expectation values of quantum $K$-theory operators acting on $\mathcal{H}_N$ denoted
\begin{equation}\label{eq:intro_vev}
    \langle \mathcal{O} \rangle_{QK_T}.
\end{equation}
Physically, this quantity is computed by the $B$-twisted index $\mathcal{I}$ on $S^2 \times_B S^1$ with a Wilson line insertion \cite{Jockers:2018sfl}. To further elucidate the geometric interpretation of this quantity, we introduce an angular momentum refinement graded by a parameter $q$ on $S^2 \times_B S^1$ and slice open the path integral on $T^2$ along an equator of $S^2$ with a complete set of states/boundary conditions $|\lambda\rangle \in \mathcal{H}_N$ \cite{bullimore2016boundaries, Bullimore:2020jdq}. The setup is illustrated in figure \ref{fig:factorisation}. The expectation value may then be computed as
\begin{equation}\label{eq:outlineTI}
\begin{split}
    \langle \mathcal{O} \rangle &= \lim_{q \to 1} \sum_{\lambda} H_{\lambda}^{(\mathcal{O})} (q,t_i) H_{\lambda}(q^{-1},t_i) \\
    &= \sum_{\lambda} \langle \mathcal{O} | \lambda \rangle \langle \lambda | 0 \rangle.
\end{split}    
\end{equation}
In the first line $H_{\lambda}$ are partition functions of the theory on the hemisphere geometry $HS^2$. Geometrically they compute the (virtual) Euler characteristic of the space of quasimaps to $X_N$, based at $\lambda$, \textit{i.e.} the vertex functions of Okounkov \cite{okounkov2015lectures}, but normalised by appropriate perturbative contributions \cite{Bullimore:2020jdq}. In the second line $\langle \mathcal{O} |$ denotes the state created on $\mathcal{H}_{N}$ by the path integral on the bulk $HS^2$ in the presence of a line operator and $\langle \mathcal{O} | \lambda \rangle$ is the overlap with a boundary condition that is associated with massive vacua $\lambda$. The hemisphere partition function $H_{\lambda}^{(\mathcal{O})}$ may then be viewed as the transition matrix between the line operator and fixed point bases of $\mathcal{H}_N$.

In the $q \to 1$ limit the inner products can be interpreted as eigenvalues of quantum $K$-theory ring operators acting on the localised $K$-theory Hilbert space. We will show that the $q \to 1$ asymptotics of the hemisphere partition functions are explicitly given by 
\begin{equation}
    \langle \mathcal{O} | \lambda \rangle = \lim_{q \to 1} H_{\lambda}^{(\mathcal{O})}(q) = \left.\frac{\hat{a}^{1/2}(T^*R)}{\hat{a}^{1/2}(\mu_{\mathbb{C}})\hat{a}^{1/2}(\Delta_{\mathfrak{g}_{\mathbb{C}}})}\right\rvert_{\lambda} \frac{\mathcal{O}_{\lambda}}{\sqrt{\det \partial^2 \mathcal{W}_{\lambda}}}e^{\frac{1}{\epsilon}\mathcal{W}_{\lambda}},
\end{equation}
where $\mathcal{W}_{\lambda}$ and $\mathcal{O}_{\lambda}$ respectively denote the values of the effective twisted superpotential of the theory and the line operator on the Bethe root corresponding to the massive vacuum $\lambda$. The normalisation is given by the $\hat{a}$-genus of various weight spaces associated to the Higgs branch geometry that we explain in detail in the main body. We also later give a Higgs branch geometric interpretation to the superpotential $\mathcal{W}$.

We now turn to the main physical object of study, which is the $B$-twisted index with fluxes for flavour symmetries $\mathcal{I}_{S^2}^B(t_1,t_2,\zeta;\fn_1,\fn_2,\fn_C)$. That is, background fluxes for $T$ are turned on, with strengths $( \mathfrak{n}_1, \mathfrak{n}_2,\mathfrak{n}_C) \in \Gamma(T) \cong \bZ^3$ in the co-character lattice of $T$. In the absence of the angular momentum refinement, it follows directly from the Coulomb branch localisation formula that
\begin{equation}\label{eq:intro_CB_localisation}
    \cI_{S^2}^B = \langle e^{( \fn_1t_1\partial_{t_1}+\fn_1t_1\partial_{t_2}-\fn_C\zeta\partial_\zeta) \mathcal{W}} \rangle = \sum_{\lambda} \left.\frac{\hat{a}(T^*R)}{\hat{a}(\mu_{\mathbb{C}})\hat{a}(\Delta_{\mathfrak{g}_{\mathbb{C}}})}\right\rvert_{\lambda} \frac{e^{(\fn_1t_1\partial_{t_1}+\fn_1t_1\partial_{t_2}-\fn_C\zeta\partial_\zeta) \mathcal{W}}}{\det \partial^2 \mathcal{W}_{\lambda}},
\end{equation}
where the latter is a sum over evaluations of gauge fugacities at the Bethe roots $\lambda$. 

The above may equivalently be understood as the expectation value of an operator in $QK_T(X_N)$, i.e. of a Wilson line operator. This is explained as follows. Since the theory is topologically twisted, the flux can be concentrated at a pole of the $S^2$ (and wrapping the $S^1$), and is thus equivalent to an insertion of a $\textit{'t Hooft}$ line operator, as opposed to the Wilson line operators considered above. By introducing the angular momentum refinement corresponding to $q= e^{-\epsilon}$, we take the 't Hooft line operator to act on one of the hemisphere partition functions, along the $S^1$ at the pole of the $HS^2$:
\begin{equation}
    \mathcal{I}_{S^2}^B(t_1,t_2,\zeta;\fn_1,\fn_2,\fn_C) = \lim_{q \to 1}\sum_{\lambda}  \left( \hat{p}_C^{\mathfrak{n}_C}\hat{p}_1^{\mathfrak{n}_1}\hat{p}_2^{\mathfrak{n}_2} \cdot H_{\lambda}(q,t_1,t_2,\zeta) \right) \, H_{\lambda}(q^{-1}, t_1,t_2,\zeta)\,,
\end{equation}
where $(\hat{p}_{1},\hat{p}_{2},\hat{p}_C)$ are 't Hooft line operators for $T$ with unit flux. In the hemisphere partition function, 't Hooft line operator insertions along the $S^1$ at the tip of the $HS^2$ lead to a shift of parameters:
\bea
    \hat{p}_{1} \cdot H_{\lambda}(q,t_1,t_2,\zeta)  =  H_{\lambda}(q,t_1q,t_2,\zeta)\,,\quad &\hat{p}_{2} \cdot H_{\lambda}(q,t_1,t_2,\zeta)  =  H_{\lambda}(q,t_1,t_2q,\zeta)\,,\\
    \hat{p}_{C} \cdot H_{\lambda}(q,t_1,t_2,\zeta)  &=  H_{\lambda}(q,t_1,t_2,\zeta q^{-1})\,.
\eea 
In this case we see that the leading $q\to 1$ asymptotics of a vertex/hemisphere partition function with such an insertion is proportional to
\begin{equation}\label{eq:intro_asymptotics_vertex}
    \lim_{q\to 1} \hat{p}_{1,2} \cdot H_{\lambda} \sim e^{\mathcal{W}/\epsilon + t_{1,2} \partial_{t_{1,2}}\mathcal{W}}\,,\quad \lim_{q\to 1} \hat{p}_C \cdot H_{\lambda} \sim e^{\mathcal{W}/\epsilon - \zeta \partial_\zeta\mathcal{W}}\,,
\end{equation}
This computes the spectrum of the operators $(\hat{p}_{1},$ $\hat{p}_{2},$ $\hat{p}_C)$. In the $q\rightarrow 1$ limit, they are equivalent to multiplication by the conjugate momenta
\begin{equation}
    p_{1,2} \equiv e^{t_{1,2} \partial_{t_{1,2}}\mathcal{W}}\,,\quad p_C \equiv e^{-\zeta \partial_{\zeta}\mathcal{W}}
\end{equation}

We further argue that the action of $\hat{p}_i$ on a vertex function is equivalent to the insertion of a \textit{Wilson} line operator $\mathcal{L}_i$:
\begin{equation}
    \lim_{q\to 1}\hat{p}_{1,2} \cdot H_{\lambda}= \lim_{q\to1}H_{\lambda}^{(\mathcal{L}_{1,2})}\,,\quad \lim_{q\to 1}\hat{p}_C \cdot H_{\lambda}= \lim_{q\to1}H_{\lambda}^{(\mathcal{L})}\, 
\end{equation}
For $\hat{p}_C$ this is immediate as these are simply Wilson lines for the gauge symmetry. For the other (Higgs branch) flavour symmetries, this follows from certain difference equations (in flavour fugacities) that these partition functions obey \cite{aganagic2017quasimap, Aganagic:2016jmx, Ferrari:2023hza}, which are a generic feature of 3d theories with at least $\mathcal{N}=2$ supersymmetry. In the $q\rightarrow 1$ limit, this implies that $(\hat{p}_{1},$ $\hat{p}_{2},$ $\hat{p}_C)$ are equivalent, \textit{up to} an element of the Bethe ideal $\cI$, to a Wilson line operator and thus an element of the twisted chiral ring.  The above statements are non-trivial for the Higgs branch flavour symmetries as generically $p_{1,2} = e^{ t_{1,2} \partial_{t_{1,2}}\mathcal{W}}$ may include denominators in the gauge fugacities.

We may now interpret the twisted index with flux in light of the comments surrounding equations \eqref{eq:intro_vev} and \eqref{eq:outlineTI}:
\begin{equation}
\begin{aligned}
    \mathcal{I}_{S^2}^B(t_1,t_2,\zeta;\fn_1,\fn_2,\fn_C) &=
    \lim_{q\to 1}  \sum_{\lambda} H^{(\mathcal{L}_1^{ \mathfrak{n}_1}\mathcal{L}_2^{ \mathfrak{n}_2}\mathcal{L}^{\mathfrak{n}_C})}_{\lambda}(q,t_1,t_2,\zeta) H_{\lambda}(q^{-1},t_1,t_2,\zeta) \\
    &=   \sum_{\lambda} \langle \mathcal{L}_1^{ \mathfrak{n}_1}\mathcal{L}_2^{ \mathfrak{n}_2}\mathcal{L}^{\mathfrak{n}_C} | \lambda \rangle  \langle \lambda | 0 \rangle \\
    &= \langle \mathcal{L}_1^{ \mathfrak{n}_1}\mathcal{L}_2^{ \mathfrak{n}_2}\mathcal{L}^{\mathfrak{n}_C} \rangle_{QK_T(X_N)} \\
\end{aligned}
\end{equation}
In summary, the twisted index with flux computes the expectation value of certain elements of the quantum equivariant $K$-theory of $X_N$.

We then turn to holography. We are interested in static magnetically charged BPS black holes in M-theory on AdS$_4\times S^7$. By the usual holography arguments, the entropy of such black holes should be captured by the twisted index with fluxes in the large $N$ limit $\lim\limits_{N \to \infty} \mathcal{I}_{S^2}^B$. Using the above, introducing such fluxes corresponds to the shifts $\hat{p}_1^{\mathfrak{n}_1}\hat{p}_2^{\mathfrak{n}_2}\hat{p}_C^{\mathfrak{n}_C}$ and therefore to line operator insertions---we thus make a correspondence between black hole entropy and the geometric quantity:
\begin{equation}
    \lim_{N \to \infty} \langle \mathcal{L}_1^{\mathfrak{n}_1} \mathcal{L}_2^{\mathfrak{n}_2} \mathcal{L}^{\mathfrak{n}_C} \rangle_{QK_T(X_N)}.
\end{equation}

To reproduce the black hole entropy, we consider the large $N$ limit of this expression for $X_N = \text{Hilb}^N(\mathbb{C}^2)$. We find it more convenient to use the equality of this geometric quantity to the sum over Bethe roots \eqref{eq:intro_CB_localisation}, and work in a continuum limit where the gauge fugacities $\{s_a\}$ are characterised by an eigenvalue density $\rho$. The twisted superpotential becomes a functional $\cW[\rho]$, and the Bethe equations are the extremisation equations of $\cW$ with respect to $\rho$. There is a solution for $\rho$, that we call the black hole solution, such that when substituted into the summand of \eqref{eq:intro_CB_localisation} one obtains
\begin{equation}
    \lim_{N \to \infty}\lim_{q \to 1}\langle \mathcal{L}_1^{\mathfrak{n}_1} \mathcal{L}_2^{\mathfrak{n}_2} \mathcal{L}^{\mathfrak{n}_C} \rangle_{QK_T} =  \exp N^\frac{3}{2}\frac{\sqrt{2}}{3}\sqrt{\wt\Delta_1\wt\Delta_2\wt\Delta_3\wt\Delta_4}\sum_{i=1}^4\frac{\wt\fn_i}{\wt\Delta_i}\,.
\end{equation}
As mentioned in \eqref{eq: entropy function}, this is the known asymptotic formula of the twisted index \cite{benini2016black,Hosseini:2016tor,Hosseini:2016ume, Colombo:2024mts} and the black hole entropy function, but now interpreted in terms of expectation values in $QK_T$ of the Hilbert scheme of a large number of points $N$ in the complex plane $\mathbb{C}^2$.

The black hole solution $\rho$ may be interpreted as identifying a dominant Bethe root in the large $N$ limit.  In section \ref{sec:BHpartition} we develop new numerical techniques to identify this black hole solution at large but finite $N$ and evolve it to a classical vacuum on the Hilbert scheme. This turns out to correspond to a particular triangular Young diagram, as shown in Fig \ref{fig:eg triangular partitions}. This method allows us to make the conjecture that the hemisphere partition function/quasimap count with the triangular Young diagram boundary condition is the dual in the enumerative geometry of the Hilbert scheme to the Cardy block/gravitational block \cite{Choi:2019dfu,Choi:2019zpz, hosseini2019gluing}. This is a precise identification of a holographic geometric dual quantity to a supersymmetric black hole entropy function.

\subsection*{Discussion}
We have built a bridge between holography calculations in the physics literature and the enumerative geometry of the Hilbert scheme of points in the complex plane. We believe that the work raises a number of interesting additional questions. Firstly, it would be interesting to investigate other 3d $\mathcal{N}=4$ theories that flow to holographic theories in the IR, for example the necklace quiver of \cite{Hosseini:2016ume}---we expect the arguments of this paper to extend straightforwardly. In addition, there is a relation between the quantum $K$-theory ring and classical integrable systems \cite{koroteev2021quantum} whereby the spectrum of the ring realises the phase space of a classical integrable system. In this context, our computations correspond to expectation values of conjugate momenta. It would be interesting to investigate the classical integrable system associated to $\text{Hilb}^N(\mathbb{C}^2)$ and elucidate its phase space geometry when $N$ is large. In particular, one might expect relationships between the equivariant volume functional on the gravity side and the generating function of the corresponding classical phase space. Finally, it would be interesting to investigate other closed $3$-manifold indices from this enumerative perspective, such as the $S^3$ partition function or partition functions on Seifert manifolds \cite{closset2019three}.

\subsection*{Acknowledgements}
SC and DZ conducted part of this research while visiting the Okinawa Institute of Science and Technology (OIST) through the Theoretical Sciences Visiting Program (TSVP). DZ is supported by a Junior Research Fellowship from St John's College, University of Oxford. ZZ is supported by NSFC No.~12175237, the Fundamental Research Funds for the Central Universities, and funds from the Chinese Academy of Sciences. SC and DZ would like to thank Masazumi Honda and RIKEN for their hospitality while the manuscript was finalised. Finally, the authors would like to thank Nick Dorey and Hunter Dinkins for many helpful discussions.

\section{The Hilbert Scheme}
We begin with a discussion and definition of the Higgs branch geometry of the Hilbert scheme of $N$ points in the plane. We then discuss the enumerative geometry and its relationship to 3d $\mathcal{N}=4$ twisted indices.

\subsection{Geometry of the Hilbert scheme}
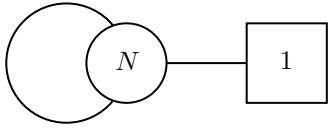
\begin{figure}
    \centering

\tikzset{every picture/.style={line width=0.75pt}} 

\begin{tikzpicture}[x=0.75pt,y=0.75pt,yscale=-1,xscale=1]

\draw   (110,70) .. controls (110,58.95) and (118.95,50) .. (130,50) .. controls (141.05,50) and (150,58.95) .. (150,70) .. controls (150,81.05) and (141.05,90) .. (130,90) .. controls (118.95,90) and (110,81.05) .. (110,70) -- cycle ;
\draw  [draw opacity=0] (123.02,89.24) .. controls (117.52,95.82) and (109.25,100) .. (100,100) .. controls (83.43,100) and (70,86.57) .. (70,70) .. controls (70,53.43) and (83.43,40) .. (100,40) .. controls (109.4,40) and (117.78,44.32) .. (123.29,51.08) -- (100,70) -- cycle ; \draw   (123.02,89.24) .. controls (117.52,95.82) and (109.25,100) .. (100,100) .. controls (83.43,100) and (70,86.57) .. (70,70) .. controls (70,53.43) and (83.43,40) .. (100,40) .. controls (109.4,40) and (117.78,44.32) .. (123.29,51.08) ;  
\draw    (150,70) -- (190,70) ;
\draw   (190,50) -- (230,50) -- (230,90) -- (190,90) -- cycle ;

\draw (205,63) node [anchor=north west][inner sep=0.75pt]    {$1$};
\draw (123,63) node [anchor=north west][inner sep=0.75pt]    {$N$};

\end{tikzpicture}
    \caption{\small The $\cN=4$ ADHM quiver.}
    \label{fig:adhm_quiver}
\end{figure}

We briefly review the Higgs branch geometry of the Hilbert scheme of $N$ points in the plane, $\text{Hilb}^N(\mathbb{C}^2)$. We refer the reader to \cite{nakajima1999lectures} for a more comprehensive treatment. The Hilbert scheme parametrises codimension $N$ ideals in $\mathbb{C}[x,y]$
\begin{equation}
    X_N = \text{Hilb}^N(\mathbb{C}^2) = \{\mathcal{J} \subset \mathbb{C}[x,y] \, : \, \text{dim}\, \mathbb{C}[x,y]/\mathcal{J} = N \}.
\end{equation}
The variety admits a torus action $T = \mathbb{C}^{\times}_{t_1}\times \mathbb{C}^{\times}_{t_2}$ induced by the coordinate action on $\mathbb{C}^2$
\begin{equation}\label{torus action}
    (x,y) \to (t_1 x, t_2 y),
\end{equation}
this action has fixed points labelled by Young diagrams $\lambda$ with $N$ boxes. The fixed ideals are
\begin{equation}
    \mathcal{J}_{\lambda} = \{ x^{\lambda_a} y^a \, :\, a=1,2,\ldots, l(\lambda)\}.
\end{equation}
We write $l(\lambda)$ for the length of a partition $\lambda$ and $|\lambda| = N$. In fact, the Hilbert scheme is a symplectic variety and the symplectic form is scaled with weight $-1$ under the diagonal group action $t_3 \equiv (t_1 t_2)^{-1}$. To see this more explicitly we note that the stability condition for the moment map associated to the quiver in figure \ref{fig:adhm_quiver} may be expressed as
\begin{equation}
    \mu_{\mathbb{C}}^{-1}(0)^{\text{stable}} = \{ (A,B,I,J) \, : \, J = 0,\, I \,\text{cyclic for } A,B\}.
\end{equation}
We may then define a polynomial ideal invariant under $G$ by
\begin{equation}
    \mathcal{J}_{A,B,I} = \{ p[x,y] \in \mathbb{C}[x,y] \, : \, p(x,y)I = 0 \} \,.
\end{equation}
Nakajima proves in Theorem 1.14 of \cite{nakajima1999lectures}  that this map is indeed one-to-one. The upshot is that the Hilbert scheme may be viewed as a symplectic quotient
\begin{equation}
    X_N = \mu^{-1}_{\mathbb{C}}(0)^{\text{stable}} / G_{\mathbb{C}} \,.
\end{equation}

\paragraph{Gauge theory Higgs branch.}

$\text{Hilb}^N(\mathbb{C}^2)$ may be interpreted as the Higgs branch of a certain 3d $\mathcal{N}=4$ gauge theory and therefore as a symplectic singularity. The quiver description encodes the field content of a gauge theory with gauge group $G=U(N)$: the 3d ADHM theory with one flavour. It has a vector multiplet, a hypermultiplet $(I,J)$ in the fundamental representation, and a hypermultiplet $(A,B)$ in the adjoint representation. In the language of 3d $\cN=2$ supersymmetry, $A$, $B$, $I$, $J$ are chiral multiplets, while the $\cN=4$ vector multiplet consists of an adjoint chiral multiplet $C$ and an $\cN=2$ vector multiplet. The theory has R-symmetries $R_H$ and $R_C$ acting on the hypermultiplet and vector multiplet scalars respectively, while there are additional flavour symmetries $F_H$ and $F_C$ acting on the adjoint hypermultiplet and monopole operators respectively. The charges of the scalar components of each $\cN=2$ chiral multiplet are collected in Table \ref{tab: charges}. We shall mostly use $F_1=(F_H+R_H-R_C)/2$ and $F_2=(R_H-R_C-F_H)/2$ as the basis for Higgs branch flavour symmetries (from the point of view of a 3d $\mathcal{N}=2$ theory) since it corresponds to the torus action \eqref{torus action}. The fugacities corresponding to $F_1$ and $F_2$ will be denoted as $t_1$ and $t_2$ respectively, while that corresponding to $F_C$ is $\zeta$. The fugacities $t$ and $z$ corresponding to $\frac{R_H-R_C}{2}$ and $F_H$ that were used in \cite{Crew:2020psc} can be written as $t=t_1t_2$, $z=(t_1/t_2)^{1/2}$. 

\begin{table}[h]
\centering
\small
\begin{equation*}
\arraycolsep=5mm
\begin{array}{c | c | c c | c c | c c}
    \toprule 
        & G & R_H & R_C & \frac{R_H-R_C}{2} & F_H & F_1 & F_2 \\
    \midrule
     I  & \square & 1 & 0 & \frac{1}{2} & 0 & \frac{1}{2} & \frac{1}{2} \\[.2em]
     J  & \overline{\square} & 1 & 0 & \frac{1}{2} & 0 & \frac{1}{2} & \frac{1}{2} \\[.2em]
     A  &  \text{adj} & 1 & 0 & \frac{1}{2} & 1 & 1 & 0 \\[.2em]
     B  &  \text{adj} & 1 & 0 & \frac{1}{2} & -1 & 0 & 1 \\[.2em]
     C  &  \text{adj} & 0 & 2 & -1 & 0 & -1 & -1 \\
    \bottomrule
\end{array}
\end{equation*}
\normalsize
\vspace{-5mm}
\caption{\small Charges of scalar components of $\cN=2$ chiral multiplets in ADHM with one flavour.
\label{tab: charges}}
\end{table}

\paragraph{$K$-theory ring.}
The equivariant $K$-theory ring is generated by tautological bundles $\mathcal{V}$ and the topologically trivial bundle $\mathcal{W}$ of rank $N$ and rank $1$ respectively. Kirwan surjectivity \cite{mcgerty2018kirwan} enables us to realise the $K$-theory ring as the polynomial quotient ring
\begin{equation}
    K_{T}(X_N) = \mathbb{Z}[\{s_a^{\pm 1}\}_{i=1,\ldots, N}, t_1^{\pm 1}, t_2^{\pm 1}, t_3^{\pm 1}]^{S_N} / \mathcal{I} \,.
\end{equation}
Here, $S_N$ is the symmetric group and we consider polynomials invariant under the $S_N$ action on the variables $\{s_a\}$. We have included the redundant generator $t_3 = (t_1 t_2)^{-1}$. The ideal $\mathcal{I}$ is generated by the relations 
\begin{equation}\label{eq: classical BAE}
    (s_a-t_1^\frac{1}{2}t_2^\frac{1}{2} )\prod_{\substack{b=1 \\ b \neq a}}^N (s_b - t_1^{-1} s_a)(s_b - t_2^{-1} s_a)(s_b - t_3^{-1} s_a) = 0,\, \quad a=1,\ldots, N.
\end{equation}
The solutions to these equations are given by
\begin{equation}\label{eq:classical soln BAE}
    s_a = t_1^{i_a-\frac{1}{2}}t_2^{j_a-\frac{1}{2}}\,,
\end{equation}
where $a=1,\ldots,N$ run over the boxes of a Young tableaux $\lambda$, and correspond to an evaluation at the fixed point on $X_N$ corresponding to $\lambda$. The number of solutions is then given by the partition function $p(N)$. Here, $i_a$ and $j_a$ denote the row and column indices of the box $a \in \lambda$, numbered from the top left in the `English' convention. Our conventions for partitions and Young diagrams follow those of \cite{Crew:2020psc}. We denote by $\mathcal{V}_{\lambda}$ the evaluation of a vector bundle $\mathcal{V}$ at a fixed point $\lambda$. The image of the tautological class under the Kirwan map is
\begin{equation}
\begin{split}
    \mathcal{V} &= s_1 + \ldots + s_N .
\end{split}    
\end{equation}
and the determinant line bundle is given by
\begin{equation}\label{eq:detlinebundle}
    \mathcal{L} := \det \mathcal{V} = s_1 s_2 \ldots s_N.
\end{equation}
The tangent bundle of $\mathcal{M}_H$ is given by the cohomology of the complex
\begin{equation}
    0 \rightarrow \mathfrak{g}_{\mathbb{C}} \rightarrow T^*R \rightarrow \mathfrak{g}_{\mathbb{C}}^\ast \rightarrow 0,
\end{equation}
where $\mathfrak{g}_{\mathbb{C}}$ corresponds to the gauge group quotient and $\mathfrak{g}_{\mathbb{C}}^\ast$ the complex moment map condition. Thus in terms of $K$-theory classes:
\begin{equation}\label{eq:tangentcomplex}
    T^*X_N= T^*R - \text{ch} \,\mathfrak{g}_{\mathbb{C}} -  t \,\text{ch} \,\mathfrak{g}_{\mathbb{C}}^{\ast}.
\end{equation}
The hypermultiplet weight space may be read off from the above quiver description and expressed as 
\begin{equation}\label{eq:hypers}
    T^*R = t^{1/2}\sum_{a=1}^N s_a + t^{1/2}\sum_{a=1}^N s_a^{-1} + z^{-1}t^{1/2} \sum_{a,b=1}^N \frac{s_a}{s_b} + z t^{1/2} \sum_{a,b=1}^N  \frac{s_a}{s_b}.
\end{equation}
The character of $\mathfrak{g}_{\mathbb{C}}$ for $\text{Hilb}^N(\mathbb{C}^2)$ may be expressed as
\begin{equation}
    \text{ch} \, \mathfrak{g}_{\mathbb{C}} = \sum_{a,b=1}^N\frac{s_a}{s_b}.
\end{equation}

Therefore we find that the character of the tangent bundle is given by
\begin{equation}\label{eq:preevaltangent}
    TX_N = (t_1t_2)^{1/2}\sum_{a=1}^N s_a + (t_1t_2)^{1/2} \sum_{a=1}^N s_a^{-1} -(1-t_1)(1-t_2)\sum_{a,b=1}^N  \frac{s_a}{s_b}.
\end{equation}
The tangent bundle may be pushed forward to fixed points by evaluating on solutions to the classical Bethe equations \eqref{eq:classical soln BAE} corresponding to a fixed point partition $\lambda$ giving \cite{nakajima1999lectures}
\begin{equation}
    T_{\lambda}X_N = \sum_{s\in \lambda} z^{-l_{\lambda}(s) -a_{\lambda}(s) - 1}t^{-\frac{1}{2}l_{\lambda}(s) + \frac{1}{2}a_{\lambda}(s) - \frac{1}{2}} + z^{l_{\lambda}(s) +a_{\lambda}(s) +1}t^{\frac{1}{2}l_{\lambda}(s) - \frac{1}{2}a_{\lambda}(s) - \frac{1}{2}},
\end{equation}
where $a_{\lambda}(s)$ and $l_{\lambda}(s)$ denote the arm and leg lengths respectively of a box $s \in \lambda$. Our partition conventions are summarised in Figure \ref{fig:YTeg}.

\paragraph{Fixed point Hilbert space.}
We consider a vector space spanned by the fixed points of $\text{Hilb}^N(\mathbb{C}^2)$:
\begin{equation}
    \mathcal{H}_N = \text{sp}_{\mathcal{R}} |\lambda \rangle \,,
\end{equation}
This may be realised geometrically as the localisation of the  $K$-theory ring: $K_{T}(X_N)^{\text{loc.}} := K_{T}(X_N)\otimes \mathcal{R}$, where $\mathcal{R}$ denotes the representation ring or equivalently the equivariant $K$-theory of a point
\begin{equation}
    \mathcal{R} = K_{T}(\text{pt.}) = \mathbb{Z}[t_1^{\pm 1}, t_2^{\pm 1}, t_3^{\pm 1}].
\end{equation}
Later, we will realise the equivariant localised $K$-theory physically in the setup of the hemisphere partition function. There is an action of the $K$-theory ring on $\mathcal{H}_N$ defined diagonally by
\begin{equation}
    \mathcal{V} | \lambda \rangle = \mathcal{V}_{\lambda} | \lambda \rangle.
\end{equation}
where $\mathcal{V}_{\lambda}$ denotes the evaluation of $V$ at the fixed point $\lambda$ solving equations \eqref{eq:classical soln BAE}. In the following we will upgrade this to an action of quantum $K$-theory on $\mathcal{H}_N$.

\subsection{Quantum \texorpdfstring{$K$}{}-theory}\label{subsec:quantumKtheory}
In this subsection we discuss the relationship between the quantum $K$-theory ring, rings of line operators in 3d $\mathcal{N}=4$ gauge theories and integrability.

\paragraph{Integrability.}
The main algebraic object we consider is the quantum toroidal algebra $\mathfrak{gl}_1$, denoted $\mathcal{E}$. The representation theory of this algebra was studied in \cite{feigin2013representations,feigin2015quantum,feigin2017integrals}. We are interested in Fock modules of this algebra $\mathcal{F}(u)$. The geometric $R$-matrix construction \cite{maulik2012quantum,smirnov2016rationality,negut2012moduli} allows us to identify the Fock module $\mathcal{F}(u)$ with the fixed point Hilbert space discussed above. Namely
\begin{equation}
    \mathcal{F}(u) = \bigoplus_{N=1}^{\infty} \mathcal{H}_N.
\end{equation}
The algebra $\mathcal{E}$ is a Hopf algebra \cite{feigin2013representations} with a universal $\mathcal{R}$-matrix from which we may define a transfer matrix in the Fock space by 
\begin{equation}
    T(u) = \text{Tr}_{\mathcal{F}(u)} \mathcal{R}.
\end{equation}
This transfer matrix is diagonalised by the Bethe equations  \cite{feigin2015quantum}
\begin{equation}\label{eq:bethe_eqns}
    (-\zeta)^{-1}\, t_1^{-\frac{1}{2}}t_2^{-\frac{1}{2}}\frac{s_a-t_1^{\frac{1}{2}}t_2^{\frac{1}{2}}}{s_a-t_1^{-\frac{1}{2}}t_2^{-\frac{1}{2}}}\prod_{I=1}^3\prod_{\substack{b=1\\ b\neq a}}^N \frac{s_b - t_I^{-1} s_a}{s_b - t_I s_a} = 1, \quad a=1,\ldots, N. 
\end{equation}
We note that by clearing denominators in this equation and then setting $\zeta\rightarrow 0$ gives the classical relations \eqref{eq: classical BAE}, which are solved by the classical solutions \eqref{eq:classical soln BAE}. 

\paragraph{The quantum $K$-theory ring.}
The $K$-theory ring may be deformed by the Bethe equations as follows
\begin{equation}\label{eq:q_k_ring}
    QK_{T}(X_N) = \mathbb{Z}[\{s_a^{\pm 1}\}_{a=1,\ldots, N}, t_1^{\pm 1}, t_2^{\pm 1}, t_3^{\pm 1}] / \mathcal{I_{\zeta}} \,,
\end{equation}
where now the ideal $I_{\zeta}$ is generated by the Bethe equations \eqref{eq:bethe_eqns}. This ring is that of Wilson line operators in the 3d gauge theory - the ring structure can be found by computing correlators in the twisted index on $S^2\times S^1$ or hemisphere partition function $HS^2 \times S^1$ \cite{Jockers:2018sfl}. The ring structure is deformed by $\zeta$ via contributions of higher degree quasimaps on this geometry, analogously to the 2d case \cite{Vafa:1991uz}.

Defining the complexified gauge holonomies $u_a$ via $s_a=e^{  i u_a}$, the Bethe equations \eqref{eq:bethe_eqns} may also be expressed as the vacuum conditions
\begin{equation}
    \partial_{u_a} \mathcal{W} = 0 \mod 2\pi\bZ\quad \Leftrightarrow \quad e^{s_a \frac{\partial}{\partial_{s_a}} \mathcal{W}} = 1, \quad \text{for } a=1,\ldots,N
\end{equation}
$\mathcal{W}$ is effective twisted superpotential for the effective abelian theory in the IR for the ADHM theory, and is a function of the complexified gauge $s_a$ and flavour holonomies $t_i$. It is thus also the Yang-Yang potential of the above integrable system \cite{nekrasov2009quantum,nekrasov2009supersymmetric}. Explicitly, 
\begin{equation}\label{cW}
\begin{aligned}
    \mathcal{W} = &-\sum_{a=1}^N \log s_a \log(-1)^{N-1}\zeta 
    -\sum_{a \neq b}^N \text{Li}'_2\Big(\frac{s_a}{s_b}\Big)
    -\sum_{a,b=1}^N\text{Li}'_2\Big(\frac{s_a}{s_b}t_3^{-1}\Big)\\
    &+\sum_{a=1}^N \left[\text{Li}'_2\Big(s_a\,t_1^{\frac{1}{2}}t_2^{\frac{1}{2}}\Big)+\text{Li}'_2\Big(s_a^{-1}\,t_1^{\frac{1}{2}}t_2^{\frac{1}{2}}\Big)\right] 
    +\sum_{a, b=1}^N \left[\text{Li}'_2\Big(\frac{s_a}{s_b}t_1\Big) + \text{Li}'_2\Big(\frac{s_a}{s_b}t_2\Big)\right],
\end{aligned}
\end{equation}
where we have introduced the notation for the modified dilogarithm:
\begin{equation}
    \text{Li}'_2(x) := \text{Li}_2(x) + \frac{1}{4}\log^2(-x).
\end{equation}
These vacuum equations capture the ring relations of the chiral ring by the fact that the chiral ring is protected under renormalisation. 

Geometrically, we may write the superpotential as:
\begin{equation}\label{eq:geometricSP}
    \mathcal{W} = -\log \mathcal{L} \log\zeta  - \sum_{w \in \mu_{\bC}} \text{Li}'_2(w) - \sum_{ w \in \Delta_{\mathfrak{g}_{\bC}}} \text{Li}'_2(w) + \sum_{w \in T^{*}R} \text{Li}'_2(w).
\end{equation}
where $\Delta_{\mathfrak{g}_{\bC}}$ denotes the set of roots of $\mathfrak{g}_{\bC}$: $\text{ch}\,\Delta_{\mathfrak{g}_{\bC}} = \prod_{a\neq b} s_{ab}$, $\mu_{\bC}$ is shorthand for the weights under which the complex moment map transforms: $\text{ch}\,\mu_{\bC}= \text{ch}\, t \mathfrak{g}_{\bC}^* = \prod_{a, b = 1}^N t s_{ab}$ and $\mathcal{L}$ denotes the determinant line bundle\footnote{For more general theories, this term in the superpotential would include a sum over the Picard group of the Higgs branch.} \eqref{eq:detlinebundle} on $\text{Hilb}^N(\mathbb{C}^2)$.

That the above yields the Bethe equations \eqref{eq:bethe_eqns} can be readily checked by using the fact that
\begin{equation}
    s \frac{d}{ds} \Li_2(s) = -\log(1-s),
\end{equation}
so that for symplectic pairs of $T$-weights $(w,t w^{-1})$, with $w = s_a^{n_a}\ldots $ we have
\begin{equation}
    e^{s_a \partial_{s_a} \left( \text{Li}'_2(w) + \text{Li}'_2(tw^{-1}) \right)} = \left( \frac{\hat{a}(w)}{\hat{a}(t w^{-1})}\right)^{n_a} = \left( t^{-\frac{1}{2}} w  \frac{1-tw^{-1}}{1-w}\right)^{n_a}.
\end{equation}
Our conventions for $\hat{a}$ genera can be found in Appendix \ref{appendix:asymptotics}.

We denote a quantum $K$-theory class associated to a tautological bundle $\mathcal{V}$ as $\hat{\mathcal{V}}$. The quantum $K$-theory ring acts naturally on the localised $K$-theory ring $K_{T}(X_N)^{\text{loc.}} := K_{T}(X_N)\otimes \mathcal{R}$ which as a vector space is $\mathcal{H}_N = \text{sp}_{\mathcal{R}} |\lambda \rangle$. It acts diagonally as:
\begin{equation}
    \hat{\mathcal{V}}|\lambda \rangle = \mathcal{V}_{\lambda}|\lambda \rangle ,
\end{equation}
where now $\mathcal{V}_{\lambda}$ denotes evaluation of $\mathcal{V}$ at the quantum Bethe root associated to $\lambda$. Later in this article, we physically realise this action in the gauge theory.

\paragraph{Supersymmetric indices.}

We consider the 3d $\mathcal{N}=4$ ADHM quiver theory, topologically twisted by $R_C$ (\textit{i.e.} B-twisted), on a hemisphere geometry $HS^2 \times S^1$. The path integral produces the space of states $\mathcal{H}_N$ on the torus $T^2$ boundary. We are interested in computing the graded index
\begin{equation}
    \mathcal{I}_{HS^2} = \text{Tr}_{HS^2}\,\left[ (-1)^F q^{J} t_1^{F_1}  t_2^{F_2} \zeta^{F_C}\right].
\end{equation}
The grading by angular momentum is implemented by a twisted periodicity on the matter fields parameterised by $q = e^{-\ep}$, see \textit{e.g.} \cite{Zhang:2022wwy}. Alternatively, it can be implemented by a deformation of the metric, so that $HS^2$ fibres non-trivially as one completes a cycle on $S^1$ \cite{Benini:2013yva}. In this latter picture, the boundary torus corresponds to $E_{\tau} = \bC^*/q^{\bZ}$, where $q=e^{2 \pi i \tau}$ and $\tau$ is the complex structure of the torus. The flavour grading is implemented by introducing complexified holonomies as described above.

In this setup, one may also insert an element of the ring of Wilson lines $\mathcal{O}$, corresponding to an element of $QK_T(X)$, at the tip of $HS^2$, wrapping the remaining $S^1$ cycle. In the $q\to 1$ limit, inserting an operator $\mathcal{O}$ at the pole and computing the index yields the quantity $\langle 0 | \hat{\mathcal{O}}|0 \rangle = \langle 0 | \mathcal{O} \rangle$ in the Hilbert space $\mathcal{H}_{N}$ described above.

\paragraph{Boundary conditions.}
We must also specify a boundary condition on the torus boundary of $HS^2 \times S^1$. In this work, we consider exceptional Dirichlet boundary conditions $\mathcal{B}_{\lambda}$ which are labelled by vacua---in our case partitions $\lambda$ with $|\lambda|=N$ \cite{Crew:2020psc}, we will recap their definition in the following. 

Boundary conditions on the boundary torus may be considered states $|\lambda \rangle_q$ living in the elliptic cohomology of $X_N$ \cite{Bullimore:2021rnr}. We will momentarily take the $q \to 1$ limit of elliptic cohomology where we recover the localised $K$-theory ring $K_{T}(X_N)^{\text{loc.}}$ and may consider the exceptional Dirichlet boundary conditions as giving the localised fixed point basis, up to a normalisation we specify. We now recall from \cite{Crew:2020psc} the definition of this boundary condition for $\text{Hilb}^N(\mathbb{C}^2)$. We fix a chamber $\mathfrak{C} = \{|z|>1\}$ for the equivariant parameters. We then split the weight space of all $\mathcal{N}=4$ hypermultiplets $T^*R$ according to this choice as
\begin{equation}
    T^*R = Q_{\lambda}^+ + \bar{Q}_{\lambda}^0 + Q_{\lambda}^0  + Q_{\lambda}^{-} \,.
\end{equation}
In our case we find\footnote{These are the weight spaces relevant for the later hemisphere partition function evaluation.}
\begin{equation}\label{eq:adhmbc}
\begin{aligned}
    Q_{\lambda}^+ + \bar{Q}_{\lambda}^0 &= t^{1/2}\sum_{a \in \lambda} s_a + z^{-1} t^{1/2}\sum_{a,b \in I_1}  \frac{s_a}{s_b} + z t^{1/2}\sum_{a,b \in I_2} \frac{s_a}{s_b} \\
    Q_{\lambda}^- + Q_{\lambda}^0 &= t^{1/2}\sum_{a \in \lambda} s_a^{-1} + z t^{1/2}\sum_{a,b \in I_1}  \frac{s_b}{s_a} + z^{-1}t^{1/2}\sum_{a,b \in I_2} \frac{s_b}{s_a}
\end{aligned}
\end{equation}
and
\begin{equation}\label{eq:massive_weights}
    Q_{\lambda}^0 = t^{1/2} s_{(1,1)}^{-1} + \sum_{ \substack{a,b| i_a = i_b +1 \\\cap \, j_a = j_b}}  z t^{1/2} \frac{s_b}{s_a} + \sum_{ \substack{a,b| a \in \lambda^B \\ \, \cap\, i_a = i_b \\ \cap \, j_a = j_b+1}}  z t^{1/2} \frac{s_b}{s_a}
\end{equation}
where $I_1$ and $I_2$ are subsets of pairs of boxes in young diagrams given by
\begin{equation}
\begin{aligned}
    I_1 &= \{ (a,b) \in \lambda \times \lambda \, : \, (b \in \lambda^B) \cap (i_a > i_b) \; \text{or} \; b \notin \lambda^B \}, \\
    I_2 &= \{ (a,b) \in \lambda \times \lambda \, : \, (a \in \lambda^B) \cap (i_b \le i_b)\},
\end{aligned}
\end{equation}
and $\lambda_B$ denotes the set of boxes along the bottom-most edge of the Young tableau. We note that the weight spaces $Q_{\lambda}^+$, $\bar{Q}_{\lambda}^0$ are paired with $Q_{\lambda}^-$, $Q_{\lambda}^0 $ respectively under $\omega \to t \omega^{-1}$. Thus, this defines a holomorphic Lagrangian splitting of the space $T^*R$. Further, $Q_{\lambda}^0$ correspond precisely to the $\cN=2$ chiral multiplets which obtain non-zero expectation values in the vacuum $\lambda$.

Note that the set of weights in $Q_{\lambda}^0$ define a tree in the Young tableau $\lambda$ \cite{Crew:2020psc}. Solving the set of equations $w = 1$ for $w \in Q_{\lambda}^0$ specifies uniquely:
\begin{equation}\label{eq:boundaryvev}
    s_a = t_1^{i_a-\frac{1}{2}}t_2^{j_a-\frac{1}{2}},
\end{equation}
which are solutions to the classical Bethe equations \eqref{eq:classical soln BAE}, and also the fixed point evaluation of the fugacities; they are the values that the gauge fugacities must take in order for scalar components of chirals in $Q_{\lambda}^0$ to take their non-zero values in the vacuum $\lambda$. More precisely, the first term in \eqref{eq:massive_weights} sets $s_{(1,1)} = t^{\frac{1}{2}}$. Then, the remaining terms correspond to edges linking boxes $a$ and $b$ in the tree, which determines the value of $s_a$ from the value of $s_b$.

Physically to describe $\cB_{\lambda}$, the split specifies which chirals we give Dirichlet boundary conditions and which we give Neumann. In particular, the chiral multiplets with weights in $Q_{\lambda}^+ + \bar{Q}_{\lambda}^0$ are given Neumann boundary conditions, and those in $Q_{\lambda}^0  + Q_{\lambda}^{-}$ are given Dirichlet boundary conditions. The chirals in $Q_{\lambda}^{-}$ are set to zero at the boundary, whilst those in $Q_{\lambda}^0$ are given a non-zero expectation value. These chirals correspond precisely to those which obtain a non-zero expectation value in the vacuum $\lambda$ on the Higgs branch. Turning on these expectation values at the boundary breaks the boundary symmetry which descends from the bulk $U(N)$ gauge symmetry, setting the boundary values of the gauge fields in accordance with \eqref{eq:boundaryvev}, whilst preserving $F_{1,2}$. Thus, these are what are referred to as \textit{exceptional} Dirichlet boundary conditions in \cite{bullimore2016boundaries}.

More details on the physical construction may be found in the appendix of \cite{Crew:2020psc}. In particular, these exceptional Dirichlet boundary conditions are chosen in order to have a Higgs branch image which coincides with the holomorphic attracting Lagrangian submanifold of a fixed point $\lambda$ under Morse flow induced by the equivariant parameter $z$. They realise an exact holomorphic factorisation \cite{Bullimore:2020jdq} of closed $3$-manifold partition functions, which we return to momentarily.

With this choice of chamber and split the tangent bundle \eqref{eq:preevaltangent} also splits into positive and negative directions after evaluation at a fixed point \eqref{eq:boundaryvev} according to
\begin{equation}
    T_{\lambda}X_N = T_{\lambda}^+ X_N + T_{\lambda}^- X_N ,
\end{equation}
where we have
\begin{equation}
\begin{split}
    T_{\lambda}^+X_N &= \sum_{s\in \lambda} z^{a_{\lambda}(s) + l_{\lambda}(s) +1}t^{\frac{1}{2}(l_{\lambda}(s)-a_{\lambda}(s)+1)}, \\  T_{\lambda}^- X_{N} &= \sum_{s\in \lambda}z^{-a_{\lambda}(s)-l_{\lambda(s)}-1}t^{\frac{1}{2}(a_{\lambda}(s)+l_{\lambda}(s)+1)}\,.
\end{split}    
\end{equation}

\paragraph{Localisation.}
In the work \cite{Crew:2023tky} by two of the authors, it is described how to compute the above physical quantities via localisation in terms of Higgs branch geometry, which we now briefly review.
The hemisphere partition function with a Wilson line operator insertion in the presence of an exceptional Dirichlet boundary condition $\mathcal{B}_{\lambda}$ is given more explicitly by
\begin{equation}\label{eq:Jacksonintegral}
    H_{\lambda}^{(\mathcal{O})} = \int d_q s\, \mathcal{O}(x,s)\, e^{\frac{\varphi(s,\zeta)}{\log q}} \hat{a}\left(\frac{1-t^{-1}q}{1-q}(Q_{\lambda}^+ + \bar{Q}_{\lambda}^0 - t\mathfrak{g}_{\mathbb{C}})\right).
\end{equation}
The $\hat{a}$ genus is defined multiplicatively for collections of weights---our conventions can be found in appendix \ref{appendix:asymptotics}. In the above prescription the exponential pre-factor is given by
\begin{equation}
    \varphi(s,\zeta) = \log \zeta \log \mathcal{L} \,,
\end{equation}
where $\mathcal{L}$ is the determinant line bundle \eqref{eq:detlinebundle}, given by $\mathcal{L}=s_1s_2\ldots s_N$ in coordinates. The measure  $d_q s$ denotes a formal Jackson $q$-integral that instructs us to sum over $q$-shifts of the Chern root evaluations \eqref{eq:classical soln BAE}.
Namely
\begin{equation}
    \int d_qs\, f(s_a) = \sum_{\vec{k} \in \mathbb{Z}^N} f(t_1^{i_a -1/2}t_2^{j_a - 1/2} q^{k_a})
\end{equation}
The inserted Wilson line $\mathcal{O}(x,s)$ is an element of the representation ring $K_{T\times U(N)}(\text{pt.})$ i.e. a polynomial in gauge and flavour fugacities (and their inverses) symmetrised over the gauge group. The weight spaces entering above are the same that enter the boundary condition prescription in the previous subsection. Geometrically the partition function computes the equivariant Euler character of the symmetrised virtual structure sheaf on the space of quasimaps to the Higgs branch with a descendant insertion $\chi_T(\mathsf{QM},\mathcal{O}\otimes \mathcal{V})$ with a particular physically motivated normalisation. More details on the relationship between the hemisphere partition function and the enumerative geometry of Higgs branches are reviewed in \cite{Crew:2023tky, Dedushenko:2023qjq}.

We now consider evaluating the hemisphere partition function. One may equivalently express the integrand of \eqref{eq:Jacksonintegral} in terms of regularised $q$-Pochhammer symbols as follows. The necessary zeta function regularisation\footnote{Our regularisation differs slightly to the usual zeta function regularisation for 3d $\mathcal{N}=2$ theories as in \textit{e.g.} \cite{Tanaka:2014oda} as it makes use of a phase ambiguity present only in the regularisation of 1-loop determinants of 3d $\mathcal{N}=4$ multiplets.} is discussed in further detail in appendix \ref{appendix:asymptotics}. We first substitute the split \eqref{eq:adhmbc} into our formula \eqref{eq:Jacksonintegral}, upon which we see that the integrand before evaluation is given by 
\bea\label{eq:qJackson_integrand}
&e^{\frac{\log\zeta}{\log q}\sum_{a\in\lambda}\log(s_a)}
\frac{\prod_{a,b\,\in\lambda}\left(t_1t_2\,s_{ab};q\right)'_\infty}{\prod_{a,b\,\in\,\lambda}(s_{ab}\,q;q)'_\infty}
\prod_{a\in\lambda}\frac{\left(t_1^{-\frac{1}{2}}t_2^{-\frac{1}{2}}s_a q;q\right)'_\infty}{\left(t_1^{\frac{1}{2}}t_2^{\frac{1}{2}}s_a;q\right)'_\infty}\\
&\prod_{\substack{a,b\,\in\lambda\;\text{s.t.} \\ (b\,\in\lambda^B)\cap (i_a > i_b) \\ \text{or}\, (b\,\notin \lambda^B)}}\frac{\left(t_1^{-1}s_{ab}\,q;q\right)'_\infty}{\left(t_2\,s_{ab};q\right)'_\infty}\prod_{\substack{a,b\,\in\lambda\,\text{s.t.} \\ (a\,\in\lambda^B)\cap (i_b\leq i_a)}}\frac{\left(t_2^{-1}s_{ab}\,q;q\right)'_\infty}{\left(t_1\,s_{ab};q\right)'_\infty}\eea
In the above, we have modified $q$-Pochhammer symbols defined as:
\begin{equation}
    (a;q)'_{\infty} = e^{-\mathcal{E}[-\log(- a)]}(a;q)_{\infty}
\end{equation}
where
\begin{equation}
    \mathcal{E}[x] = \frac{\ep}{24}-\frac{x}{4} + \frac{x^2}{4\ep}
\end{equation}
and recall $q = e^{-\ep}$.

The Jackson $q$-integral instructs us to evaluate the integrand on $q$-shifted solutions to the classical Bethe equations \eqref{eq:classical soln BAE}, $s_a \mapsto  t_1^{i_a-\frac{1}{2}}t_2^{j_a-\frac{1}{2}} q^{k_a}$ where $k_a \in \bZ$, and summing over $k \in \bZ^N$. It turns out that only $k$ forming a reverse plane partition over $\lambda$ gives non-zero contributions. The result of the evaluation is
\begin{equation}
    H_{\lambda}^{(\mathcal{O})}(q,t) = e^{\frac{\varphi(t,\zeta)}{\log q}}\hat{a}\left( \frac{1-t^{-1}q}{1-q}N_{\alpha}^{+}\right)V_{\lambda}^{(\mathcal{O})}(q,t),
\end{equation}
where $\mathsf{V}_{\lambda}^{(\mathcal{O})}(q,t)$ is the vertex function (see for example \cite{smirnov2016rationality}) normalised such that
\begin{equation}
    \mathsf{V}_{\lambda}(q,t) = 1 + O(\zeta)\,.
\end{equation}
The explicit expression in terms of a vortex sum over reverse plane partitions may be found in \textit{e.g.} \cite{Crew:2020psc} for the evaluated index however we will not make use of this form in the following.

\paragraph{Integral form.}
We now re-write the $q$-Jackson integral \eqref{eq:Jacksonintegral} as a contour integral amenable to saddle point analysis. We note first the identity
\begin{equation}\label{eq:0weight_contribution}
    \frac{(t;q)_{\infty}'}{(q;q)_{\infty}'}  \left. \frac{(q w^{-1}q^n;q)_{\infty}'}{(tw^{-1}q^n;q)_{\infty}'}\right|_{w=1} =  (qt^{-1})^{\frac{n}{2}}\frac{(t;q)_n}{(q;q)_n}.
\end{equation}
where the first fraction on the left-hand side comes from the abelian (diagonal) part of the vector multiplet contribution. The $q$-Jackson integrand \eqref{eq:qJackson_integrand} contains a product of such terms, one for each $w \in Q_{\lambda}^0$. We note also the identity 
\begin{equation}\label{eq:0weight_replacement}
    \underset{w=q^{-n}}{\text{Res}}\, \frac{1}{2\pi i s} \frac{(q;q)_{\infty}''}{(qt^{-1};q)_{\infty}'} \frac{(qt^{-1}w;q)_{\infty}'}{(w;q)_{\infty}'} = \mp (qt^{-1})^{\frac{n}{2}}\frac{(t;q)_n}{(q;q)_n},
\end{equation}
if $w = s^{\pm 1} (...)$. Thus we can replace each term of the form \eqref{eq:0weight_contribution} from $Q_{\lambda}^0$,  with the above, and trade the $q$-Jackson integral for a contour integral. In the above the double prime notation indicates the regularisation $(q;q)'' = e^{-\mathcal{E}[-\log(-1)]} (q;q)$.

The integral form of the hemisphere partition function is
\begin{equation}\label{eq:contourintegralform}
    H_{\lambda} = \int_{\Gamma} \prod_{a=1}^N\frac{ds_a}{2\pi is_a} e^{\frac{\log\zeta}{\log q}\sum_{a\in\lambda}\log(s_a)} \, \Phi,
\end{equation}
where the integrand is
\begin{equation}
    \Phi =  \left[ \frac{(q;q)_{\infty}''}{(qt^{-1};q)_{\infty}'}  \right]^N \prod_{a\neq b} \frac{(t s_a/s_b;q)'_{\infty}}{(q s_a/s_b;q)'_{\infty}} \prod_{w \in Q_{\lambda}^{+} + Q_{\lambda}^0} \frac{(qt^{-1}w;q)_{\infty}'}{(w;q)_{\infty}'} 
\end{equation}
and the contour $\Gamma$ encloses the poles $s_a =  t_1^{i_a-\frac{1}{2}}t_2^{j_a-\frac{1}{2}} q^{k_a} $, $a =1,\ldots,N$.

\paragraph{Classical limit.}
We now consider the classical limit\footnote{In the sense of semi-classical quantisation of the variety obtained from the spectrum of the quantum $K$-theory ring.} $\epsilon \to 0$ with $q = e^{-\epsilon}$ of the integral form of the hemisphere partition function. The main asymptotic result we require is that the regularised $\hat{a}$-genus of a non-zero weight satisfies to leading order
\begin{equation}
    \lim_{\epsilon \to 0}\hat{a}\left( \frac{1-t^{-1}q}{1-q}w\right) = \hat{a}(w)^{1/2}\hat{a}(tw^{-1})^{1/2} e^{ \frac{1}{\epsilon} \left( \Li_2'(w) + \Li_2'(t w^{-1}) + \frac{\pi^2}{6}\right) + \ldots}
\end{equation}
where $\ldots$ indicates $O(\ep)$ terms. We will often drop the ellipses when taking limits in the following. Thus if we take a polarisation $T^*W = W +  t  W^*$, where $R$ is some collection of weights, the weights pair and we have
\begin{equation}
    \lim_{\epsilon \to 0} \prod_{w \in W}\hat{a}^{\frac{1}{2}}\left( \frac{1-t^{-1}q}{1-q}w\right) = \prod_{w\in T^*W}\hat{a}(w)e^{\frac{1}{\epsilon}\left(\Li_2(w) + \frac{1}{4}\log^2(-w) + \frac{\pi^2}{6}\right) + \ldots}
\end{equation}
Using this, we have the asymptotics of the parts of the integrand $\Phi$:
\begin{equation}
\begin{aligned}
    \prod_{w \in Q_{\lambda}^{+} + Q_{\lambda}^0} \frac{(qt^{-1}w;q)_{\infty}'}{(w;q)_{\infty}'}  &\rightarrow e^{\frac{\pi^2}{6\ep}(N^2+N)} \prod_{w\in T^*R}\hat{a}(w)^{\frac{1}{2}} e^{\frac{1}{\epsilon}\Li'_2(w)},\\
    \prod_{a\neq b} \frac{(t s_a/s_b;q)'_{\infty}}{(q s_a/s_b;q)'_{\infty}} &\rightarrow e^{-\frac{\pi^2}{6\ep}(N-N^2)} \frac{e^{-\frac{1}{\ep}\left( \Li'_2(s_a/s_b) + \Li'_2( t s_a/s_b)\right)}}{\left[ \hat{a}(ts_a/s_b) \hat{a}(s_a/s_b)\right]^{\frac{1}{2}}} \\
\end{aligned}
\end{equation}
and
\begin{equation}\label{eq:diagonal_asymptotics}
    \left[ \frac{(q;q)_{\infty}''}{(qt^{-1};q)_{\infty}'}  \right]^N  \rightarrow \left(\frac{2\pi}{\ep}\right)^{\frac{N}{2}}e^{\frac{ N \pi^2}{3 \ep}} e^{\frac{N \pi i }{2}} e^{\frac{N \pi^2}{4\ep}} \hat{a}(t)^{-\frac{N}{2}}e^{-\frac{1}{\ep} N \Li_2'(t)}
\end{equation}
We thus see that the leading $q \to 1$ asymptotics of the integral \eqref{eq:contourintegralform} is given by
\begin{equation}\label{eq:limitcontourintegral}
    \lim_{\epsilon \to 0 }H_{\lambda} 
    \sim \int_{\Gamma}  \prod_{a=1}^N\frac{ds_a}{2\pi is_a}  \left(-\frac{2\pi}{\ep}\right)^{\frac{N}{2}}  \, \frac{\hat{a}(T^*R)^{1/2}}{\hat{a}(\mu_{\mathbb{C}})^{1/2}\hat{a}(\Delta_{\mathfrak{g}})^{1/2}} e^{\frac{1}{\epsilon} \mathcal{W}} 
\end{equation}
where the superpotential is as in \eqref{eq:geometricSP}.
\begin{equation}
    \mathcal{W} = \sum_{w \in T*R } \Li_2'(w) - \sum_{w \in \Delta_{\mathfrak{g}_{\bC}} + \mu_{\bC}} \Li_2'(w).
\end{equation}
We have dropped a factor of $e^{\frac{ N \pi^2}{4\ep}}$, coming from \eqref{eq:diagonal_asymptotics}, since it is a constant and glues to $1$ under $\ep \rightarrow -\ep$.

We conjecture, following also \cite{dinkins2022exotic}, that the $\ep \rightarrow 0 $ asymptotics of $H_{\lambda}$ are dominated by the saddle point approximation for a saddle point (which we also label by $\lambda$), which solves the Bethe equations \eqref{eq:bethe_eqns} and reduces to the classical solutions $s_a = t_1^{i_a-1/2}t_2^{j_a-1/2}$ corresponding to a Young Tableaux, as $\zeta \rightarrow 0$. The saddle point formula thus gives
\begin{equation}\label{eq: saddle pt hemisphere}
    \langle 0 | \hat{\mathcal{O}} | \lambda \rangle = \lim_{\epsilon \to 0} H_{\lambda}^{(\mathcal{O})} = \left.\frac{\hat{a}^{1/2}(T^*R)}{\hat{a}^{1/2}(\mu_{\mathbb{C}})\hat{a}^{1/2}(\Delta_{\mathfrak{g}_{\mathbb{C}}})}\right\rvert_{\lambda} \frac{\mathcal{O}_{\lambda}}{\sqrt{\det (- \partial^2 \mathcal{W}_{\lambda})}}e^{\frac{1}{\epsilon}\mathcal{W}_{\lambda}},
\end{equation}
where $(\partial^2 \mathcal{W})_{ab}$ denotes the Hessian matrix of $\mathcal{W}$ with elements $\partial_{u_a} \partial_{u_b} W$ and $\rvert_{\lambda}$ denotes evaluation of the Chern roots on the quantum Bethe root.

\paragraph{Factorisation.}
We may thus compute expectation values of quantum $K$-theory classes by inserting a complete set of states 
\begin{equation}\label{eq: vev factorisation}
    \langle 0 | \hat{\mathcal{O}} | 0 \rangle =  \sum_{\lambda} \langle 0| \hat{\mathcal{O}} | \lambda \rangle \langle \lambda | 0 \rangle 
\end{equation}
We then use the asymptotics result above and the fact that reversing the orientation corresponds to $\epsilon \to -\epsilon$ to find
\begin{equation}\label{eq:indexformula}
    \langle 0 | \mathcal{O} | 0 \rangle  = \sum_{\lambda} \mathcal{O}_{\lambda} \left.\frac{\hat{a}(T^*R)}{\hat{a}(\mu_{\mathbb{C}})\hat{a}(\Delta_{\mathfrak{g}_{\mathbb{C}}})}\right\rvert_{\lambda} \frac{1}{\det(-\partial^2 \mathcal{W}_{\lambda})},
\end{equation}
where we evaluate on solutions to the quantum Bethe equations. In the following we will interpret this physically as factorising the twisted index and the above formula corresponds to the localisation formula in (7.26) of \cite{closset2019three} for the twisted index.

\section{The Twisted Index}
In this section, we discuss the twisted index on $S^2 \times_B S^1$ and its factorisation into hemisphere partition functions in the presence of a line operator insertion.

\paragraph{The twisted index.}
We now consider the topologically twisted index on geometry $S^2 \times_B S^1$, without angular momentum refinement ($q$). The twist is carried out with respect to the $R_C$ R-symmetry. Following \cite{benini2015topologically}, the index is defined as
\begin{equation}
    \mathcal{I}^B_{S^2}(t_1,t_2,\zeta;\fn_1,\fn_2,\fn_C) = \text{Tr}_{\mathcal{H}_{S^2}^B} \left[ (-1)^Ft_1^{F_1}t_2^{F_2}\zeta^{F_C}\right],
\end{equation}
where $\cH_{S^2}^B$ denotes states on $S^2$ in the presence of a topological twist and additional flavour fluxes $(\fn_{1},$ $\fn_{2},$ $\fn_C)$, which are annihilated by a pair of supercharges $Q_+^{1 \dot{1}}, Q_-^{2\dot{2}}$.  The Dirac quantisation conditions for the fluxes impose $\fn_{1,2},\fn_C\in\mathbb{Z}$, $\fn_1+\fn_2\in 2\mathbb{Z}$. The index may be localised to give an integral expression, which takes the form of a Jeffreys-Kirwan integral:
\begin{equation}\label{eq:cb_localisation_twisted_index}
\begin{aligned}
    \mathcal{I}^B_{S^2}=& \frac{(-1)^N}{N!}\sum_{\fm\in\mathbb{Z}^N}\int_\text{JK}\prod_{a=1}^N\frac{ds_a}{2\pi is_a}\;(-\zeta)^{\sum_{a=1}^N\fm_a}\bigg(\prod_{a=1}^N s_a\bigg)^{\fn_C}\prod_{\substack{a,b=1\\ a\neq b}}^N\Bigg(\frac{s_{ab}^\frac{1}{2}}{1-s_{ab}}\Bigg)^{\fm_a-\fm_b-1}\\
    &\prod_{a,b=1}^N\Bigg(\frac{s_{ab}^\frac{1}{2}t_1^\frac{1}{2}}{1-s_{ab}t_1}\Bigg)^{\fm_a-\fm_b+\fn_1+1}\Bigg(\frac{s_{ab}^\frac{1}{2}t_2^\frac{1}{2}}{1-s_{ab}t_2}\Bigg)^{\fm_a-\fm_b+\fn_2+1}\Bigg(\frac{s_{ab}^\frac{1}{2}t_3^\frac{1}{2}}{1-s_{ab}t_3}\Bigg)^{\fm_a-\fm_b-\fn_1-\fn_2-1}\\
    &\prod_{a=1}^N\Bigg(\frac{s_a^\frac{1}{2}t_3^{-\frac{1}{4}}}{1-s_at_3^{-\frac{1}{2}}}\Bigg)^{\fm_a+\frac{1}{2}(\fn_1+\fn_2)+1}\Bigg(\frac{s_a^{-\frac{1}{2}}t_3^{-\frac{1}{4}}}{1-s_a^{-1}t_3^{-\frac{1}{2}}}\Bigg)^{-\fm_a+\frac{1}{2}(\fn_1+\fn_2)+1}\,,
\end{aligned}
\end{equation}
where $s_{ab} := s_a/s_b$ and $t_3=(t_1t_2)^{-1}$ as before. Turning on the flavour fluxes $\fn_{1,2}$ is equivalent to topologically twisting by the R-symmetry $R_C-\fn_1F_1-\fn_2F_2$.

In the previous work \cite{Crew:2020psc} of two of the authors, it is shown that the $B$-twisted index with angular momentum refinement factorises exactly into hemisphere partition functions.\footnote{More precisely, this factorisation was demonstrated with a zeta function regularisation of $\hat{a}$ producing modified $q$-Pochhammers $(a;q)'_{\infty} := e^{-\mathcal{E}[-\log(a)]}(a;q)_{\infty}$. In this work we have used the regularisation $(a;q)'_{\infty} :=  e^{-\mathcal{E}[-\log(-a)]}(a;q)_{\infty}$. In the new regularisation, for each term
\begin{equation}
    \frac{(qt^{-1}w ;q)'_{\infty} }{(w;q)_{\infty}'} 
\end{equation}
appearing in the $q$-Jackson integral, we obtain an extra factor of $e^{\frac{\pi i }{2\ep} \log(qt^{-1})}$. The contribution from the new regularisation of the vector multiplet contributions is the reciprocal of this. Thus, overall, the new regularisation produces an extra factor of $e^{\frac{ N \pi i }{2\ep} \log(qt^{-1})} $. With flux for $t$, we have $t \rightarrow t q^{\fn_t/2}$. Under the twisted index gluing, gluing the new $H_{\lambda}$ will just produce a factor of $(-1)^N e^{\frac{N \fn_t \pi i}{2}}$ relative to gluing the index from \cite{Crew:2020psc}. One may redefine the twisted index to absorb this constant factor. } The twisted index without the angular momentum deformation is then recovered in the limit:
\begin{equation}
    \mathcal{I}^B_{S^2}(t_i, \zeta;\fn_i, \fn_C)  = \lim_{q\rightarrow 1}\sum_{\lambda} H_{\lambda}\left(t_i q^{\frac{\fn_i}{2}}, \zeta q^{\frac{\fn_C}{2}} ;q \right) H_{\lambda}\left(t_i q^{-\frac{\fn_i}{2}}, \zeta q^{-\frac{\fn_C}{2}} ;q^{-1} \right).
\end{equation}

\paragraph{Geometric interpretation.}
The twisted index may be re-expressed in terms of Higgs branch geometry as follows. We first recall the twisted superpotential is given in terms of tangent weights as 
\begin{equation}
    \mathcal{W} = -\log \zeta \log \mathcal{L} + \sum_{w \in T^*R} \Li'_2(w) - \sum_{w \in \mu_{\mathbb{C}}}  \Li'_2(w) - \sum_{w \in \Delta_{\mathfrak{g}_{\mathbb{C}}}} \Li_2'(w) \,,
\end{equation}
with $\mathcal{L} = s_1 \ldots s_N$. Each weight $w$ in this expression is a monomial in $t_1,t_2,t_3$ and $s_a$. The operator we are then interested in is
\begin{equation}
    \left( \sum_{a=1}^N \mathfrak{m}_a s_a \partial_{s_a} + \mathfrak{n}_1 t_1 \partial_{t_1}+ \mathfrak{n}_2 t_2 \partial_{t_2}  - \mathfrak{n}_C \zeta \partial_{\zeta}\right)\cdot \mathcal{W}\,.
\end{equation}
The twisted index may be expressed as
\begin{equation}
    \mathcal{I}_{S^2}^B = \frac{1}{N!}\sum_{\mathfrak{m} \in \mathbb{Z}^N} \oint  \prod_{a=1}^N\frac{ds_a}{2 \pi is_a} \frac{\hat{a}(T^*R)}{\hat{a}(\mu_{\mathbb{C}})\hat{a}(\Delta_{\mathfrak{g}_{\mathbb{C}}})} \exp\left(\left(\sum_{a=1}^N \mathfrak{m}_a s_a \partial_{s_a} + \mathfrak{n}_1 t_1 \partial_{t_1}+ \mathfrak{n}_2 t_2 \partial_{t_2}  - \mathfrak{n}_C \zeta \partial_{\zeta}\right)\cdot \mathcal{W}\right)\,.
\end{equation}
With a specific choice of covector in the JK prescription, the fluxes which contribute are bounded as $\fm_i<M-1$ where $M$ is some large integer, see \cite{benini2015topologically} for further details. We can then perform the geometric sum over $\mathbb{Z}^N$ to give
\begin{equation}
    \mathcal{I}_{S^2}^B = \frac{1}{N!}\oint \prod_{a=1}^N \frac{ds_a}{2 \pi i s_a} \frac{1}{1-e^{s_a \partial_{s_a} \mathcal{W}}}\frac{\hat{a}(T^*R)}{\hat{a}(\mu_{\mathbb{C}})\hat{a}(\Delta_{\mathfrak{g}_{\mathbb{C}}})}\exp\left(\left(\mathfrak{n}_1 t_1 \partial_{t_1}+ \mathfrak{n}_2 t_2 \partial_{t_2}  - \mathfrak{n}_C \zeta \partial_{\zeta}\right)\cdot \mathcal{W}\right)\,.
\end{equation}
The residue evaluation of this integral then involves contributions from the Bethe equations \eqref{eq:bethe_eqns}. Evaluating this by the residue theorem we find
\begin{equation}\label{eq:twistedindexresidues}
    \mathcal{I}_{S^2}^B(\fn_1,\fn_2,\mathfrak{n}_C) = \sum_{\lambda} \left.\frac{\hat{a}(T^*R)}{\hat{a}(\mu_{\mathbb{C}})\hat{a}(\Delta_{\mathfrak{g}_{\mathbb{C}}})}\right\rvert_{\lambda}\frac{1}{\det (-\partial^2 \mathcal{W}_{\lambda})} \mathcal{O}_{\lambda}(\fn_1,\fn_2,\mathfrak{n}_C),
\end{equation}
where $\mathcal{O}_{\lambda}(\fn_1,\fn_2,\mathfrak{n}_C)$ denotes the flux-adding operator corresponding to the flux insertion, evaluated on the Bethe root. This localisation formula is in agreement with the index formula \eqref{eq:indexformula} derived by factorisation above. 

\subsection{Flux \& Line Operators}
We now discuss the relationship between background flux in the twisted index and line operator or $QK_T(X)$ class insertions in more detail. In the previous section we showed that the Coulomb branch localisation formula may be interpreted geometrically in terms of the Higgs branch geometry as
\begin{equation}\label{index eval on roots}
    \mathcal{I}_{S^2}^B(\fn_1,\fn_2,\mathfrak{n}_C) = \sum_{\lambda} \left.\frac{\hat{a}(T^*R)}{\hat{a}(\mu_{\mathbb{C}})\hat{a}(\Delta_{\mathfrak{g}_{\mathbb{C}}})}\right\rvert_{\lambda}\frac{1}{\det (-\partial^2 \mathcal{W}_{\lambda})} \mathcal{O}_{\lambda}(\fn_1,\fn_2,\mathfrak{n}_C)
\end{equation}
with
\begin{equation}
    \mathcal{O}(\fn_1,\fn_2,\mathfrak{n}_C) = \exp\left(\left(\mathfrak{n}_1 t_1 \partial_{t_1}+ \mathfrak{n}_2 t_2 \partial_{t_2}  - \mathfrak{n}_C \zeta \partial_{\zeta}\right)\cdot \mathcal{W}\right)
\end{equation}
This operator is a flux-adding operator for the flavour symmetries of the theory. It may be considered as a 't Hooft line operator, since in the twist the flux may be localised to a point on $S^2$, and wrapping the $S^1$ \cite{closset2019three}. We will refer to these operators as 't Hooft line operators from hereon out.

Let us first consider the case of topological flux only so that $\mathfrak{n}_C$ is turned on and $\mathfrak{n}_{1,2} = 0$. In that case we have
\begin{equation}
    \mathcal{O}(\mathfrak{n}_C) = \exp \left( -\mathfrak{n}_C \zeta \partial_{\zeta} \mathcal{W} \right) = \left( s_1 s_2 \ldots s_N\right)^{\mathfrak{n}_C},
\end{equation}
hence inserting $\mathcal{O}(\mathfrak{n}_C)$ is equivalent to a descendant insertion of $\mathfrak{n}_C$ powers of the determinant line bundle $\mathcal{L}$. Hence, according to the discussion in section \ref{subsec:quantumKtheory}, the $B$-twisted index with $\mathfrak{n}_C$ units of flux computes the vacuum expectation value of $\mathcal{L}^{\mathfrak{n}_C}$ in the quantum $K$-theory ring:
\begin{equation}
    \mathcal{I}_{S^2}^{B}(\mathfrak{n}_C) = \langle \mathcal{L}^{\otimes \mathfrak{n}_C} \rangle 
\end{equation}
In section \ref{subsec:quantumKtheory} we discussed how the twisted index/expectation value may also be sliced open and factorised according to
\begin{equation}
    \langle \mathcal{L}^{\otimes \mathfrak{n}_C} \rangle = \sum_{\lambda}\langle \mathcal{L}^{\otimes \mathfrak{n}_C} | \lambda \rangle \langle \lambda | 0 \rangle = \lim_{q \to 1} \sum_{\lambda} H_{\lambda}^{(\mathcal{L}^{\otimes \mathfrak{n}_C})}(q,t)H_{\lambda}(q^{-1},t)
\end{equation}
where $H^{(\mathcal{L}^{\otimes \mathfrak{n}_C})}_{\lambda}(q,t)$ denotes a hemisphere partition function \eqref{eq:Jacksonintegral} with the Wilson line $\mathcal{L}=s_1s_2\ldots s_N$ inserted or, in the enumerative geometry language, a vertex function with descendant insertion. 

It is an immediate consequence of the localisation formula \eqref{eq:Jacksonintegral} that this line operator insertion may be realised by a $q$-difference operator acting on the hemisphere partition function. Namely
\begin{equation}
    H^{(\mathcal{L}^{\mathfrak{n}_C})}_{\lambda}(q,t) = \hat{p}_{C}^{\mathfrak{n}_C} \cdot H_{\lambda}(q,t),
\end{equation}
where $\hat{p}_C$ is a $q$-difference operator acting by $\hat{p}_C: \zeta \to q \zeta$. The expression then follows from the fact:
\begin{equation}
    \hat{p}_{C} \cdot e^{\varphi(s,\zeta) / \log q} = (s_1s_2 \ldots s_N)e^{\varphi(s,\zeta) / \log q}.
\end{equation}
We now consider the $q\to 1$ limit of this action. Recall that 
\begin{equation}
    \lim_{q\to 1}H_\lambda(q,t) = \left.\frac{\hat{a}^{1/2}(T^*R)}{\hat{a}^{1/2}(\mu_{\mathbb{C}})\hat{a}^{1/2}(\Delta_{\mathfrak{g}_{\mathbb{C}}})}\right\rvert_{\lambda} \frac{1}{\sqrt{\det (-\partial^2 \mathcal{W}_{\lambda})}}e^{\frac{1}{\epsilon}\mathcal{W}_{\lambda}},
\end{equation}
and thus
\begin{equation}
    \lim_{q\to 1}\hat{p}_C^{\mathfrak{n}_C}\cdot H_\lambda(q,t) = \left.\frac{\hat{a}^{1/2}(T^*R)}{\hat{a}^{1/2}(\mu_{\mathbb{C}})\hat{a}^{1/2}(\Delta_{\mathfrak{g}_{\mathbb{C}}})}\right\rvert_{\lambda} \frac{1}{\sqrt{\det(- \partial^2 \mathcal{W}_{\lambda})}}e^{\mathcal{W}_{\lambda}/\epsilon - \mathfrak{n}_C \zeta \partial_{\zeta} \mathcal{W}_{\lambda}},
\end{equation}
In the classical limit the $q$-difference operator then acts on the hemisphere partition function as multiplication $p_C = e^{-\zeta \partial_{\zeta}\mathcal{W}}$ so that
\begin{equation}
    \lim_{q\to 1}\hat{p}_C\cdot H_{\lambda}(q,t) = \lim_{q\to 1} e^{-\zeta\partial_{\zeta}\mathcal{W}_{\lambda}}H_{\lambda}(q,t),
\end{equation}
or in the language of overlaps in $K$-theory:
\begin{equation}
    p_{C}: \langle \lambda | 0 \rangle \to e^{-\zeta \partial_{\zeta}\mathcal{W}_{\lambda} }\langle \lambda |0 \rangle = \langle \lambda | \mathcal{L} \rangle.
\end{equation}

Above, we discussed the identification of the $q$-difference operator conjugate to the topological fugacity with a Wilson line operator insertion. For the other (Higgs branch) global symmetries, corresponding to fugacities $t_1, t_2$, it turns out that this is also true. The identification follows from certain difference equations (in flavour fugacities) that these partition functions are expected to obey \cite{Ferrari:2023hza, Ferrari:2024ksu} on physical grounds.

In the $q \rightarrow 1$ limit, these imply that insertions of $p_{1,2} = e^{t_{1,2} \partial_{t_{1,2}} \mathcal{W}}$ in the unrefined twisted index or hemisphere partition function/vertex function  are equivalent to insertions of Wilson lines, due to the identifications imposed by the vacuum/Bethe equations. That is, the 't Hooft line operators $(p_1$, $p_2,$ $p_C)$ are all equivalent, \textit{up to} an element of the Bethe ideal $\cI$ (which implies they are the same when evaluated on a solution to the Bethe equations), to Wilson line operators, and thus to elements of $QK_T(X_N)$, the twisted chiral ring. Thus, their correlation functions and expectation values are the same on the twisted backgrounds we consider in this work. The statements are non-trivial for the Higgs branch flavour symmetries as generically $p_{1,2} = e^{t_{1,2} \partial_{t_{1,2}} \mathcal{W}}$ may include denominators in the gauge fugacities. In appendix \ref{appendix:lineoperators}, we show this explicitly for the ADHM theory with $N = 2$ and supersymmetric QED with two flavours.\footnote{Demonstrating this explicitly for the ADHM quiver variety with $N$ generic is more difficult, and will require not only the exact form of the qKZ \cite{aganagic2017quasimap} equations obeyed by the vertex functions (which to the best of our knowledge are not explicitly known) but also difference equations for the $t$-parameter (which to the best of our knowledge are also not currently known).}

\paragraph{N.B.} Strictly speaking, the results of \cite{Ferrari:2023hza, Ferrari:2024ksu} imply that the 't Hooft flux operators are equivalent to a linear combination of Wilson line insertions in the gauge fugacity with coefficients possibly \textit{rational} in the equivariant and FI parameters. These are elements of the quantum $K$-theory of $X_N$ \textit{localised} at constant values of the parameters $t_1, t_2, \zeta$. This is $\bC[\{s_a\}^{\pm}]/I_{\zeta}$, as opposed to \eqref{eq:q_k_ring}. Physically, this makes sense as the values of these parameters are fixed constant background parameters for flavour symmetries, so throughout this paper we are more properly working in this ring. However, in an abuse of notation and nomenclature, and to make contact with the mathematical literature clear, we will continue to call the ring of physical Wilson lines $QK_T(X)$, and the operations on it (such as the inner product), should all be considered to be in the localised quantum $K$-theory ring.\\

Therefore our above conclusions for the topological symmetry generalises to the action of $(\hat{p}_1,$ $\hat{p}_2,$ $\hat{p}_C)$ for every flavour symmetry on a hemisphere partition function/vertex function. Namely the action of $(\hat{p}_1,$ $\hat{p}_2,$ $\hat{p}_C)$ are equivalent to the insertion of corresponding \textit{Wilson} line operators $(\mathcal{L}_1,$ $\mathcal{L}_2,$ $\mathcal{L})$. We have then in the $q \rightarrow 1$ limit:
\begin{equation}
\begin{split}
    \mathcal{I}^B_{S^2}(\mathfrak{n}_1,\mathfrak{n}_2,\mathfrak{n}_C) &= \sum_{\lambda} \left(e^{\left(\mathfrak{n}_1 t_1 \partial_{t_1}+ \mathfrak{n}_2 t_2 \partial_{t_2}  - \mathfrak{n}_C \zeta \partial_{\zeta}\right)\cdot \mathcal{W}}\right) \left.\frac{\hat{a}(T^*R)}{\hat{a}(\mu_{\mathbb{C}})\hat{a}(\Delta_{\mathfrak{g}_{\mathbb{C}}})}\right\rvert_{\lambda}\frac{1}{\det (-\partial^2 \mathcal{W}_{\lambda})}  \\
    &=\lim_{q\to 1} \sum_{\lambda} \left(\hat{p}_1^{\mathfrak{n}_1}\hat{p}_2^{\mathfrak{n}_2}\hat{p}_C^{\mathfrak{n}_C}\right) \cdot H_{\lambda}(q,t) H(q^{-1},t) \\
    &=\lim_{q\to 1} \sum_{\lambda} H_{\lambda}^{(\mathcal{L}_1^{\mathfrak{n}_1}\mathcal{L}_2^{\mathfrak{n}_1}\mathcal{L}^{\mathfrak{n}_C})}(q,t) H(q^{-1},t) \\
    &= \sum_{\lambda} \langle \mathcal{L}_1^{\mathfrak{n}_1}\mathcal{L}_2^{\mathfrak{n}_1}\mathcal{L}^{\mathfrak{n}_C} |  \lambda \rangle \langle\lambda| 0 \rangle \\
    &= \langle \mathcal{L}_1^{\mathfrak{n}_1}\mathcal{L}_2^{\mathfrak{n}_2}\mathcal{L}^{\mathfrak{n}_C} \rangle_{QK_T}
\end{split}    
\end{equation}
In the first line we use the geometric localisation formula \eqref{eq:twistedindexresidues}. In the second line we identify the flux insertions as the classical limit of the action of coordinate $q$-difference operators on the hemisphere partition functions/vertex functions, the operators may be thought of as flux-adding operators as in \cite{closset2016comments, closset2019three}. In the third line we identify the action of difference operators with Wilson line insertions/descendent insertions in vertex functions. We thus identify the twisted index with flux insertions as computing an expectation value of corresponding operators in the classical limit.

\paragraph{Remarks.}
We conclude this section with some comments on the classical-quantum correspondence. In the case of the cotangent bundle to the complete flag variety Koroteev et. al. \cite{koroteev2021quantum} show that the spectrum of the quantum $K$-theory may be associated with the phase space of a classical integrable system. In the complete flag case, this is the trigonometric RS system. To the best of the authors' knowledge, this classical integrable system for the Hilbert scheme of points $N$ has not yet been identified. In any case, it is expected that the functions
\begin{equation}
    \left\{ t_1,t_2,\, p_{1,2} = e^{t_{1,2}\partial_{t_{1,2}} \mathcal{W}}\right\}
\end{equation}
form canonical coordinates on the classical phase space. When quantising this classical integrable system, with quantisation parameter $q = e^{-\epsilon}$, the vertex functions $H_{\lambda}(q,t)$ form wavefunctions. The coordinates $(t_i,p_i)$ are promoted to operators that act as $q$-difference (as discussed above) as multiplication on the vertex functions/hemisphere partition functions.

\section{Large \texorpdfstring{$N$}{} Asymptotics}
We now turn to $\text{AdS}_4$ holography. In this section we bring together the results of the previous sections to reframe celebrated results in the physics literature \cite{benini2016black} in terms of the enumerative geometry of the Hilbert scheme.

\subsection{\texorpdfstring{$\text{AdS}_4$}{} Black Holes}

In this work, we are interested in the supersymmetric static magnetically charged black holes in M-theory on AdS$_4\times S^7$ constructed in \cite{Cacciatori:2009iz,DallAgata:2010ejj,Hristov:2010ri,Katmadas:2014faa,Halmagyi:2014qza}. Preserving two supercharges, they are dual to BPS states in topologically twisted ABJM theory with $k = 1$ and at large $N$ via the standard AdS/CFT duality in \cite{Aharony:2008ug}. The $k = 1$ ABJM theory is in turn dual to the low energy limit of ADHM theory with one flavour \cite{Aharony:2008ug,Kapustin:2010xq}, which has been the subject of the previous sections. 

On the field theory side, the topologically twisted indices of ADHM with one flavour are expected to reproduce the entropy of these black holes at large $N$, and this has been verified in \cite{benini2016black,Hosseini:2016ume}. In \cite{Hosseini:2016tor, Hosseini:2016ume}, a specific solution to the Bethe equations at large $N$ was found to reproduce the black hole entropy, henceforth referred to as the black hole solution. It is conjectured that the black hole solution dominates the index at large $N$, so that the index is approximately given by the summand of \eqref{index eval on roots} evaluated on the black hole solution. The index, and the twisted superpotential evaluated on the black hole solution are then
\begin{equation}\label{eq: W and I on bh sol}
\begin{aligned}
    \cW_\lambda &=iN^\frac{3}{2}\frac{2\sqrt{2}}{3}\sqrt{\wt\Delta_1\wt\Delta_2\wt\Delta_3\wt\Delta_4}\,,\\
    \log\cI^B_{S^2} \approx N^\frac{3}{2}\frac{\sqrt{2}}{3}&\sqrt{\wt\Delta_1\wt\Delta_2\wt\Delta_3\wt\Delta_4}\sum_{i=1}^4\frac{\wt\fn_i}{\wt\Delta_i}=-i\sum_{i=1}^4\wt\fn_i\partial_{\wt\Delta_i}\cW_\lambda\,. 
    \end{aligned}
\end{equation}
Here $\wt\Delta_i$ and $\wt\fn_i$ are redundant parametrisations of the chemical potentials and fluxes, which we will specify later. They satisfy the constraints $\sum_{i=1}^4\wt\fn_i=-2$ and $\sum_{i = 1}^4\wt\Delta_i = 2\pi$. The $\cI$-extremisation mechanism \cite{benini2016black, Benini:2016rke} then ensures that the logarithm of the degeneracy of BPS states is given by extremising $\log\cI^B_{S^2}$ with respect to $\wt\Delta_i$ under the constraint $\sum_{i = 1}^4\wt\Delta_i = 2\pi$.

On the gravitational side, the black hole with magnetic charges $p^\Lambda$, with $\Lambda = 0,1,2,3$, is a solution of a $4$d $\cN = 2$ gauged supergravity containing the gravity multiplet and three vector multiplets \cite{Cacciatori:2009iz}. This supergravity theory is a consistent truncation of $11$d supergravity reduced on $S^7$ \cite{Cvetic:1999xp}. It is completely specified by FI parameters and the prepotential 
\be\label{eq: sugra prepotential}
\cF = -2i\sqrt{X^0X^1X^2X^3}\,,
\ee
where $X^\Lambda$ are holomorphic sections of a line bundle on the scalar manifold. They also serve as homogeneous coordinates on the scalar manifold. Supersymmetry of the black hole solution requires that the magnetic charges satisfy $\sum_{\Lambda = 0}^3p^\Lambda = -2$. The attractor mechanism \cite{DallAgata:2010ejj,Cacciatori:2009iz,benini2016black}, ensures that the remaining BPS equations for the near-horizon geometry are equivalent to extremising
\be\label{eq: attractor mech}
\cS = i\frac{\pi}{G_{\text{N}}^{(4)}}p^\Lambda \cF_\Lambda(X)\,,\quad \cF_\Lambda \equiv \frac{\partial\cF}{\partial X^\Lambda}\,,
\ee
with respect to $X^\Lambda$, subject to the constraint $\sum_{\Lambda = 0}^3X^\Lambda = 1$, and the black hole entropy is then given by the extremal value of $\cS$. 

Tying the two sides of the correspondence together, the AdS/CFT dictionary instructs us to identify \cite{Aharony:2008ug,benini2016black}
\be
\frac{1}{G_{\text{N}}^{(4)}} = \frac{2\sqrt{2}}{3}N^\frac{3}{2}\,,\quad p^{0,1,2,3} = \fn^{4,1,2,3}\,.
\ee
The extremisation problems in \eqref{eq: W and I on bh sol} and \eqref{eq: attractor mech} then become identical if we also identify $X^{0,1,2,3} = \Delta_{4,1,2,3}/2\pi$, which is consistent with the constraints $\sum_{\Lambda = 0}^3X^\Lambda = 1$ and $\sum_{i = 1}^4\Delta_i = 2\pi$. Hence the field theory computation exactly reproduces the Bekenstein-Hawking entropy of this class of black holes.

We note that the prepotential in \eqref{eq: sugra prepotential} is directly proportional to $\cW_\lambda$ in \eqref{eq: W and I on bh sol} under the identification $X^{0,1,2,3} = \Delta_{4,1,2,3}/2\pi$ required by the entropy computation. In addition, the logarithm of the $S^3$ partition function is
\be
F_{S^3} = N^\frac{3}{2}\frac{4\pi}{3}\sqrt{2r_1r_2r_3r_4}\,,
\ee
where $r_{1,2,3,4}$ parametrise a general assignment of R-charges satisfying $\sum_{i=1}^4r_i = 2$\cite{Jafferis:2011zi}. Under the identification $r_i = \Delta_i/\pi$, $\cW_\lambda$ is also proportional to $F_{S^3}$, and the constraint $\sum_{i = 1}^4\Delta_i = 2\pi$ coincides with $\sum_{i=1}^4r_i = 2$. Furthermore, it is known that $F_{S^3}(r)\propto\text{Vol}_{S^7}(r)^{-\frac{1}{2}}$, where $\text{Vol}_{S^7}(r)$ is the volume of $S^7$ under different Sasakian metrics parametrised by $r_i$. The $r_i$ determine how the Reeb vector of the cone over $S^7$ is embedded within the $\text{U}(1)^4\subset\text{SO}(8)$ isometry of $S^7$ \cite{Jafferis:2011zi}. In summary, we have
\be
\cW_\lambda(\Delta) = -\frac{2\pi^2}{G_{\text{N}}^{(4)}}\cF\left(\frac{\Delta}{2\pi}\right)=\frac{i\pi}{2}F_{S^3}\left(\frac{\Delta}{\pi}\right)=\frac{i\pi^4}{4G_{\text{N}}^{(4)}\sqrt{3\text{Vol}_{S^7}\left(\frac{\Delta}{\pi}\right)}}\,.
\ee
The twisted superpotential, which plays a central role in the quantum $K$-theory of the Hilbert scheme $\text{Hilb}^N(\mathbb{C}^2)$, has the same functional form as the supergravity prepotential and is related to the Sasakian volume of $S^7$ when evaluated on the black hole solution.

\subsection{Large \texorpdfstring{$N$}{} Bethe Equations}\label{sec: large N sol}

In the following, we review the solution of the Bethe equations in the large $N$ limit, as derived in \cite{Herzog:2010hf, Benini:2015eyy, Hosseini:2016tor, Hosseini:2016ume}. Firstly, the $S_N$ Weyl symmetry of the Bethe equations allows us to order the gauge holonomies $u_a\equiv -i\log s_a\in (-\pi,\pi)$ such that $\mathbb{I}\text{m}\, u_a$ monotonically increases with $a$. The appropriate ansatz for $u_a$ in the large $N$ limit is
\begin{equation}
u_a = iN^\frac{1}{2}t_a + v_a\,,\quad t_a, v_a \in\bR\,,\quad t_a, v_a \sim\cO(1)\,,
\end{equation}
where $t_a$ is a monotonically increasing sequence in $a$. We assume that in the large $N$ limit, $t_a$ and $v_a$ become dense in an interval and one can approximate
\begin{equation}
t_a = t\left(\frac{a}{N}\right)\,,\quad v_a = v\left(\frac{a}{N}\right)\,,
\end{equation}
for continuous functions $t$ and $v$. We can then parametrize the holonomies by values of $t$ instead of $a$, with the two variables being related by the density
\begin{equation}\label{density def}
\rho(t)\equiv\frac{1}{N}\frac{da}{dt}\,.
\end{equation}
Note that $\rho > 0$ due to the monotonicity of $t$. The ansatz is then
\begin{equation}\label{gauge hol ansatz}
u(t) = iN^\frac{1}{2}t + v(t)\,.
\end{equation}

In the large $N$ limit, summations tend to integrals $\sum_{a=1}^N\rightarrow N\int dt\,\rho(t)$ and since $\sum_{a=1}^N 1=N$, the density is normalized as
\begin{equation}\label{density normalization}
\int dt\,\rho(t) = 1\,.
\end{equation}
We define the chemical potentials $\Delta_{C}$ and $\Delta_{1,2,3}$ as
\begin{equation}\label{chem pt def}
\zeta = e^{i\Delta_C}\,,\quad \Delta_C\in (-\pi,\pi)\,,\quad t_{I} = e^{i\Delta_{I}}\,,\quad \Delta_I\in (0,2\pi)\,,\quad I=1,2,3\,.
\end{equation}
Since $\prod_{I=1}^3t_I=1$, there is the constraint
\begin{equation}\label{chem pot constraint}
\sum_{I=1}^3\Delta_I=2\pi,\,4\pi\,.
\end{equation}
Note that the $\Delta_I$ cannot sum to $0$ or $6\pi$ since that would imply $t_I=1$, which we exclude. For brevity, we shall restrict to the case $\sum_{I=1}^3\Delta_I=2\pi$, noting that all of the following conclusions are true in the other case as well, with only minor modifications to the formulae. We also assume that 
\begin{equation}\label{ang ranges}
\re u_a -\re u_b + \Delta_I \in (0,2\pi)\quad\forall a,b\,,\qquad \pm\re u_a + \frac{\Delta_1+\Delta_2}{2} \in (0, 2\pi)\quad\forall a\,.
\end{equation}

As in \cite{Hosseini:2016tor, Hosseini:2016ume}, one obtains that the large $N$ limit of the twisted superpotential is
\begin{equation}\label{large N cW}
\frac{\cW}{iN^\frac{3}{2}}=\Delta_C\int dt\,\rho(t)\,t+\sum_{I=1}^3g_+(\Delta_I)\int dt\,\rho^2(t)+\frac{\Delta_3}{2}\int dt\,\rho(t) |t|+\cO(N)\,,
\end{equation}
where
\begin{equation}\label{eq: g_+ def}
    g_+(\Delta_I)\equiv\frac{1}{6}\Delta_I^3-\frac{\pi}{2}\Delta_I^2+\frac{\pi^2}{3}\Delta_I\,.
\end{equation}
In this limit, the Bethe equations have turned into the extremization equation of $\cW$ with respect to $\rho$. Since $\rho$ needs to satisfy the normalization constraint \eqref{density normalization}, one cannot directly take the functional derivative of \eqref{large N cW} with respect to $\rho$ without first solving the constraint. Instead, as is standard, we introduce a Lagrange multiplier $\mu$ and extremize
\begin{equation}
\frac{\cW}{iN^\frac{3}{2}}=\Delta_C\int dt\,\rho(t)\,t+\sum_{I=1}^3g_+(\Delta_I)\int dt\,\rho^2(t)+\frac{\Delta_3}{2}\int dt\,\rho(t) |t|-\mu\left(\int dt\,\rho(t) - 1\right)
\end{equation}
with respect to $\rho$ and $\mu$. Note that adding the Lagrange multiplier does not change the value of $\cW$ on solutions to the Bethe equations. The derivative with respect to $\mu$ returns the constraint, while the derivative with respect to $\rho$ gives
\begin{equation}
\Delta_C\,t + 2\sum_{I=1}^3g_+(\Delta_I)\rho(t)+\frac{\Delta_3}{2}|t|-\mu = 0\,.
\end{equation}
This determines $\rho$ to be
\begin{equation}\label{crit density}
\rho(t) = \frac{\mu - \Delta_C\,t - \frac{\Delta_3}{2}|t|}{2\sum_{I=1}^3g_+(\Delta_I)}\,.
\end{equation}

Supposing that $\rho(t)$ has finite support on $[t_-,t_+]$, where $t_+>0$ and $t_-<0$, $\rho(t_\pm)=0$ allows us to solve for
\begin{equation}\label{t limits}
t_\pm = \pm\frac{\mu}{\frac{\Delta_3}{2}\pm\Delta_C}. 
\end{equation}
In order to be consistent with $t_+ > 0$ and $t_- <0$, the signs of $\mu$ and $\frac{\Delta_3}{2}\pm\Delta_C$ must agree. This implies that 
\begin{equation}\label{delta_C range}
    |\Delta_C| < \bigg|\frac{\Delta_3}{2}\bigg|\,,\quad \sgn(\mu)=+1\,.
\end{equation}
The normalization condition \eqref{density normalization} then determines $\mu$ to be
\begin{equation}\label{mu val}
\mu = \sqrt{\frac{1}{2}\Delta_1\Delta_2(\Delta_3+2\Delta_C)(\Delta_3-2\Delta_C)}\,.
\end{equation}
When evaluated on the solution determined by \eqref{crit density}, \eqref{t limits}, and \eqref{mu val}, the value of the twisted superpotential is
\begin{equation}\label{on-shell W}
    \cW_\lambda=\frac{2i N^\frac{3}{2}}{3}\sqrt{\frac{1}{2}\Delta_1\Delta_2(\Delta_3+2\Delta_C)(\Delta_3-2\Delta_C)}\,.
\end{equation}

Having found the dominant Bethe root at large $N$, we should take the large $N$ limit of \eqref{index eval on roots} and evaluate it on this root to find the large $N$ index. The result is
\begin{equation}\label{eq: index on bh sol}
\begin{aligned}
    &\lim_{N\rightarrow\infty}\log\left[\left.\frac{\hat{a}(T^*R)}{\hat{a}(\mu_{\mathbb{C}})\hat{a}(\Delta_{\mathfrak{g}_{\mathbb{C}}})}\right\rvert_{\lambda}\frac{1}{\det (-\partial^2 \mathcal{W}_\lambda)} \mathcal{O}_\lambda(\fn_1,\fn_2,\mathfrak{n}_C)\right]\\
    &=-\mathfrak{n}_C\int dt\rho(t)\,t+\left[\sum_{I=1}^3(\fn_I+1)g_+'(\Delta_I)-\frac{\pi^2}{3}\right]\int dt\rho^2(t)+\frac{\fn_3}{2}\int dt\rho(t)|t|\\
    &= 
    \frac{N^\frac{3}{2}}{3}\sqrt{\frac{1}{2}\Delta_1\Delta_2(\Delta_3+2\Delta_C)(\Delta_3-2\Delta_C)}\left(\frac{\fn_1}{\Delta_1}+\frac{\fn_2}{\Delta_2}+\frac{\fn_3-2\mathfrak{n}_C}{\Delta_3+2\Delta_C}+\frac{\fn_3+2\mathfrak{n}_C}{\Delta_3-2\Delta_C}\right)\,,
\end{aligned}
\end{equation}
where we have introduced $\fn_3= -2-\fn_1-\fn_2$ for convenience. $g_+'$ is the derivative of $g_+$ in \eqref{eq: g_+ def}. In terms of the parameters suited to ABJM theory \cite{Hosseini:2016ume}
\bea
\wt\Delta_1 = \Delta_1\,,\quad \wt\Delta_2 = \Delta_2\,,\quad &\wt\Delta_3 = \frac{\Delta_3}{2}+\Delta_C\,,\quad \wt\Delta_4 = \frac{\Delta_3}{2}-\Delta_C\,, \\
\wt\fn_1 = \fn_1\,,\quad\wt\fn_2 = \fn_2\,,\quad &\wt\fn_3=\frac{\fn_3}{2}-\fn_C\,,\quad \wt\fn_4 = \frac{\fn_3}{2}+\fn_C\,,
\eea
which satisfy $\sum_{i=1}^4\wt\Delta_i=2\pi$ and $\sum_{i=1}^4\wt\fn_i = -2$, \eqref{on-shell W} and \eqref{eq: index on bh sol} become \eqref{eq: W and I on bh sol}, as claimed.

\section{The Black Hole Partition}\label{sec:BHpartition}

Recall from \eqref{eq: saddle pt hemisphere}, \eqref{eq: vev factorisation}, and \eqref{eq:twistedindexresidues} that the twisted index can be obtained by gluing hemisphere partition functions $H_\lambda$ in the $q\rightarrow 1$ or $\epsilon\rightarrow 0$ limit. Here, $\lambda$ not only refers to the boundary condition $\cB_\lambda$ or classical solution \eqref{eq:classical soln BAE} which correspond to a partition $\lambda$, but it also refers to a solution to the full Bethe equations \eqref{eq:bethe_eqns} which reduces to the classical solution under $\zeta\rightarrow 0$ and dominates the asymptotics of $H_\lambda$ under $\epsilon\rightarrow 0$. In the Cardy limit approach to computing the large $N$ index and thereby the entropy function, a particular Bethe root, namely the black hole solution reviewed in Section \ref{sec: large N sol}, was found to dominate the large $N$ and $\epsilon\rightarrow 0$ asymptotics of $H_\lambda$. The right-hand side of \eqref{eq: saddle pt hemisphere} evaluated on the black hole solution was known as the Cardy block in \cite{Choi:2019dfu,Choi:2019zpz}, and the entropy function was then obtained by gluing two Cardy blocks, as in \eqref{eq: vev factorisation}. In gravity, this approach is identical to the proposal of gluing two gravitational blocks in \cite{hosseini2019gluing}. 

\begin{figure}[H]
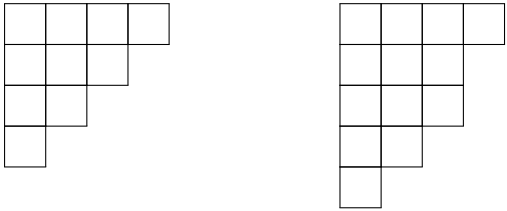

    \centering
    \begin{ytableau}
    {} & {} & {} & {} \\ 
    {} & {} & {} \\
    {} & {} \\
    {}
\end{ytableau} \hspace{2cm} \begin{ytableau}
    {} & {} & {} & {} \\ 
    {} & {} & {} \\
    {} & {} & {} \\
    {} & {} \\
    {}
\end{ytableau} 
    \caption{\small The triangular partitions of $N = 10$ (left) and $N = 13$ (right), as defined in \eqref{def triangular partitions}.}
    \label{fig:eg triangular partitions}
\end{figure}

However, up until this point, the field theory dual to the gravitational/Cardy blocks was not precisely known because the boundary condition $\cB_\lambda$ or classical solution corresponding to the black hole solution was not known. In this section, we remedy this by evolving a finite $N$ analogue of the black hole solution towards $\zeta\rightarrow 0$, and matching the result with a classical solution. The result is that the classical solution at the end of the evolution corresponds to the partition
\begin{equation}\label{def triangular partitions}
\begin{aligned}
    \lambda = (1^{m_1},2^{m_2},\ldots)\,,\quad m_i &= \begin{cases}
    1\quad \text{if} \quad 1\leq i\leq h \land i\neq N - \frac{h(h+1)}{2}\,, \\
    2\quad \text{if} \quad i = N - \frac{h(h+1)}{2}\,,\\
    0\quad \text{otherwise}
    \end{cases}\\
    h &= \max\left\{n\in\mathbb{N}\;\bigg|\; \frac{n(n+1)}{2}\leq N\right\}\,.
\end{aligned}
\end{equation}
If $N$ is a triangular number, $\lambda$ has parts $\{1,2,\ldots,h\}$, each with multiplicity one, and its Young diagram has a perfect triangular shape. Otherwise, the multiplicity of one part $N - \frac{h(h+1)}{2}$ is doubled, but the shape of the Young diagram is still roughly triangular. We shall denote the partitions of equation \eqref{def triangular partitions} as the triangular partitions. For example, the triangular partitions of $N = 10$ and $13$ are illustrated in Figure \ref{fig:eg triangular partitions}.

Our results therefore allow us to make the following conjectures, for which we give strong numerical evidence in the remainder of this paper.

\begin{conjecture}
The black hole solution of the vacuum/Bethe ansatz equations for the ADHM theory, which dominates the twisted index at large $N$ and reproduces the energy functional of static magnetically charged black holes in M-theory on $\text{AdS}_4\times S^7$, is homotopic as $\zeta \rightarrow 0$ to the classical solution \eqref{eq:classical soln BAE} corresponding to the massive vacuum $\lambda$ of the ADHM theory labelled by the unqiue \textit{triangular-shaped} Young tableau/partition above.
\end{conjecture}
Combined with the twisted index factorisation \eqref{eq: vev factorisation} and the saddle point formula for the hemisphere partition function/vertex function \eqref{eq: saddle pt hemisphere}, we also have that

\begin{conjecture}
The geometric object holographically dual to the gravitational block, \textit{i.e.} the precise geometric definition of the Cardy block for this holographic system, is the hemisphere partition function $H_{\lambda}$ with an exceptional Dirichlet boundary condition $\cB_{\lambda}$ (as described in section \ref{subsec:quantumKtheory}) corresponding to the triangular partition above.
\end{conjecture}

\subsection{Finite \texorpdfstring{$N$}{} Solutions}\label{subsec: finite N sols}

In order to provide evidence that the large $N$ black hole solution corresponds to the triangular partitions, the strategy we use is to first find the finite $N$ analogue of the continuum solution \eqref{crit density} numerically, and then evolve it to $\zeta = 0$ via homotopy continuation.

\begin{figure}[H]
    \centering        
    \includegraphics[]{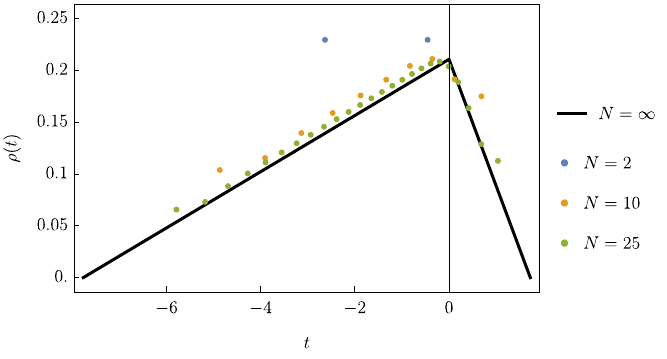}
    \caption{\small Plots of the density $\rho(t)$, where the discrete points are extracted from the numerical solutions at $N=2,10,25$, while the continuous curve is the large $N$ solution given in \eqref{crit density}. We have chosen $\Delta_1 = \frac{\pi^2}{4}$, $\Delta_2=e$, $\Delta_C=\frac{\pi}{9}$, which are in the allowed ranges \eqref{chem pt def}, \eqref{delta_C range}. One can observe that the agreement between the numerical solution and the analytic prediction is improving with increasing $N$.}
    \label{fig:finite N BH sols}
\end{figure}

We wish to numerically solve the finite $N$ Bethe equations in the form 
\begin{equation}\label{log Bethe eqn}
    \cW_a - 2\pi in_a\,,\quad n_a\in\bZ\,,\quad \forall a = 1,\ldots, N\,,
\end{equation}
where $\cW_a$ are the logarithms of the left-hand side of \eqref{eq:bethe_eqns}. At large $N$, this is preferable to solving $\exp\cW_a = 1$ since we are looking for solutions with $s_a\sim e^{N^\frac{1}{2}}$ for which the numerator and denominator of \eqref{eq:bethe_eqns} diverge quickly. A solution is found by taking the discretisation of \eqref{crit density} as an initial seed and applying a quasi-Newton method to minimize the error
\begin{equation}\label{num err}
    \frac{1}{N}\sum_{a=1}^N|\cW_a - 2\pi i n_a|^2\,.
\end{equation}
Specifically, the definition of the density $\rho$ \eqref{density def} implies
\begin{equation}
    N\int_{t_-}^tdt'\rho(t') = a(t)\,,
\end{equation}
which is a quadratic equation in $t$ that can be solved analytically for $t(a)$ when $a = 1,\ldots, N$. The initial guess $u_0$ is then taken to be
\begin{equation}
    (u_0)_a = iN^\frac{1}{2} t(a)\,,
\end{equation}
and the integers $n_a$ in \eqref{log Bethe eqn} are fixed to
\begin{equation}
    n_a = \left\lfloor \frac{\mathbb{I}\text{m}\,\cW_a(u_0)}{2\pi} \right\rceil\,,
\end{equation}
where $\lfloor\cdot\rceil$ denotes rounding to the nearest integer. This determination of $n_a$ will be correct if $u_0$ is a close to the exact solution. For $2\leq N\leq 75$, we have verified convergence of the quasi-Newton method, and the resulting solution has purely imaginary values of $u_a$ which are purely imaginary--in agreement with the initial guess. If we then identify $t(a) = u_a/iN^\frac{1}{2}$ and approximate the density \eqref{density def} by computing $da/dt$ via finite differences, the agreement with the $N\rightarrow\infty$ solution \eqref{crit density} improves as $N$ gets larger\footnote{For small $N$, one could be sceptical due to the relatively poor agreement, but the property $\re u_a = 0$ is itself a distinguishing feature which separates the putative black hole solution from other solutions of the Bethe equations, which generically have nonzero $\re u_a$.} (see Figure \ref{fig:finite N BH sols}). This is evidence that these numerical solutions are the finite $N$ analogues of the continuum solution \eqref{crit density}. 

\subsection{Tracking Solutions to the Classical Limit}

We now evolve this finite $N$ numerical solution to $\zeta\rightarrow 0$, in order to determine the corresponding classical vacuum/partition $\lambda$ to the black hole solution, and therefore the hemisphere partition function $H_{\lambda}$ dual to the gravitational block. We shall see it is the triangular partition described at the beginning of this section.

Using the parametrisation $\zeta = e^{\alpha + i\Delta_C}$, the limit is taken by varying $\alpha$ from $0$ to $-\Lambda$, where $\Lambda$ is some large positive cutoff. In the following, we shall set $\Lambda = 14$, meaning that at the endpoint one has $|\zeta|=e^{-14}\approx 10^{-6}$, which is sufficiently small for our purposes. The solution is tracked using the approach of homotopy continuation, implemented using the predictor-corrector method  of \cite{Hao:2013jqa,doi:10.1142/5763}, which we now outline. In each step $i$ of the algorithm, we start from a solution $u = u_i$ at $\alpha = \alpha_i$, and seek the solution $u_{i+1} = u_i + \Delta u$ at $\alpha_{i+1} = \alpha_i + \Delta \alpha_i$, where $\Delta \alpha_i$ is a variable step size. As a solution, $u_{i+1}$ satisfies

\begin{equation}\label{eq: LO W exp}
\begin{aligned}
    0 &= \cW_a(u_{i+1}, \alpha_{i+1}) - 2\pi i n_a \approx \left.\frac{\partial\cW_a}{\partial u_b}\right\rvert_{u = u_i,\alpha = \alpha_i} \Delta u_b + \left.\frac{\partial \cW_a}{\partial \alpha}\right\rvert_{u = u_i,\alpha = \alpha_i} \Delta \alpha_i \\
    &= J_{ab}\, \Delta u_b - (\vec{1})_a \Delta \alpha_i\,,\qquad J(u_i,\alpha_i)_{ab}\equiv \left.\frac{\partial \cW_a}{\partial u_b}\right\rvert_{u = u_i,\alpha = \alpha_i}\,,\qquad (\vec{1})_a = 1\quad \forall a = 1,\ldots, N\,. 
\end{aligned}
\end{equation}
We find that $\det J\neq 0$ along the path from $\alpha = 0$ to $\alpha = -14$. We can therefore invert $J$ and approximate
\begin{equation}\label{eq: predictor du}
    \Delta u_a \approx J(u_i,\alpha_i)^{-1}_{ab}(\vec{1})_b\,\Delta \alpha_i\,. 
\end{equation}
This is the predictor step which provides a good estimate $u_{i+1}\approx u_i + J^{-1}\vec{1}\,\Delta \alpha_i$ of the solution at $\alpha = \alpha_{i+1}$ if $\Delta \alpha_i$ is sufficiently small. In the subsequent corrector step, we start from $u_{i+1}\approx u_i + J^{-1}\vec{1}\,\Delta \alpha_i$ and perform Newton iterations
\begin{equation}
\begin{aligned}
    (u_{i+1})_a &= (u_{i+1})_a -  J(u_{i+1},\alpha_{i+1})^{-1}_{ab}(\cW_b(u_{i+1},\alpha_{i+1}) - 2\pi i n'_b)\,, \\
    n'_a &= \left\lfloor \frac{\mathbb{I}\text{m}\cW_a(u_i + J^{-1}\vec{1}\,\Delta \alpha_i, \alpha_{i+1})}{2\pi} \right\rceil\,,
\end{aligned}
\end{equation}
until either the error \eqref{num err} is below some tolerance, typically chosen to be $10^{-32}$, or the number of iterations exceeds a maximum number, typically chosen to be $5$. Recall that $\cW_a$ contains a sum of logarithms. The change from $n_a$ to $n'_a$ is to treat the case where the branch cuts of one or more of the logarithms are crossed when passing from $u_i$ to $u_i + J^{-1}\vec{1}\,\Delta \alpha_i$. If the maximum number of iterations is reached but the error is still greater than the tolerance, the step size $\Delta \alpha_i$ is halved, after which the $i^{th}$ predictor and corrector steps that we have just described are repeated. Otherwise, we have succeeded in finding the solution $u = u_{i+1}$ at $\alpha = \alpha_{i+1}$ to within the chosen tolerance, and can move on to the $(i+1)^{th}$ step of the algorithm. There is a danger that during the corrector step an intermediate solution $u_i$ at $\alpha_i$ might jump to a track that is not homotopically connected to the black hole solution. By keeping the maximum number of iterations in the corrector small and the tolerance tight, we can minimise the chances of this happening. In addition, if the corrector step is successful for $3$ consecutive steps without needing to half the step size, the step size is doubled. This allows a larger step size to be used provided that the convergence of the corrector is not compromised, and speeds up the tracking considerably.   

\begin{figure}[h]
    \centering        
    \includegraphics[width=0.75\textwidth]{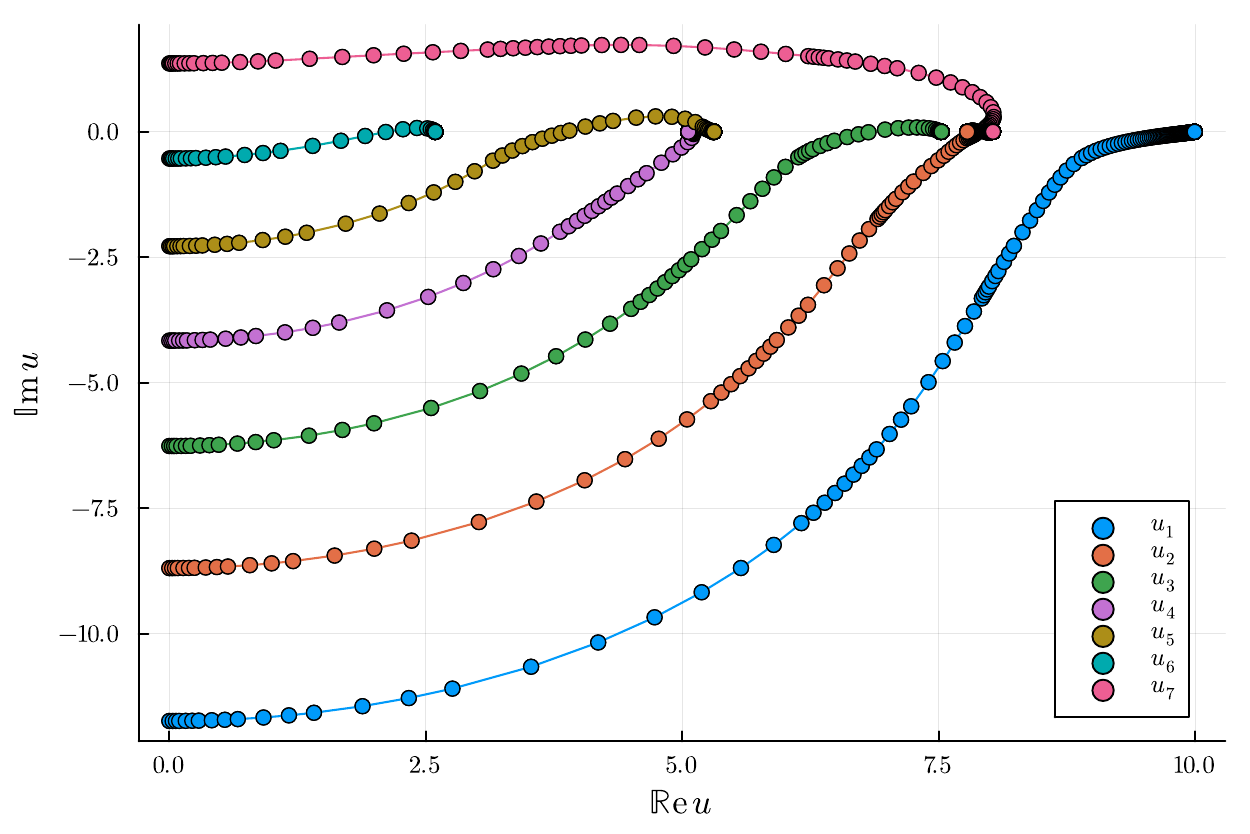}
    \caption{\small Plots of $u_a$ in the complex plane on the solutions at various intermediate values of $\alpha$ while running the predictor-corrector algorithm. The data is for $N = 7$, $\Delta_1 = \frac{\pi^2}{4}$, $\Delta_2=e$, and $\Delta_C=\frac{\pi}{9}$. }
    \label{fig:track plot}
\end{figure}

To illustrate how the solution smoothly varies as $\zeta\rightarrow 0$, we plot in Figure \ref{fig:track plot} the values of $u_a$ on the solutions at intermediate values of $\alpha$. Notice that the solutions smoothly interpolate between the purely imaginary black hole solution at $|\zeta| = 1$ and a purely real classical solution at $|\zeta|=e^{-14}$. Recall that the classical solutions \eqref{eq:classical soln BAE} are real since the chemical potentials in \eqref{chem pt def} are real.

Having reached the endpoint $\alpha = -14$, the solution is refined by performing Newton iterations until the error is smaller than an even tighter tolerance, which we have chosen to be $10^{-64}$. After sorting the numerical solution and classical solutions by increasing real parts, the matching classical solution and its corresponding partition is then picked out by minimising

\begin{equation}
    \sum_{a=1}^N\left\lvert (\text{sorted numerical solution})_a - (\text{sorted classical solution})_a \right\rvert^2\,,
\end{equation}
among all classical solutions, \textit{i.e.} all solutions of the form \eqref{eq:classical soln BAE} corresponding to a partition.  Since we have chosen a small enough value of $|\zeta|=e^{-14}$ at the endpoint, it turns out that the smallest error is several orders of magnitude smaller than the second smallest error, so that we may unambiguously identify which classical solution and partition is the result under the $\zeta\rightarrow 0$ limit. For example, with $N = 7$, $\Delta_1 = \frac{\pi^2}{4}$, $\Delta_2=e$, and $\Delta_C=\frac{\pi}{9}$, the minimum error is roughly $3.5\times 10^{-5}$, while the second smallest error is roughly $0.25$. As mentioned, the classical solutions with the minimum error turn out to correspond to the triangular partitions defined in \eqref{def triangular partitions}. We have checked this for $2\leq N\leq 28$, and we expect and conjecture the same to be true for all $N > 28$.

\subsection{Tracking in the Reverse Direction}\label{sec: reverse tracking}

To provide further evidence of the correspondence between black hole solutions and triangular partitions, it is desirable to track the classical solution of the triangular partition in the reverse direction from $\zeta = 0$ to $\zeta = e^{i\Delta_C}$, and verify that the endpoint of this homotopy is the black hole solution. For this purpose, solving the Bethe equations in the form \eqref{log Bethe eqn} is not possible since $\cW_a$ is indeterminate on the classical solutions (\textit{e.g.} there are terms $-\log\zeta = -\log 0$ and $\log\big(s_a-t_1^{1/2}t_2^{1/2}\big) = \log 0$ when $i_a = j_a = 1$). Instead, we track the solutions of 

\begin{equation}\label{eq: Bethe denom cleared}
\begin{aligned}
B_a(s,v)&\equiv\Big(s_a-t_1^\frac{1}{2}t_2^\frac{1}{2}\Big)\prod_{I=1}^3\prod_{\substack{b=1\\b\neq a}}^N(s_b-t_I^{-1}s_a)+v\,\zeta\, t_1^{\frac{1}{2}}t_2^{\frac{1}{2}}\Big(s_a-t_1^{-\frac{1}{2}}t_2^{-\frac{1}{2}}\Big)\prod_{I=1}^3\prod_{\substack{b=1\\b\neq a}}^N(s_b-t_Is_a)=0\,,\\
 a&=1,\ldots,N\,,
\end{aligned}
\end{equation}
which are simply the Bethe equations \eqref{eq:bethe_eqns} with their denominators cleared, and where $v$ is a parameter that allows us to interpolate between $\zeta = 0$ and $\zeta = e^{i\Delta_C}$ as it is tuned from $v = 0$ to $v = 1$.

One might hope that the tracking of solutions proceed analogously to the previous subsection, using the predictor-corrector method. However, the initial step from $v = 0$ to $v = \Delta v\neq 0$ poses a difficulty, which we now explain. 

Let $s^*$ be a classical solution at $v = 0$. We seek the solution at $v = \Delta v$, written as a systematic expansion
\begin{equation}
    s = s^* + \Delta s = s^* + \Delta^{\scriptscriptstyle{(1)}}s + \Delta^{\scriptscriptstyle{(2)}}s + \ldots\,,
\end{equation}
where each term $\Delta^{\scriptscriptstyle{(n)}}s$ is of order $(\Delta v)^n$. To leading order in $\Delta v$, the analogous Taylor expansion to \eqref{eq: LO W exp}

\begin{equation}\label{eq: exp B dv}
    0 = B_a(s^* + \Delta s, \Delta v) = \partial_{s_b}B_a(s^*, 0)\Delta^{\scriptscriptstyle{(1)}}s_b + \partial_uB_a(s^*,0)\Delta v + \cO\big((\Delta v)^2\big)
\end{equation}
implies

\begin{equation}\label{eq: LO s constr}
    J_{ab}\Delta^{\scriptscriptstyle{(1)}} s_b = -\partial_vB_a(s^*,0)\Delta v\,,\qquad J_{ab} \equiv \partial_{s_b}B_a(s^*,0)\,.
\end{equation}
If $\lambda$ is the partition associated with the classical solution $s^*$, we find empirically that

\begin{equation}\label{eq: cond J singular}
    \text{rank}(\lambda)\geq 2 \qquad \iff \qquad \det J = 0 \,,
\end{equation}
where rank$(\lambda)$ is defined to be the maximum positive integer such that $\lambda$ contains at least rank$(\lambda)$ parts with values greater or equal to rank$(\lambda)$. Diagrammatically, rank$(\lambda)$ is the length (in units of boxes) of a side of the largest square that can fit into the Young diagram of $\lambda$, also called the Durfee square. To be more precise, the forward implication of \eqref{eq: cond J singular} can be proven, and the proof is be found in Appendix \ref{app: singular jacobian}. Numerically, we were unable to find a counterexample to the reverse implication in \eqref{eq: cond J singular}, and we conjecture it to be true. Consequently, the initial predictor step fails when tracking the solutions associated with partitions of rank $\geq 2$, since it requires us to invert $J$. 

\begin{figure}[H]
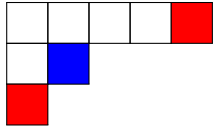

    \centering
    \begin{ytableau}
        {} & {} & {} & {} & *(red) \\
        {} & *(blue) \\
        *(red)
    \end{ytableau}
    \caption{\small Young diagram of the partition $\{5,2,1\}$ with exterior corners coloured red and interior corners coloured blue.}
    \label{fig:corner def}
\end{figure}

For these problematic partitions, we shall attempt to fix $\Delta^{\scriptscriptstyle{(1)}}s$ by analysing \eqref{eq: LO s constr} and the Taylor expansion more carefully. Now, the generic solution to \eqref{eq: LO s constr} is

\begin{equation}\label{eq: gen sol LO ds}
    \Delta^{\scriptscriptstyle{(1)}}s = \Delta^{\scriptscriptstyle{(1)}}_\times s + \Delta^{\scriptscriptstyle{(1)}}_0 s\,,\qquad \Delta^{\scriptscriptstyle{(1)}}_0 s\in \text{Nul}(J)
\end{equation}
where $\Delta^{\scriptscriptstyle{(1)}}_\times s$ is any particular vector that satisfies $J_{ab}\Delta^{\scriptscriptstyle{(1)}}_\times s_b = -\partial_vB_a(s^*,0)\Delta v$, and Nul$(J)$ is the right null space of $J$ (the right and left null spaces are not equal since $J$ is not symmetric). Due to the structure of $\partial_vB$ and $J$, which is explained in Appendix \ref{app: particular sol}, there is a specific choice of $\Delta^{\scriptscriptstyle{(1)}}_\times s$ that is particularly simple, and also helpful for later simplifications. To specify it, we define a corner box of a Young diagram to be a box that can be removed to leave a valid Young diagram. If a corner box is in the top row or left-most column, we shall call it an exterior corner. The remaining corner boxes will be known as interior corners.\footnote{This is not to be confused with the notion of inner and outer corners commonly used in the literature.} As an illustration, the exterior and interior corners of the partition $\{5,2,1\}$ are coloured red and blue respectively in Figure \ref{fig:corner def}   

With this terminology, our choice of $\Delta^{\scriptscriptstyle{(1)}}_\times s$ is specified as

\begin{equation}\label{eq: particular sol LO}
    \Delta^{\scriptscriptstyle{(1)}}_\times s_a = \begin{cases}
        -\frac{\partial_vB_a(s^*,0)\Delta v}{J_{aa}}\,,\quad &\text{if $a$ is an exterior corner} \\
        0\,,\quad &\text{otherwise}
    \end{cases}
\end{equation}

\begin{figure}[H]
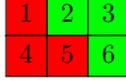

    \centering
    \begin{ytableau} 
    *(red) 1 & *(green) 2 & *(green) 3 \\ 
    *(red) 4 & *(red) 5 & *(green) 6 \end{ytableau}
    \caption{\small Young diagram of the partition $\{3,3\}$. Starting with the classical solution $\{s_1^*,\ldots, s_6^*\}$ where $s_a$ is associated with the box $a$ labelled above, one obtains $2$ infinite families of solutions to \eqref{eq: Bethe denom cleared} by scaling the gauge fugacities of the red and green boxes above independently.}
    \label{fig:eg inf solns}
\end{figure}

It then remains to determine $\Delta^{\scriptscriptstyle{(1)}}_0s$, that is the part of $\Delta^{\scriptscriptstyle{(1)}}s$ in Nul$(J)$. It turns out that examining \eqref{eq: exp B dv} at $\cO\big((\Delta v)^2\big)$ is useful for this purpose, and it is

\begin{equation}
    0=J_{ab}\Delta^{\scriptscriptstyle{(2)}}s_b + \frac{1}{2}\partial_{s_c}\partial_{s_b}B_a(s^*,0)\Delta^{\scriptscriptstyle{(1)}}s_b\Delta^{\scriptscriptstyle{(1)}}s_c + \partial_{s_b}\partial_uB_a(s^*,0)\Delta v\Delta^{\scriptscriptstyle{(1)}}s_b\,.
\end{equation}
Contracting with the left eigenvectors $v_a$ of $J$ gives
\begin{equation}\label{eq: quad ds null}
    v_a\left[\frac{1}{2}\partial_{s_c}\partial_{s_b}B_a(s^*,0)\Delta^{\scriptscriptstyle{(1)}}s_b\Delta^{\scriptscriptstyle{(1)}}s_c + \partial_{s_b}\partial_uB_a(s^*,0)\Delta v\Delta^{\scriptscriptstyle{(1)}}s_b\right]=0\,.
\end{equation}
Since the left and right null spaces of a square matrix have the same dimension, these are $\dim(\text{Nul}(J))$ quadratic equations for the same number of variables in $\Delta^{\scriptscriptstyle{(1)}}_0s$. The next step would be to find bases for the left and right null spaces of $J$, and solve \eqref{eq: quad ds null} for the coefficients of $\Delta^{\scriptscriptstyle{(1)}}_0s$ in this basis. We refer the reader to Appendix \ref{app: null vects J} for an explicit construction of such bases. Although the number of equations and variables match, it turns out that \eqref{eq: quad ds null} has an infinite number of solutions in all cases. This is a reflection of the fact that the Bethe equations \eqref{eq: Bethe denom cleared} have infinitely many solutions. For example, if $\{s_1^*,\ldots, s_6^*\}$ is the classical solution associated with the partition in Figure \ref{fig:eg inf solns}, where the gauge fugacities are attached to the boxes as labelled, then one can check that $\{c_1s_1^*,c_2s_2^*,c_2s_3^*,c_1s_4^*,c_1s_5^*,c_2s_6^*\}$ with $c_{1,2}\in\bC$ are solutions of \eqref{eq: Bethe denom cleared} at any $v$.

A common property of these infinite families of solutions is that there exist in them groups of three fugacities whose values are related via

\begin{equation}\label{eq: wheel combi}
    (s_1,s_2,s_3) = (x, t_1x, t_1t_2 x)\quad \text{or}\quad (x, t_2x, t_1t_2 x)\,,\quad x\in\bC\,.
\end{equation}
The Fock modules $\cF(u)$ of quantum toroidal $\mathfrak{gl}_1$ vanish on solutions containing \eqref{eq: wheel combi}. Such vanishing conditions are called `wheel conditions' in \cite{feigin2015quantum}, and they are satisfied by construction. Since the members of $\cF(u)$ become eigenvectors of the integrable Hamiltonians when evaluated on solutions to the Bethe equations, and the eigenvectors should not vanish, one should regard solutions that contain \eqref{eq: wheel combi} as unphysical. Therefore, we shall reject solutions to \eqref{eq: quad ds null} where $s^* + \Delta^{\scriptscriptstyle{(1)}}s$ contains \eqref{eq: wheel combi}, and hope that what remains is a single solution that can be used as the initial predictor. 

At this point, it is helpful to visualise some of these statements in terms of Young diagrams. All rank $\geq 2$ partitions correspond to classical solutions $s^*$ which contain at least one group of fugacities related by \eqref{eq: wheel combi}. Whenever there are $3$ boxes arranged in a L shape or rotated L shape like the ones in red and green in Figure \ref{fig:eg inf solns}, the corresponding fugacities in $s^*$ obey \eqref{eq: wheel combi}. If a partition contains at most $n$ such L shapes, we shall call it a $n$-wheel partition. For example, Figure \ref{fig:corner def} is a 1-wheel partition, and Figure \ref{fig:eg inf solns} is a 2-wheel partition. Therefore visually, for $s^* + \Delta^{\scriptscriptstyle{(1)}}s$ to be a physical solution, $\Delta^{\scriptscriptstyle{(1)}}s$ must vary the fugacities of one or more boxes in each L shape such that the relationships \eqref{eq: wheel combi} are destroyed.

\begin{figure}[H]
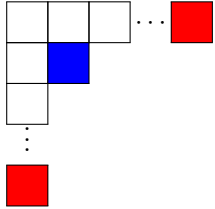

    \centering
    \begin{ytableau}
        {} & {} & {} & \none[\cdots] & *(red) \\
        {} & *(blue) \\
        {} \\
        \none[\vdots] \\
        *(red)
    \end{ytableau}
    \caption{\small Young diagram of a generic $1$-wheel partition. The exterior and interior corners are coloured red and blue respectively.}
    \label{fig:1-wheel partition}
\end{figure}

Now, we find empirically that a physical solution is contained among the solutions to \eqref{eq: quad ds null} iff $s^*$ corresponds to a $1$-wheel partition. Otherwise, for all the $n$-wheel partitions with $n\geq 2$ that were examined, all of the solutions to \eqref{eq: quad ds null} are unphysical, and one expects that higher order terms $\Delta^{\scriptscriptstyle{(\geq 2)}}s$ are needed to destroy the relations \eqref{eq: wheel combi}. A $1$-wheel partition generically looks like Figure \ref{fig:1-wheel partition}, and the physical solution to \eqref{eq: quad ds null} is 

\begin{equation}\label{eq: 1-wheel predictor}
    \Delta^{\scriptscriptstyle{(1)}}s_a = \begin{cases}
        -\frac{\partial_vB_a(s^*,0)\Delta v}{J_{aa}}\,,\quad &\text{if $a$ is an exterior corner} \\
        -\frac{2\partial_u\partial_{s_a}B_a(s^*,0)\Delta v}{\partial_{s_a}^2B_a(s^*,0)}\,,\quad &\text{if $a$ is an interior corner}\\
        0\,,\quad &\text{otherwise}
    \end{cases}
\end{equation}
Its derivation is covered in Appendix \ref{app: phys sol}.

In this work, we shall restrict our attention to rank $1$ partitions and rank $2$ $1$-wheel partitions, leaving the tracking of the classical solutions of higher-wheel partitions to future work, since it would involve examining corrections to even higher order. In the rank $1$ case, a conventional predictor-corrector method can be used throughout since $\det J\neq 0$ everywhere. For the rank $2$ $1$-wheel partitions, the initial predictor step has to be replaced by \eqref{eq: 1-wheel predictor}, after which the conventional method can take over since $\det J\neq 0$. Among the triangular partitions defined in \eqref{def triangular partitions}, those for $2\leq N\leq 7$ are within the restricted set of partitions that we can treat, and we tracked their corresponding classical solutions from $v = 0$ to $v = 1$ in \eqref{eq: Bethe denom cleared}. It turns out that the resulting solutions at $v = 1$ precisely match the black hole solutions found in Section \ref{subsec: finite N sols}. This is a nontrivial check of the correspondence between black hole solutions and triangular partitions.

\appendix
\addtocontents{toc}{\protect\setcounter{tocdepth}{1}}

\section{\texorpdfstring{$\hat{a}$}{}-genus Asymptotics \& Regularisation}\label{appendix:asymptotics}

In this appendix, we consider the asymptotics and regularisation of the $\hat{a}$-genus of a particular weight combination. The $\hat{a}$-genus of a weight $w$ is defined by
\begin{equation}
    \hat{a}(w) := \frac{1}{w^{-1/2}-w^{1/2}},
\end{equation}
and extended multiplicatively to sums of weights as $\hat{a}(w_1+w_2) = \hat{a}(w_1)\hat{a}(w_2)$. We often write $\hat{a}(W)$ as shorthand for $\hat{a}\left(\sum_{w\in W}w\right)$ where $W$ is a weight space. The weight combination that appears frequently in the main body of this work is
\begin{equation}
    \hat{a}\left( \frac{1-t^{-1}q}{1-q} w\right).
\end{equation}
In this appendix we are interested in the asymptotics of this quantity as $q \to 1$. We first use the multiplicative property to give:
\begin{equation}
\begin{split}
    \hat{a}(w) &= \exp\left(-\log\left(w^{-1/2}-w^{1/2}\right)\right), \\   
    \hat{a}\left( \frac{1}{1-q}w\right) &= \exp\left(-\sum_{i=0}^{\infty} \log\left(q^{i/2}w^{-1/2}-q^{i/2}w^{1/2}\right)\right), 
\end{split}    
\end{equation}
and the quantity we are interested in is then given by
\begin{equation}
    \hat{a}\left( \frac{1-t^{-1}q}{1-q}w\right) = \exp\left( - \sum_{i=0}^{\infty} \log \frac{q^{-i/2}w^{-1/2}-q^{i/2}w^{1/2}}{t^{1/2}q^{-1/2}q^{-i/2}w^{-1/2}-t^{-1/2}q^{1/2}q^{i/2}w^{1/2}}\right).
\end{equation}
The exponent of the right-hand side may be split:
\begin{equation}\label{eq:aroofexpanded}
\begin{aligned}
        \sum_{i=0}^{\infty} \Big(&-\log\left(-q^{-i/2}w^{-1/2}\right) - \log\left(1-q^{i}w\right) \\
        &+ \log\left(-t^{1/2}q^{-1/2}q^{-i/2}w^{-1/2}\right)+\log\left(1-t^{-1}q q^i w\right)\Big).
\end{aligned}
\end{equation}
The first and third terms above require zeta function regularisation. This is equivalent to using the following identity to re-write the summations:
\begin{equation}
    \exp\left(\sum_{i=0}^{\infty} \log f(i)\right) = \left.\exp \left( \frac{d}{ds} \sum_{i=0}^{\infty} f(i)^s \right) \right|_{s=0}.
\end{equation}
These terms of the exponent individually diverge as $\sim s^{-2}$ at $s=0$, but the divergence is cancelled in the sum. More precisely, we have
\begin{equation}
    -\frac{d}{ds}\sum_{i=0}^{\infty}(-q^{-i/2}w^{-1/2}) = \frac{2}{s^2\log q} + \left( - \frac{\log q}{24} + \frac{1}{4}\log(-w) - \frac{1}{4} \frac{\log^2 (-w)}{\log q}\right) + \ldots,
\end{equation}
and
\begin{equation}
    \frac{d}{ds} \sum_{i=0}^{\infty} (-t^{1/2}q^{-1/2}w^{-1/2}q^{-i/2}) = -\frac{2}{s^2 \log q} + \left( \frac{\log q}{24} - \frac{1}{4} \log(-t^{-1}q w) + \frac{1}{4} \frac{\log^2(-t^{-1}qw)}{\log q}\right) + \ldots.
\end{equation}
where $\ldots$ indicate higher order terms in $s$. The combination gives in total at $s=0$:
\begin{equation}
    \frac{1}{4}\left( \log (-w) - \log(-t^{-1}q w) - \frac{\log^2(-w)}{\log q} + \frac{\log^2 (-t^{-1}q w)}{\log q}\right).
\end{equation}

This regularisation allows us to re-write the $\hat{a}$-genus in terms of $q$-Pochhammer symbols. Using the definition
\begin{equation}
    \left(x;q\right)_{\infty} := \exp \sum_{i=0}^{\infty}\log(1-q^i x), 
\end{equation}
we find
\begin{equation}
    \hat{a}\left( \frac{1-t^{-1}q}{1-q} w \right) = e^{-\mathcal{E}[-\log(-t^{-1}qw)]-\mathcal{E}[-\log(-w)]}\frac{(t^{-1}qw;q)_{\infty}}{(w;q)_{\infty}},
\end{equation}
with the regularisation term is given by
\begin{equation}
    \mathcal{E}[x] = \frac{\epsilon}{24} - \frac{x}{4} + \frac{x^2}{4 \epsilon},
\end{equation}
where we set $q = e^{-\epsilon}$.

\paragraph{Asymptotics.}
We now consider the asymptotics as $\epsilon \to 0$. In this limit the regularisation terms contribute the following
\begin{equation}
    \frac{1}{\epsilon}\left( \frac{1}{4}\log^2 (-w) - \frac{1}{4}\log^2 (-t^{-1}w)\right) + \frac{1}{4}\log (-w) + \frac{1}{4}\log(-t^{-1}w) + \ldots
\end{equation}
We now consider the contribution to the asymptotics of the second and fourth terms in \eqref{eq:aroofexpanded}. We Taylor expand the logarithm and perform the geometric series over $i$. Setting $q=e^{-\epsilon}$ and expanding using the definition of the dilogarithm
\begin{equation}\label{dilog def}
    \Li_2(x) = \sum_{n=1}^{\infty}\frac{x^n}{n^2},
\end{equation}
we find that the contribution of the second and fourth terms is
\begin{equation}
    \frac{1}{\epsilon}\left( \Li_2(w) - \Li_2(t^{-1}w)\right) - \frac{1}{2}(\log(1-w)+\log(1-t^{-1}w)) + \ldots.
\end{equation}
We thus find in total the asymptotic expansion
\begin{equation}\label{eq:aroofasymp1}
\begin{split}
    \hat{a}\left(\frac{1-t^{-1}q}{1-q} w\right) = 
    & \exp \Bigg( \frac{1}{\epsilon}\left( \Li_2(w) - \Li_2(t^{-1}w) + \frac{1}{4}\log^2(-w) -  \frac{1}{4}\log^2(-t^{-1}w)\right) \\&+ \frac{1}{2}\left(-\log(1-w)-\log(1-t^{-1}w) + \frac{1}{2}\log(-w) + \frac{1}{2}\log(-t^{-1}w)\right) + \ldots \Bigg),
\end{split}
\end{equation}
as $\epsilon \to 0$ with $q=e^{-\epsilon}$.

\paragraph{Symplectic pairing.}
Now let us consider the asymptotics of the $\hat{a}$-genus of a weight space $T$:
\begin{equation}\label{eq:aroofasymp2}
    \hat{a}\left( \frac{1-t^{-1}q}{1-q}\,T \right).
\end{equation}
Equation \eqref{eq:aroofasymp1} gives us the sub-leading contribution to the limit $\epsilon \to 0$ as
\begin{equation}
    \prod_{w \in T} \frac{\left(-t^{-1} w\right)^{1/4}\left(- w\right)^{1/4}}{(1-w)^{1/2}(1-t^{-1}w)^{1/2}},
\end{equation}
which we may re-write as
\begin{equation}
    \prod_{w\in T} \left(\frac{1}{(w^{-1/2}-w^{1/2})(t^{-1/2}w^{1/2}-t^{1/2}w^{-1/2})} \right)^{1/2},
\end{equation}
which may be re-expressed as
\begin{equation}
    \prod_{w\in T + t T^{\vee}}\hat{a}^{1/2}(w),
\end{equation}
where the weights in $T$ and $T^{\vee}$ are paired under $w \leftrightarrow w^{-1}$. In the context of the main body of this work, the pairing between $T$ and $tT^{\vee}$ corresponds to the symplectic pairing of matter weights as a result of $\mathcal{N}=4$ supersymmetry. Let us now consider the $O(1/\epsilon)$ part of the exponential. We use the dilogarithm identity 
\begin{equation}\label{dilog inv identity}
    \Li_2(a) + \Li_2(a^{-1}) = -\frac{1}{2} \log^2(-a) - \frac{\pi^2}{6},
\end{equation}
to flip the argument of the second dilogarithm in \eqref{eq:aroofasymp1} giving
\begin{equation}
    \frac{1}{\epsilon}\left( \Li_2(w) + \Li_2(t w^{-1})+ \frac{1}{4}\log^2(-w)  + \frac{1}{4} \log^2(-t w^{-1}) + \frac{\pi^2}{6}\right).
\end{equation}
This is again a sum of terms respecting the $w \leftrightarrow t w^{-1}$ pairing. In conclusion, the asymptotic expansion of \eqref{eq:aroofasymp2} as $\epsilon \to 0$ is
\begin{equation}
\begin{split}
    \prod_{w \in T}\hat{a}\left( \frac{1-t^{-1}q}{1-q}w\right) = \hat{a}^{1/2}\left(W\right)\exp \left( \frac{1}{\epsilon} \sum_{w \in W }f(w) + \ldots \right)
\end{split}
\end{equation}
where $W = T + t T^{*}$ and $f(w) := \Li_2(w) + \frac{1}{4} \log^2 (-w) + \frac{\pi^2}{12}$, which we often abbreviate to $f(w) = \Li_2'(w) + \pi^2 / 12$.

\subsection{Remark on Factorisation}
We conclude the appendix with a brief remark that it is possible to factorise the twisted index \eqref{eq:cb_localisation_twisted_index} `under the integral' sign by using a fusion identity for the $\hat{a}$-genus. This is closely related to the asymptotic result above. We begin with the $\hat{a}$-genus of a weight $w$ raised to some integer power $m$ (such terms appear in the twisted index). We may then introduce a $q$-refinement: 
\begin{equation}
    \hat{a}(w)^m \to \hat{a}(w q^{-m/2})\hat{a}(w q^{-m/2 + 1})\ldots \hat{a}(w q^{m/2}) .
\end{equation}
The following fusion identity then follows from the regularisation of the $\hat{a}$-genus considered in the previous subsection:
\begin{equation}
\begin{split}
    &\hat{a}\left( \frac{1-t^{-1}q}{1-q}w q^m\right)\hat{a}\left( \frac{1-t^{-1}q^{-1}}{1-q^{-1}}w q^{-m}\right) =\\ &\hat{a}(w q^{-m/2})\hat{a}(w q^{-m/2 + 1})\ldots \hat{a}(w q^{m/2})\hat{a}(tw^{-1} q^{-m/2})\ldots \hat{a}(tw^{-1} q^{m/2}),
\end{split}    
\end{equation}
this is the property of the $\hat{a}$-roof genus that allows factorisation.

\section{'t Hooft \& Wilson Line Operator Equalities}\label{appendix:lineoperators}

In this section, we prove for two instances, namely ADHM theory with $N = 2$ and SQED[2], that the 't Hooft operator insertions $p_{1,2} = e^{t_{1,2}\partial_{t_{1,2}}\cW}$ are identical to a combination of Wilson line insertions when evaluated on solutions to the Bethe equations \eqref{eq:bethe_eqns}, and are hence elements of the quantum equivariant $K$-theory. Further, they yield the same expectation values when computed inserted in the twisted background $S^2 \times_B S^1$. This serves as evidence that the same is true for ADHM theory with generic $N$.

\subsection{SQED[2]}

For $\cN=4$ supersymmetric QED with N hypermultiplets, the effective twisted superpotential is given by:
\begin{equation}\label{WSQED2}
    \mathcal{W} = - \log s \log \zeta  - N \text{Li}'_2(t)
    +\sum_{i=1}^N \left[\text{Li}'_2\Big(s x_i t^{\frac{1}{2}}\Big)+\text{Li}'_2\Big(s x_i t^{\frac{1}{2}}\Big)\right].
\end{equation}
where the flavour symmetry has a maximal torus $U(1)^{N-1}$, for which we introduce fugacities $x_1,\ldots x_{N-1}$, and define $x_N =(x_1\ldots x_{N-1})^{-1}$. Also, $t$ is the fugacity for the same anti-diagonal R-symmetry $(R_H-R_C)/2$ as in the main body, which can be regarded as an $\cN=2$ flavour symmetry. 

For simplicity, take $N=2$, the general case can be treated similarly. We take $x_1 := x$, $x_2 = x^{-1}$. The vacuum equations are given by:
\begin{equation}\label{eq:sqed2}
    e^{s \partial_s \cW} = \zeta^{-1}\frac{\left(s x-t^{\frac{1}{2}}\right) \left(\frac{s}{x}-t^{\frac{1}{2}}\right)}{ \left(1-s t^{\frac{1}{2}} x\right) \left(1-\frac{s t^{\frac{1}{2}}}{x}\right)} = 1.
\end{equation}
This imposes a quadratic relation on $s$, and so the twisted chiral ring, which is equal to $QK_T(T^*P^1)$, can be generated by the elements $\cO \in \{ 1, s\}$.  The Bethe roots can be determined analytically in this case as
\begin{equation}
\begin{aligned}
   s &=& \frac{t^{\frac{1}{2}}(1+x^2)(1-\zeta) \pm \sqrt{ t (1+x^2)^2(1-\zeta)^2-4 (t x-\zeta  x) (x-\zeta  t x)}}{2 (x-\zeta  t x)}
\end{aligned}
\end{equation}

The 't Hooft operator for the $x$ flavour symmetry is given by:
\begin{equation}
    p_x = e^{x \partial_x \cW } = \frac{\left(s x-t^{\frac{1}{2}}\right) \left(1-\frac{s t^{\frac{1}{2}}}{x}\right)}{\left(1-s t^{\frac{1}{2}} x\right) \left(\frac{s}{x}-t^{\frac{1}{2}}\right)}
\end{equation}
and that of the $\cN=2$ flavour symmetry by:
\begin{equation}
    p_t = e^{ 2 t \partial_t \cW } =\frac{(1-t)^2}{\left(1-s t^{\frac{1}{2}} x\right) \left(1-\frac{t^{\frac{1}{2}}}{s x}\right) \left(1-\frac{s t^{\frac{1}{2}}}{x}\right) \left(1-\frac{t^{\frac{1}{2}} x}{s}\right)}
\end{equation}
Note the additional power of $2$ in the definition of $p_t$. This is due to the Dirac quantisation condition; the flux for $(R_H-R_C)/2$ must be even as there are half-charged hypermultiplets under this symmetry. Thus, this is the minimum 't Hooft operator for this symmetry.

The goal is to express both 't Hooft operators in the form:
\begin{equation}
    p_x = A_x + B_x s , \quad p_t = A_t + B_t s 
\end{equation}
$\textit{up to}$ terms which vanish on the vacuum equations \eqref{eq:sqed2}. This demonstrates that in the $q \rightarrow 1$ limit, the 't Hooft operator insertions in the twisted index and hemisphere partition function are equivalent to the insertion of Wilson lines. In this simple case, it is possible to find the interpolating (linear) polynomial in $s$ between the values of $p_x$ or $p_t$ evaluated at the two roots of the vacuum equation using \texttt{Mathematica}.\footnote{For the ADHM example this is computationally intractable, so we turn to other methods. See the next subsection.}

One may explicitly verify that:
\begin{equation}
\begin{aligned}
    p_x  = & \, \frac{ \left(x^2-1\right) (x^2 \zeta ^2 -1)t +\zeta  (t-1)^2 x^2}{\zeta  \left(t x^2-1\right)^2}
    -\frac{(\zeta +1)  t^{\frac{1}{2}} x \left(x^2-1\right) (\zeta  t-1)}{\zeta  \left(t x^2-1\right)^2} \, s\\
    &+ \frac{t^{\frac{1}{2}} \left(x^2-1\right) \left(s t^{\frac{1}{2}}-x\right) \left(-(\zeta +1) s t^{\frac{1}{2}} x+\zeta  t x^2+1\right)}{\left(t x^2-1\right)^2 \left(s-t^{\frac{1}{2}} x\right)} \left(e^{s \partial_s \cW} -1\right).
\end{aligned}
\end{equation}
The second line clearly vanishes on the vacuum equations, and so the 't Hooft operator is equal to the insertion of the linear combination of Wilson line operators on the first line, which can be interpreted as a linear combination of $QK_T(T^*\bP^1)$ insertions.

Further:
\begin{equation}
\begin{aligned}
    p_t = &\,\frac{x^2 (t-\zeta ) \left(t \left(t x^4+t x^2+\zeta  (x (t-x+1)-1) (x (t+x+1)+1)+t-2 x^2\right)-x^2\right)}{\zeta  \left(t-x^2\right)^2 \left(t x^2-1\right)^2} \\
    &+ \frac{(\zeta +1)  \sqrt{t} \left(t^2-1\right) x^3 \left(x^2+1\right) (\zeta  t-1)}{\zeta  \left(t-x^2\right)^2 \left(t x^2-1\right)^2} \, s \\
    &+ \frac{e^{s \partial_s \cW} -1}{\left(t-x^2\right)^2 \left(t x^2-1\right)^2 \left(s x-\sqrt{t}\right) \left(s-\sqrt{t} x\right)} \Bigg((\zeta +1) s^3 \left(t^2-1\right) t^{3/2} x^4 \left(x^2+1\right)+\ldots \\
    &\ldots +s^2 t x^3 \left(\zeta -\left(t^3 \left(x^4+\zeta  \left(x^2+1\right)^2+x^2+1\right)\right)-t^2 \left(x^4+\zeta  x^2+1\right) \right)\\
    &\ldots +s^2 t x^3 \left(t \left(\zeta +\zeta  x^4+x^2\right) + x^2 \left(\zeta +(\zeta +1) x^2+2\right)+1\right)\\
    &\ldots +s \sqrt{t} x^2 \left(x^2+1\right) \left(t \left(-\zeta +\zeta  x^2 \left(t^3+t^2+t-x^2-1\right)+t \left(t x^4+(t-1) x^2+t\right)-x^2\right)-x^2\right)\\
    &\ldots +t x^3 \left(x^2-t \left(t x^4+t x^2+\zeta  (x (t-x+1)-1) (x (t+x+1)+1)+t-2 x^2\right)\right) \Bigg),
\end{aligned}
\end{equation}
which demonstrates the equality of the 't Hooft operator $p_t$ to a linear combination of Wilson line or $QK_T(T^*\bP^1)$ insertions on the vacuum equations.

\subsection{ADHM with \texorpdfstring{$N = 2$}{}}
 
In ADHM theory at $N = 2$, the 't Hooft operator insertions can be massaged into the form
\bea\label{eq: raw 'thooft}
p_1 &= e^{t_1\partial_{t_1}\cW} = -\frac{t_1^2\big(1-t_3^{-1}\big)^2}{(1-t_1)^2}\frac{s_1^\frac{1}{2}s_2^\frac{1}{2}(s_2-t_3s_1)(s_1-t_3s_2)}{(s_2-t_1s_1)(s_1-t_1s_2)\Big(s_1-t_3^\frac{1}{2}\Big)^\frac{1}{2}\Big(s_2-t_3^\frac{1}{2}\Big)^\frac{1}{2}\Big(s_1-t_3^{-\frac{1}{2}}\Big)^\frac{1}{2}\Big(s_2-t_3^{-\frac{1}{2}}\Big)^\frac{1}{2}}\,,\\
p_2 &= e^{t_2\partial_{t_2}\cW} = -\frac{t_2^2\big(1-t_3^{-1}\big)^2}{(1-t_2)^2}\frac{s_1^\frac{1}{2}s_2^\frac{1}{2}(s_2-t_3s_1)(s_1-t_3s_2)}{(s_2-t_2s_1)(s_1-t_2s_2)\Big(s_1-t_3^\frac{1}{2}\Big)^\frac{1}{2}\Big(s_2-t_3^\frac{1}{2}\Big)^\frac{1}{2}\Big(s_1-t_3^{-\frac{1}{2}}\Big)^\frac{1}{2}\Big(s_2-t_3^{-\frac{1}{2}}\Big)^\frac{1}{2}}\,.
\eea
The square root of the product of the two Bethe equations \eqref{eq:bethe_eqns} is
\be
\zeta^{-1}t_3^\frac{1}{2}\frac{\Big(s_1 - t_3^{-\frac{1}{2}}\Big)^\frac{1}{2}\Big(s_2 - t_3^{-\frac{1}{2}}\Big)^\frac{1}{2}}{\Big(s_1 - t_3^{\frac{1}{2}}\Big)^\frac{1}{2}\Big(s_2 - t_3^{\frac{1}{2}}\Big)^\frac{1}{2}} = 1\,.
\ee
Therefore, when evaluated on solutions to the Bethe equations, \eqref{eq: raw 'thooft} is equal to
\bea
p_1 &= -\frac{t_1^2t_3^{-\frac{1}{2}}\zeta\big(1-t_3^{-1}\big)^2}{(1-t_1)^2}\frac{s_1^\frac{1}{2}s_2^\frac{1}{2}(s_2-t_3s_1)(s_1-t_3s_2)}{(s_2-t_1s_1)(s_1-t_1s_2)\Big(s_1-t_3^{-\frac{1}{2}}\Big)\Big(s_2-t_3^{-\frac{1}{2}}\Big)}\,,\\
p_2 &= -\frac{t_2^2t_3^{-\frac{1}{2}}\zeta\big(1-t_3^{-1}\big)^2}{(1-t_2)^2}\frac{s_1^\frac{1}{2}s_2^\frac{1}{2}(s_2-t_3s_1)(s_1-t_3s_2)}{(s_2-t_2s_1)(s_1-t_2s_2)\Big(s_1-t_3^{-\frac{1}{2}}\Big)\Big(s_2-t_3^{-\frac{1}{2}}\Big)}\,.
\eea
In the twisted index, the common factor $s_1^\frac{1}{2}s_2^\frac{1}{2}$ above contributes $(s_1s_2)^\frac{\fn_1+\fn_2}{2}$ from $p_1^{\fn_1} p_2^{\fn_2}$. Due to the Dirac quantisation condition $\fn_1 + \fn_2\in 2\bZ$, this is an integer power of the determinant line bundle $\cL = s_1 s_2$ and is already an element of quantum $K$-theory. It remains to show that the fractions
\bea\label{eq: frac in p}
\frac{(s_2-t_3s_1)(s_1-t_3s_2)}{(s_2-t_1s_1)(s_1-t_1s_2)\Big(s_1-t_3^{-\frac{1}{2}}\Big)\Big(s_2-t_3^{-\frac{1}{2}}\Big)} &= \frac{n}{d_1}\,,\\
\frac{(s_2-t_3s_1)(s_1-t_3s_2)}{(s_2-t_2s_1)(s_1-t_2s_2)\Big(s_1-t_3^{-\frac{1}{2}}\Big)\Big(s_2-t_3^{-\frac{1}{2}}\Big)} &= \frac{n}{d_2}\,,
\eea
coincide with a symmetric polynomial in $s_{1,2}$ when evaluated on Bethe roots. The Bethe equations with denominators cleared are
\bea
B_1 = \Big(s_1-t_3^{-\frac{1}{2}}\Big)(s_2-t_1^{-1}s_1)&(s_2-t_2^{-1}s_1)(s_2-t_3^{-1}s_1)\\
&+\zeta t_3^{-\frac{1}{2}}\Big(s_1 - t_3^\frac{1}{2}\Big)(s_2-t_1s_1)(s_2-t_2s_1)(s_2-t_3s_1) = 0\,,\\
B_1 = \Big(s_2-t_3^{-\frac{1}{2}}\Big)(s_1-t_1^{-1}s_2)&(s_1-t_2^{-1}s_2)(s_1-t_3^{-1}s_2)\\
&+\zeta t_3^{-\frac{1}{2}}\Big(s_2 - t_3^\frac{1}{2}\Big)(s_1-t_1s_2)(s_1-t_2s_2)(s_1-t_3s_2) = 0\,.
\eea
Our strategy is to show that the numerator $n$ in \eqref{eq: frac in p} can be written as
\be\label{eq: numerator decomp}
n = q_1 d_1 + a_1 B_+ + b_1B_- = q_2 d_2 + a_2 B_+ + b_2B_-\,,\quad B_+ = B_1+B_2\,,\quad B_-=\frac{B_1-B_2}{s_1-s_2}\,, 
\ee
where $q_{1,2}, a_{1,2}, b_{1,2}$ are symmetric polynomials in $s_{1,2}$ whose coefficients are rational functions of $t_{1,2,3}$ and $\zeta$. They must be symmetric since $n$ and $d_{1,2}$ are symmetric polynomials by their definition in \eqref{eq: frac in p}, and since $B_1\leftrightarrow B_2$ under $s_1\leftrightarrow s_2$, $B_\pm$ are also symmetric. In this way, $n/d_1$ is equal to $q_1$ on Bethe roots, and $n/d_2$ is equal to $q_2$. \footnote{More generally, $a_{1,2}$ and $b_{1,2}$ could be rational functions of $s_{1,2}$, as long as their denominators do not vanish on the Bethe roots. Then it would still be true that the fractions $n/d_{1,2}$ are equal to $q_{1,2}$ on the Bethe roots. However, it turns out here that we can prove \eqref{eq: numerator decomp} while restricting $a_{1,2}$ and $b_{1,2}$ to be polynomials. We do not rule out the possibility that $a_{1,2}$ and $b_{1,2}$ might have to be rational functions for $N>2$.} Note that $B_1 - B_2$ has $s_1 - s_2$ as a factor, so that $B_-$ is a polynomial. In addition, one can show that $B_-$ does not vanish when $s_1 = s_2$, meaning that $B_\pm$ only vanish on physical Bethe roots on the Coulomb branch of the gauge theory where $s_1\neq s_2$, so that the gauge group is broken to $U(1)$. The trick of dividing by $s_1 - s_2$ is taken from \cite{Jiang:2017phk}. Since $n$, $d_{1,2}$, $B_\pm$ are symmetric polynomials, we can express them in terms of the elementary symmetric polynomials
\be
e_1 = s_1 + s_2\,,\quad e_2 = s_1 s_2\,.
\ee
Now \eqref{eq: numerator decomp} is equivalent to saying that $n$ is a member of the polynomial ideals generated by $\{d_1$, $B_+$, $B_-\}$ and $\{d_2$, $B_+$, $B_-\}$, i.e.
\be\label{eq: ideal membership}
n\in I_1 = \langle d_1\,, B_+\,, B_- \rangle \quad \land \quad n\in I_2 = \langle d_2\,, B_+\,, B_- \rangle\,.
\ee
The standard algorithmic way to determine whether \eqref{eq: ideal membership} is true (the ideal membership problem) is to compute Gr\"obner bases for $I_1$, $I_2$, and reduce $n$ with respect to the bases \cite{compAlgGeom}. Then \eqref{eq: ideal membership} is true if and only if the remainder is zero under the division algorithm. Choosing lexicographic ordering in $\{e_1, e_2\}$, we compute the Gr\"obner bases for $I_1$ and $I_2$ using \texttt{ExtendedGroebnerBasis} in Mathematica. The result turns out to be the same for $I_1$ and $I_2$:
\be
I_1 = I_2 = \langle e_2(e_2 - 1)\,,\quad (1+t_3^{-1})e_2 - t_3^{-\frac{1}{2}}e_1e_2\,,\quad t_3^{-1}e_1^2 - (1+t_3^{-1})^2e_2\rangle\,,
\ee
implying that they are actually the same ideal. Since the numerator $n$ is precisely a multiple of the third basis element
\be
n = -t_3^2\left[t_3^{-1}e_1^2 - (1+t_3^{-1})^2e_2\right]\,,
\ee
the remainder of reducing $n$ with respect to the Gr\"obner basis is zero and \eqref{eq: ideal membership} is true. We have thus proven that \eqref{eq: numerator decomp} must be true for some polynomials $q_{1,2}, a_{1,2}, b_{1,2}$. To find these polynomials explicitly, we can use the conversion matrix returned by \texttt{ExtendedGroebnerBasis}, which gives the basis element $t_3^{-1}e_1^2 - (1+t_3^{-1})^2e_2$ in terms of the starting bases $\{d_1, B_+, B_-\}$ and $\{d_2, B_+, B_-\}$. The result is  
\bea
&q_1 = t_1^{-1} t_3^{2} \zeta ^{-3} \left(1-t_1\right)^{-1} \left(1+t_1\right){}^{-3} \left(1+t_2\right)^{-1} \left(t_3^{-1}-1\right){}^{-3} \left(t_3^{-\frac{1}{2}}+\zeta \right)^{-1}\bigg\{\zeta ^3 \left(t_1^2-1\right){}^2\times\\
&\left(1+t_2\right) \left(t_3^{-1}-1\right){}^2 \left(1+t_3^{-1}\right) \left(t_3^{-\frac{1}{2}}+\zeta \right)+\left(1+t_3^{-\frac{1}{2}} \zeta \right)^2 \bigg[t_1 \left(1+t_2\right)+t_3^{-\frac{1}{2}} \zeta  \left(1+t_1^2+t_1^3+\frac{1+2 t_1}{t_3}\right)\\
&+\zeta ^2 \left(1-t_1-t_1^2+t_1^3
   \left(1+t_2 \left(2+t_1\right) \left(1+t_2\right)\right)\right)+t_1^3 t_3^{-\frac{3}{2}} \zeta ^3 \left(1+t_2\right)\bigg] \left[2t_3^{-1} e_1^2-\left(1+t_1\right)
   \left(1+t_2\right) \left(1+t_3^{-1}\right) e_2\right]\\
   &+t_3^{-\frac{1}{2}} \left(t_1-1\right) \left(t_3^{-1}-1\right) \bigg[t_3^{-1}(1+t_2)+t_3^{-1}\zeta  \left(1+t_2+t_1 \left(t_2-1\right) \left(t_1^2+t_2+2 t_1
   \left(1+t_2\right)\right)\right)\\
   &+\zeta ^2 \left(1+t_2+t_1 \left(2-t_1-t_1^2-t_2 \left(1+3 t_1+t_1^2+t_1^3\right)-t_2^2 \left(1+2 t_1\right)+t_2t_3^{-2}
   \left(2+t_1\right)\right)\right)\\
   &+t_3^{-\frac{1}{2}} \zeta ^3 \left(1+t_1^2+t_1^3-t_2 \left(1+t_1-t_1^2+t_1^3+t_1^4\right)-t_2 t_3^{-1}\left(1+2 t_1
   \left(1+t_1\right)\right)+t_1t_3^{-3}\right)\\
   &+\zeta ^4 \left(1-t_2-t_1 \left(1+t_2+t_1 \left(1-t_2+t_1 \left(1+t_2\right) \left(-1-\left(2+t_1\right)
   t_2+\left(1+t_1\right) t_2^2\right)\right)\right)\right)+t_1^3 t_3^{-\frac{3}{2}} \zeta ^5 \left(1+t_2\right)\bigg] e_1\bigg\}
\eea

\bea
&a_1 = 2^{-1}t_1^{-\frac{1}{2}} t_3^{\frac{3}{2}} \zeta^{-3} \left(1-t_1\right)^{-1} \left(1+t_1\right)^{-3} \left(1+t_2\right)^{-1} \left(1-t_3^{-1}\right)^{-3} \left(t_3^{-\frac{1}{2}}+\zeta \right)^{-1}\bigg\{t_2^\frac{1}{2}\left(1+t_3^{-\frac{1}{2}} \zeta \right)\times\\
&\bigg[t_1 \left(1+t_2\right)+t_3^{-\frac{1}{2}} \zeta  \left(1+t_1^2+t_1^3+t_3^{-\frac{1}{2}} \left(1+2 t_1\right)\right)+\zeta ^2
   \left(1-t_1-t_1^2+t_1^3 \left(1+t_2 \left(2+t_1\right) \left(1+t_2\right)\right)\right)\\
   &+t_1^3 t_3^{-\frac{3}{2}} \zeta ^3 \left(1+t_2\right)\bigg] \left[\left(1+t_1\right)^2
   e_2-t_1 e_1^2\right] + t_1^\frac{1}{2}\left(t_3^{-1}-1\right) \left(1+t_1 t_3^{-\frac{1}{2}} \zeta \right) \bigg[t_3^{-1}(1+t_2)\\
   &+t_3^{-\frac{1}{2}} \zeta  \left(1+t_2+t_1 \left(t_2-1\right)
   \left(t_1^2+t_2+2 t_1 \left(1+t_2\right)\right)\right)+\zeta ^2 \left(1-t_1\right) \left(1+t_1\right){}^2\\
   &+t_2 \zeta ^2 \left(1+t_1\right) \left(1-2
   t_1-t_1^3\right)-t_3^{-2}+t_3^{-3}(2+t_1)+t_1^2 t_3^{-\frac{3}{2}} \zeta ^3 \left(1+t_2\right) \left(-1+t_1 \left(-1+t_2\right)\right)\bigg] e_1\\
   &+t_1^\frac{1}{2}\left(1+t_1\right) \left(t_3^{-1}-1\right) \bigg[-t_3^{-\frac{3}{2}} \left(1+t_2\right)-t_3^{-1}\zeta  \left(2 \left(1+t_2\right)+t_1 \left(t_2-1\right)
   \left(t_1^2+t_2+2 t_1 \left(1+t_2\right)\right)\right)\\
   &+t_3^{-\frac{1}{2}} \zeta ^2 \left(-3-2 t_1+3 t_1^2+2 t_1^3+t_2 \left(-3+2 t_1+3
   t_1^2+t_1^4\right)-\frac{2+t_1}{t_3^3}\right)+\zeta ^3 \bigg(\left(-1+t_1\right) \left(1+t_1\right)^2\\
   &+t_2 \left(-1+t_1-t_1^3+t_1^4\right)+t_3^{-2}(-1+3 t_1^2+t_1^3)+2
   t_1t_3^{-3}-t_1t_3^{-4}\bigg)+t_1^3 t_3^{-\frac{5}{2}} \zeta ^4 \left(1+t_2\right)\bigg]\bigg\}
\eea

\begin{equation}
\begingroup
\allowdisplaybreaks
\begin{aligned}
&b_1 = 2^{-1} t_1^{-\frac{1}{2}} t_3^{\frac{5}{2}}\zeta^{-3} \left(1-t_1\right)^{-1} \left(1+t_1\right)^{-3} \left(1+t_2\right)^{-1} \left(1-t_3^{-1}\right)^{-3}\left(t_3^{-\frac{1}{2}}+\zeta \right)^{-1}\bigg\{t_1^\frac{1}{2} t_3^{-\frac{1}{2}} \left(1+t_3^{-\frac{1}{2}} \zeta \right)\times\\
&\bigg(t_1 \left(1+t_2\right)+t_3^{-\frac{1}{2}} \zeta  \left(1+t_1^2+t_1^3+\frac{1+2 t_1}{t_3}\right)+\zeta ^2
   \left(1-t_1-t_1^2+t_1^3 \left(1+t_2 \left(2+t_1\right) \left(1+t_2\right)\right)\right)\\
&+t_1^3 t_3^{-\frac{3}{2}} \zeta ^3 \left(1+t_2\right)\bigg) e_1^3+2 \left(1+t_1\right) \bigg[t_3^{-1}t_1^{3/2} \left(1+t_2\right)^2 e_2+t_1^\frac{1}{2} t_3^{-\frac{1}{2}} \zeta \left(1+t_1\right) \big(-2+t_1+t_1^2\\
&-t_2 \left(1-4t_1+t_1^2\right)+t_2 t_3^{-1}\left(2+t_1+t_1^2\right)+2 t_2t_3^{-2}\big) e_2+t_1^\frac{1}{2} \zeta^2 \left(1+t_2\right) \bigg(-1+t_1 \bigg(-1+t_1\\
&+t_1^2+t_2\left(-1+t_1\right)^2 \left(1+t_1\right)+t_2t_3^{-1} \left(2+t_1\right) \left(1+t_1^2\right)+t_2t_3^{-2} \left(1+4 t_1+t_1^2\right)\bigg)\bigg) e_2\\
&+t_2^\frac{1}{2} \zeta ^3\left(1+t_1\right) \bigg(t_1 \left(1-t_1\right) \left(1+t_2\right) \left(1-t_3^{-1}\right)^2 \left(t_1t_3^{-1}-1\right)+\bigg(-1+t_3^{-1}+t_1^2 \bigg(1+t_2\bigg(1\\
&+t_2+t_1 \left(-4-3 t_2+t_1 \left(1+t_2\right) \left(2+\left(1+t_1\right) t_2+2 t_2t_3^{-1}\right)\right)\bigg)\bigg)\bigg) e_2\bigg)+t_1^{7/2} t_3^{-3}\zeta ^4\left(1+t_2\right){}^2 e_2\bigg]\\
&+t_1^\frac{1}{2} \bigg[-t_3^{-1}\left(1+t_2\right) \left(-1+t_1 \left(2+t_2\right)\right)-t_3^{-\frac{1}{2}} \zeta  \Big(-1-t_2+2 t_1 \left(t_2-1\right)+t_1^2 \Big(2+t_1\\
&+t_2
   \left(3-t_1+t_1^2\right)+t_2^2 \left(2+5 t_1+t_1^2\right)+t_2^3 \left(1+3 t_1\right)\Big)\Big)-\zeta ^2 \bigg(-1-t_2+t_1 \bigg(-1+2 t_2+t_2^2\\
   &+t_1^3t_3^{-2}
   \left(t_2-1\right)+t_1^2 t_3^{-1}\left(2+3 t_2\right) \left(1+t_2\right){}^2+t_1 \left(1+t_2^2\right)+t_1^2 \left(1+t_2\right) \left(1-5 t_2+t_2^2\right)\\
   &+t_1^3t_3^{-1}
   \left(1+2 t_2+2 t_2^2+3 t_2^3\right)\bigg)\bigg)-t_3^{-\frac{1}{2}} \zeta ^3 \bigg(-2+t_1 \bigg(1+3 t_2+t_1 \left(t_2-1\right){}^2+t_1^3t_3^{-2}
   \left(t_2^2-1\right)\\
   &-t_1^2 \left(1+4 t_2+3 t_2^2\right)+t_1^2 t_3^{-2}\left(4+5 t_2+3 t_2^2\right)+t_1^3 \left(1+3 t_2
   \left(1+t_2+t_2^2\right)\right)\bigg)\bigg)\\
   &-t_1^3t_3^{-2} \zeta ^4 \left(1+t_2\right) \left(-1+t_1 \left(1+t_2 \left(2+t_1 \left(t_2-1\right)\right)\right)\right)\bigg]
   e_1^2+\bigg[-t_2^\frac{1}{2} \left(1+t_1\right)^2 \left(1+t_3^{-\frac{1}{2}} \zeta \right) \bigg(t_1 \left(1+t_2\right)\\
   &+t_3^{-\frac{1}{2}} \zeta  \left(1+t_1^2+t_1^3+t_3^{-1}(1+2
   t_1)\right)+\zeta ^2 \left(1-t_1-t_1^2+t_1^3 \left(1+t_2 \left(2+t_1\right) \left(1+t_2\right)\right)\right)+t_1^3 t_3^{-\frac{3}{2}} \zeta ^3 \left(1+t_2\right)\bigg)
   e_2\\
   &-t_1^\frac{1}{2} \left(t_1-1\right) \left(t_3^{-1}-1\right) \bigg(t_3^{-\frac{3}{2}} \left(1+t_2\right)+t_1t_3^{-1} \zeta  \left(t_2-1\right) \left(t_1^2+t_2+2 t_1
   \left(1+t_2\right)\right)\\
   &+t_3^{-\frac{1}{2}} \zeta ^2 \left(1+t_2\right) \left(\left(1+t_1\right){}^2-\frac{2+4 t_1+2 t_1^2+t_1^3}{t_3}+\frac{t_1
   \left(2+t_1\right)}{t_3^2}\right)+\zeta ^3 \bigg(\left(t_1-1\right) \left(1+t_1\right)^2\\
   &+t_2 \left(1+t_1\right) \left(-1+2 t_1+t_1^3\right)-t_3^{-2}(-1+t_1^2+t_1^3)-2t_3^{-3}
   \left(1+t_1\right)+t_1t_3^{-4}\bigg)+t_1^3 t_3^{-\frac{5}{2}} \zeta ^4 \left(1+t_2\right)\bigg)\bigg] e_1\bigg\}
\end{aligned}
\endgroup
\end{equation}
The other set of polynomials $q_2$, $a_2$ and $b_2$ can be obtained from $q_1$, $a_1$ and $b_1$ respectively by exchanging $t_1\leftrightarrow t_2$.

\section{Numerical Details}

In this appendix, we collect some details in the initial predictor step away from $\zeta = 0$ that were omitted in Section \ref{sec: reverse tracking} for brevity. 

\subsection{Singularity of the Jacobian}\label{app: singular jacobian}

For the Jacobian $J_{ab} = \partial_{s_b}B_a(s^*,0)$, we want to prove that if $s^*$ corresponds to a partition with rank $\geq 2$, then $\det J = 0$. Since the derivative of $B_a$ is evaluated at $v=0$, only the first term of \eqref{eq: Bethe denom cleared} contributes to $J$, i.e.  
\be\label{eq: J as deriv}
J_{ab}= \partial_{s_b}B_a(s,0)\big|_{s=s^*} = \partial_{s_b}\left[\Big(s_a-t_1^\frac{1}{2}t_2^\frac{1}{2}\Big)\prod_{\substack{b=1\\b\neq a}}^N(s_b-t_1^{-1}s_a)(s_b-t_2^{-1}s_a)(s_b-t_1t_2s_a)\right]_{s=s^*}\,.
\ee
If $s^*$ corresponds to a partition with rank $\geq 2$, we first show that the homogeneous part of $B_a(s,0)$ is zero on the classical solution, i.e. 
\be\label{rescale null vect cond}
\prod_{\substack{b=1\\b\neq a}}^N(s_b^*-t_1^{-1}s_a^*)(s_b^*-t_2^{-1}s_a^*)(s_b^*-t_1t_2s_a^*)=0\,,\qquad\forall a\in \{1,\ldots,N\}\,.
\ee
For all boxes $a$ in the partition other than the top-left box, there are always neighbouring boxes $b$ either above or to the left of $a$ which satisfy $s_b^*-t_1^{-1}s_a^*=0$ or $s_b^*-t_2^{-1}s_a^*=0$. Thus \eqref{rescale null vect cond} automatically holds for these boxes. When $a$ is the top-left box, the fact that the partition has rank $\geq 2$ implies that there is a box $b$ to the bottom right of $a$ with $s_b^* - t_1t_2s_a^*=0$. Now that \eqref{rescale null vect cond} holds, due to the homogeneity of this factor, any overall rescaling of $s^*$ still satisfies \eqref{rescale null vect cond}. This means that $B_a(e^\epsilon s^*,0) = 0$ for arbitrary $\epsilon$. Expanding to linear order in $\epsilon$ implies that $J_{ab}s_b^* = 0$, i.e. that $s^*$ is a right null vector of $J$, and consequently $\det J = 0$.

\subsection{Null Vectors of the Jacobian}\label{app: null vects J}

\paragraph{Right null vectors.} 

\begin{figure}[H]
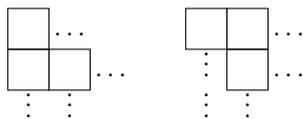

    \centering
    \begin{ytableau}
    {} & \none[\dots] \\
    {} & {} & \none[\dots] \\
    \none[\vdots] & \none[\vdots]
    \end{ytableau}\qquad 
    \begin{ytableau}
    {} & {} & \none[\dots] \\ 
    \none[\vdots] & {} & \none[\dots] \\
    \none[\vdots] & \none[\vdots]
    \end{ytableau}
    \caption{\small Shapes of the minimal subsets $S$ which satisfy \eqref{eq: indep rescaling}.}
    \label{fig: min rescaling S}
\end{figure}

Let us examine the right null space of the Jacobian and determine a spanning set of vectors. We shall first generalise the argument in Appendix \ref{app: singular jacobian} to find more null vectors. Starting from the classical solution $s^*$, one can scale a subset of the fugacities $s_a$ for $a\in S\subset \{1,\ldots,N\}$ independently while preserving \eqref{rescale null vect cond}, as long as the following is true:
\bea\label{eq: indep rescaling}
\prod_{\substack{b\in S\\b\neq a}}(s_b^*-t_1^{-1}s_a^*)(s_b^*-t_2^{-1}s_a^*)(s_b^*-t_1t_2s_a^*)&=0\,,\qquad\forall a\in S\subset\{1,\ldots,N\}\,,\\
\prod_{\substack{b\in S'\\b\neq a}}(s_b^*-t_1^{-1}s_a^*)(s_b^*-t_2^{-1}s_a^*)(s_b^*-t_1t_2s_a^*)&=0\,,\qquad\forall a\in S'=\{1,\ldots,N\}\setminus S\,.
\eea
Then $s_a = e^{\epsilon_1} s_a^*$ for $a\in S$, and $s_a = e^{\epsilon_2} s_a^*$ for $a\in S'$ still satisfies \eqref{rescale null vect cond} and $B_a(s,0)=0$. As in Appendix \ref{app: singular jacobian}, $B_a(s,0)=0$ at linear order in $\epsilon_{1,2}$ implies that the vectors
\be
(v_1)_a = \begin{cases}
    s_a^*\,,\quad &a\in S\,,\\
    \,0\,,\quad &a\in S'\,,
\end{cases} \qquad 
(v_2)_a = \begin{cases}
    \,0\,,\quad &a\in S\,,\\
    s_a^*\,,\quad &a\in S'\,,
\end{cases}
\ee
are right null vectors of $J$. To span the entire null space, one must construct as many of these subsets $S$ as possible. When trying to do so, one sees that the smallest subsets are made up of $3$ boxes arranged in either of the L shapes in Figure \ref{fig: min rescaling S}.

After tessellating the partition with as many of the L shapes in Figure \ref{fig: min rescaling S} as possible, each of the L shapes must be completed by boxes below and to the right such that the entire partition is covered by subsets. For example, such a cover and the associated null vectors are shown in Figure \ref{fig: eg max cover}. 

\begin{figure}[H]
    \centering
    \begin{ytableau}
    *(red) 1 & *(red) 2 & *(green) 3 & *(green) 4 & *(green) 5 \\
    *(cyan) 6 & *(red) 7 & *(green) 8 & *(green) 9 \\
    *(cyan) 10 & *(cyan) 11 \\
    *(cyan) 12
    \end{ytableau}
    \caption{\small A division of the partition $\{5,4,2,1\}$ into a maximum number of subsets satisfying \eqref{eq: indep rescaling}, which are coloured red, green, and blue. With $s_a$ being associated with the box labelled $a$, the zero eigenvectors coming from the $3$ independent rescalings are $(s_1^*,s_2^*,0,0,0,0,s_7^*,0,0,0,0,0)$, $(0,0,s_3^*,s_4^*,s_5^*,0,0,s_8^*,s_9^*,0,0,0)$, and $(0,0,0,0,0,s_6^*,0,0,0,s_10^*,s_{11}^*,s_{12}^*)$. }
    \label{fig: eg max cover}
\end{figure}

It turns out that the null vectors arsing from rescalings, which we have discussed thus far, are insufficient to span Nul$(J)$. We also need null vectors of the form
\be\label{eq: ind box vect}
(v_b)_a = \delta_{ab}\,,
\ee
which are associated with individual boxes $b$. To determine which boxes $b$ are allowed, we examine the eigenvalue equation

\begin{equation}\label{eq: explicit jacobian}
\begin{aligned}
    &0=\partial_cB_a(s^*,0)(v_b)_c = \partial_bB_a(s^*,0)=
\delta_{ba}\Bigg\{\prod_{\substack{c=1\\ c\neq a}}^N(s_c^*-t_1^{-1}s_a^*)(s_c^*-t_2^{-1}s_a^*)(s_c^*-t_1t_2s_a^*)\\
&-(s_a^*-t_1^\frac{1}{2}t_2^\frac{1}{2})\sum_{\substack{c=1\\c\neq a}}^N[t_1^{-1}(s_c^*-t_2^{-1}s_a^*)(s_c^*-t_1t_2s_a^*)+t_2^{-1}(s_c^*-t_1^{-1}s_a^*)(s_c^*-t_1t_2s_a^*)\\
&+t_1t_2(s_c^*-t_1^{-1}s_a^*)(s_c^*-t_2^{-1}s_a^*)]\prod_{\substack{d = 1 \\ d\neq a,c}}^N(s_d^*-t_1^{-1}s_a^*)(s_d^*-t_2^{-1}s_a^*)(s_d^*-t_1t_2s_a^*)\Bigg\}\\
&+(1-\delta_{ab})(s_a^*-t_1^\frac{1}{2}t_2^\frac{1}{2})[(s_b^*-t_2^{-1}s_a^*)(s_b^*-t_1t_2s_a^*)+(s_b^*-t_1^{-1}s_a^*)(s_b^*-t_1t_2s_a^*)\\
&+(s_b^*-t_1^{-1}s_a^*)(s_b^*-t_2^{-1}s_a^*)]\prod_{\substack{c=1\\c\neq a,b}}^N(s_c^*-t_1^{-1}s_a^*)(s_c^*-t_2^{-1}s_a^*)(s_c^*-t_1t_2s_a^*)\,,\quad \forall a=1,\ldots, N\,.
\end{aligned}
\end{equation}

In the case $a\neq b$, only the last two lines of \eqref{eq: explicit jacobian} are present. If $a$ is not any of the neighbours of $b$ drawn below,

\begin{equation*}
    \begin{ytableau}
    a & \none[] & \none[] \\ 
    \none[] & b & a \\
    \none[] & a & \none[]
\end{ytableau}
\end{equation*}
since \eqref{rescale null vect cond} holds, there must exist $c\neq a,b$ such that $(s_c^*-t_1^{-1}s_a^*)(s_c^*-t_2^{-1}s_a^*)(s_c^*-t_1t_2s_a^*)=0$, and the product in the last line above is zero. The non-trivial requirements on $b$ come from the cases when $a$ is one of the neighbours drawn above, and each of the cases shall be considered individually. When $a$ is to the top-left of $b$, either $i_a=j_a=1$ and $s_a^*-t_1^\frac{1}{2}t_2^\frac{1}{2}=0$, or otherwise there is always a box $c$ above or to the left of $a$ for which $s_c^*-t_1^{-1}s_a^*=0$ or $s_c^*-t_2^{-1}s_a^*=0$. The last two lines in \eqref{eq: explicit jacobian} are zero either way, and case where $a$ is to the top-left of $b$ does not provide any constraints on the position of $b$. Moving on to the case when $a$ is the right neighbour of $b$, if $a$, $b$ are not in the top row, there exists a box $c$ on top of $a$ satisfying $s_c^*-t_1^{-1}s_a^*=0$, making the product in the last line of \eqref{eq: explicit jacobian} zero. Otherwise if $a$, $b$ are in the top row, imposing that the last two lines of \eqref{eq: explicit jacobian} are zero is equivalent to imposing that there must be a box $c$ satisfying $s_c^*-t_1t_2s_a^*=0$. In other words, either $i_b > 1$ in which case there are no additional constraints, or $i_b = 1$ and there must exist boxes arranged in relation to $b$ as

\begin{equation*}
\begin{ytableau}
 b & a & {} \\
 \none[\cdots] & {} & {}
\end{ytableau}
\end{equation*}
Lastly, when $a$ is the bottom neighbour of $b$, if $j_b > 1$, there exists another box $c$ such that $s_c^*-t_2^{-1}s_a^*=0$ and the last two lines of \eqref{eq: explicit jacobian} are zero. Otherwise if $j_b = 1$, imposing that the last two lines of \eqref{eq: explicit jacobian} is zero is equivalent to imposing that there must be a box $c$ satisfying $s_c^*-t_1t_2s_a^*=0$. In other words, either $j_b > 1$ in which case there are no additional constraints, or $j_b = 1$ and there must exist boxes arranged in relation to $b$ as

\begin{equation*}
    \begin{ytableau}
        b & \none[\vdots] \\ 
        a & {} \\ 
        {} & {} 
    \end{ytableau}
\end{equation*}
We still have to consider the case $a = b$, where the first $3$ lines of \eqref{eq: explicit jacobian} must be zero. The analysis is similar and one can show that there are no new constraints. Taking together the constraints $\forall a$, the boxes $b$ in the partition $\lambda$ corresponding to null vectors of the form \eqref{eq: ind box vect} belong to the set

\begin{equation}\label{eq: ind null boxes}
    S_{R,\text{ind}}=\{b\in \lambda\; |\; (i_b = 1 \implies \exists(2, j_b+2)\in \lambda\,)\, \land \,(\,j_b = 1 \implies \exists(i_b+2, 2)\in \lambda)\}\,.
\end{equation}
In the above, and also in subsequent expressions, $(i_a, j_a)$ denotes a box with row number $i_a$ and column number $j_a$. 

To recap, the null space of $J$ contains two sets of vectors. The first are vectors generating independent rescalings of the classical solution, corresponding to subsets which cover the partition. The second are vectors corresponding to individual boxes in \eqref{eq: ind null boxes}. We conjecture that the null space is completely spanned by these vectors, i.e.
\be\label{eq: right null span}
\text{Nul}(J)=\text{span}_\bC\{ v_S,\; v_b\; |\; S\in\cC, b\in S_{R,\text{ind}}\}\,,
\ee
where $\cC$ is a cover of the partition with the largest number of subsets. For the partition in Figure \ref{fig: eg max cover}, $\cC$ would be the set of $3$ coloured subsets. Note that the vectors in \eqref{eq: right null span} may not be linearly independent since some subsets $S$ might be entirely made out of individual boxes in $S_{R,\text{ind}}$. In that case, the vectors corresponding to these subsets $S$ can be removed from the spanning set. It has been numerically tested that \eqref{eq: right null span} holds for all partitions up to and including $N = 16$.

\paragraph{Left null vectors.} It turns out that for the left null vectors, it is sufficient to consider vectors corresponding to individual boxes $a$, 
\be
(v_a)_c = \delta_{ac}\,,
\ee
for which the eigenvalue equation is
\be\label{left null eqn}
(v_a)_c\partial_{s_b}B_c(s^*,0) = \partial_{s_b}B_a(s^*,0) = 0\,,\quad \forall b\in \{1,\ldots, N\}\,.
\ee
As in the previous analysis, one should understand the restrictions on the boxes $a$ due to \eqref{left null eqn}. Essentially, one finds that either $i_a=j_a=1$ and the box $(2,2)$ exists in the partition, or at least $2$ of the $3$ neighbouring boxes drawn below must exist in the partition.
\be\label{Bu0 neighbours}
\begin{ytableau}
    \none[] & {} & \none[]\\
    {} & a & \none[]\\
    \none[] & \none[] & {}
\end{ytableau}
\ee
This is automatic for the interior boxes with $i_a > 1$ and $j_a > 1$. Else if $i_a = 1$ and $j_a > 1$ (since $i_a=j_a=1$ was already considered as a special case), the box with coordinates $(i_a+1,j_a+1)$ must exist in the partition. Similarly if $j_a = 1$ and $i_a > 1$, $(i_a+1,j_a+1)$ must exist in the partition. More compactly, the boxes $a$ satisfying \eqref{left null eqn} in the partition $\lambda$ belong to
\be\label{left null boxes def}
S_{L,\text{ind}}=\{a\in \lambda\; |\; (i_a = 1 \lor j_a = 1)\implies \exists (i_a + 1,j_a + 1)\in \lambda\}\,.
\ee
Note that all right null vectors of the form \eqref{eq: ind box vect} are also left null vectors since $S_{R,\text{ind}}\subset S_{L,\text{ind}}$, but the converse is not true. It has been checked for all partitions up to $N = 16$ that 
\be\label{eq: left null span}
\text{Nul}\big(J^T\big)=\text{span}_\bC\{v_a\; |\; a\in S_{L,\text{ind}}\}\,.
\ee
We have also checked that the dimensions of the left and right null spaces are equal, as expected for a square matrix $J$.

\subsection{Choice of Particular Solution}\label{app: particular sol}

In this section, we derive the particular solution \eqref{eq: particular sol LO} to \eqref{eq: LO s constr}. Firstly, let us find the boxes $a$ for which $\partial_uB_a(s^*,0)\neq 0$. Note that
\bea\label{dBdu nonzero}
\partial_vB_a(s^*,0)=\zeta t_1^\frac{1}{2}t_2^\frac{1}{2}\Big(s_a^*-t_1^{-\frac{1}{2}}t_2^{-\frac{1}{2}}\Big)\prod_{\substack{b=1\\b\neq a}}^N(s_b^*-t_1s_a^*)(s_b^*-t_2s_a^*)(s_b^*-t_1^{-1}t_2^{-1}s_a^*)\neq 0 \\
\iff \prod_{\substack{b=1\\b\neq a}}^N(s_b^*-t_1s_a^*)(s_b^*-t_2s_a^*)(s_b^*-t_1^{-1}t_2^{-1}s_a^*)\neq 0\,,
\eea
because on the classical solutions, $s_a^*\neq t_1^{-\frac{1}{2}}t_2^{-\frac{1}{2}}$ for any $a$. For the latter condition in \eqref{dBdu nonzero} to hold, $a$ cannot have neighbouring boxes below, to the right, or to the top-left, i.e.
\begin{equation*}\label{cond dBdu nonzero}
\begin{ytableau}
    \times & \none[] & \none[] \\
    \none[] & a & \times \\
    \none[] & \times & \none[]
\end{ytableau}
\end{equation*}
where the crossed boxes must be absent in the Young diagram associated with $s^*$. Since there are no boxes below or to the right of $a$, it can be removed to leave a valid Young diagram; i.e. it is a corner box as defined below \eqref{eq: gen sol LO ds}. In addition, the condition that there is no box to the top-left of $a$ implies that it must be an exterior corner. 

Next, we show that for exterior corners $a$, $J_{ba} = 0$ 
for all $b\neq a$. Explicitly, the matrix entries are the same as in \eqref{eq: explicit jacobian} but with $a$ and $b$ exchanged. Since $a$ is an exterior corner, there is no box $b\neq a$ such that $s_a^*-t_1^{-1}s_b^*=0$, $s_a^*-t_2^{-1}s_b^*=0$ or $s_a^*-t_1t_2s_b^*=0$. On the other hand, as for any box in the Young diagram, $b$ must have a neighbouring box $c\neq a, b$ (as we have just established, $c\neq a$ because $a$ is an exterior corner) on its left or above it such that $s_c^*-t_1^{-1}s_b^*=0$ or $s_c^*-t_2^{-1}s_b^*=0$, or $b$ is the top-left box with $s_b^*-t_1^\frac{1}{2}t_2^\frac{1}{2}=0$. This means
\be
\Big(s_b^*-t_1^\frac{1}{2}t_2^\frac{1}{2}\Big)\prod_{\substack{c=1\\c\neq a,b}}^N(s_c^*-t_1^{-1}s_b^*)(s_c^*-t_2^{-1}s_b^*)(s_c^*-t_1t_2s_b^*)=0\,,
\ee
and therefore $J_{ba}=0$.

If we choose $\Delta^{\scriptscriptstyle{(1)}}_\times s$ to have nonzero entries only in the fugacities corresponding to the exterior corners $a$, $J_{ba} = 0$ for $b\neq a$ implies that \eqref{eq: LO s constr} simplifies to $J_{aa}\Delta^{\scriptscriptstyle{(1)}}_\times s_a = -\partial_vB_a(s^*,0)\Delta v$ for each exterior corner $a$. This is clearly solved by \eqref{eq: particular sol LO}.

\subsection{Derivation of the Physical Solution}\label{app: phys sol}

In this section, we go through the steps to simplify \eqref{eq: quad ds null}, before solving it in the case where $s^*$ corresponds to a $1$-wheel partition. If we decompose $\Delta^{\scriptscriptstyle{(1)}}s$ as in \eqref{eq: gen sol LO ds}, \eqref{eq: quad ds null} becomes

\bea\label{2nd order eqn}
0=\,&\frac{1}{2}\partial_{s_b}\partial_{s_c}B_a(s^*,0)\Delta^{(1)}_\times s_b\Delta^{(1)}_\times s_c +\partial_{s_b}\partial_{s_c}B_a(s^*,0)\Delta^{(1)}_\times s_b\Delta^{(1)}_{0} s_c\\
&+\frac{1}{2}\partial_{s_b}\partial_{s_c}B_a(s^*,0)\Delta^{(1)}_{0} s_b\Delta^{(1)}_{0} s_c+\partial_v\partial_{s_b}B_a(s^*,0)\Delta v\big(\Delta^{(1)}_\times s_b+\Delta^{(1)}_{0} s_b\big)\,,\quad \forall a\in S_{L,\text{ind}}\,,
\eea
where we have used the basis \eqref{eq: left null span} for the left null space of $J$. Due to the judicious choice \eqref{eq: particular sol LO} that was made for $\Delta^{(1)}_\times s$, there are various simplifications in \eqref{2nd order eqn}. To see this, one can analyse the explicit formula for the matrix of second derivatives contracted with vectors $v$, $\wt v$:
\bea\label{dBdsds contracted}
&\partial_{s_c}\partial_{s_b}B_a(s^*,0)\wt v_cv_b=\wt v_a\sum_{\substack{b=1\\ b\neq a}}^N[(v_b-t_1^{-1}v_a)(s_b^*-t_2^{-1}s_a^*)(s_b^*-t_1t_2s_a^*)+(s_b^*-t_1^{-1}s_a^*)(v_b-t_2^{-1}v_a)(s_b^*-t_1t_2s_a^*)\\
&+(s_b^*-t_1^{-1}s_a^*)(s_b^*-t_2^{-1}s_a^*)(v_b-t_1t_2v_a)]\prod_{\substack{c=1\\ c\neq a,b}}^N(s_c^*-t_1^{-1}s_a^*)(s_c^*-t_2^{-1}s_a^*)(s_c^*-t_1t_2s_a^*)+(v\leftrightarrow \wt v)\\
&+(s_a^*-t_1^\frac{1}{2}t_2^\frac{1}{2})\sum_{\substack{b=1\\b\neq a}}^N[(v_b-t_1^{-1}v_a)(\wt v_b-t_2^{-1}\wt v_a)(s_b^*-t_1t_2s_a^*)+(v_b-t_1^{-1}v_a)(s_b^*-t_2^{-1}s_a^*)(\wt v_b-t_1t_2\wt v_a)\\
&+(\wt v_b-t_1^{-1}\wt v_a)(v_b-t_2^{-1}v_a)(s_b^*-t_1t_2s_a^*)+(s_b^*-t_1^{-1}s_a^*)(v_b-t_2^{-1}v_a)(\wt v_b-t_1t_2\wt v_a)\\
&+(\wt v_b-t_1^{-1}\wt v_a)(s_b^*-t_2^{-1}s_a^*)(v_b-t_1t_2v_a)+(s_b^*-t_1^{-1}s_a^*)(\wt v_b-t_2^{-1}\wt v_a)(v_b-t_1t_2v_a)]\\
&\times\prod_{\substack{c=1\\c\neq a,b}}^N(s_c^*-t_1^{-1}s_a^*)(s_c^*-t_2^{-1}s_a^*)(s_c^*-t_1t_2s_a^*)\\
&+(s_a^*-t_1^\frac{1}{2}t_2^\frac{1}{2})\sum_{\substack{b=1\\b\neq a}}^N[(v_b-t_1^{-1}v_a)(s_b^*-t_2^{-1}s_a^*)(s_b^*-t_1t_2s_a^*)+(s_b^*-t_1^{-1}s_a^*)(v_b-t_2^{-1}v_a)(s_b^*-t_1t_2s_a^*)\\
&+(s_b^*-t_1^{-1}s_a^*)(s_b^*-t_2^{-1}s_a^*)(v_b-t_1t_2v_a)]\times\sum_{\substack{c=1\\c\neq a,b}}^N[(\wt v_c-t_1^{-1}\wt v_a)(s_c^*-t_2^{-1}s_a^*)(s_c^*-t_1t_2s_a^*)\\
&+(s_c^*-t_1^{-1}s_a^*)(\wt v_c-t_2^{-1}\wt v_a)(s_c^*-t_1t_2s_a^*)+(s_c^*-t_1^{-1}s_a^*)(s_c^*-t_2^{-1}s_a^*)(\wt v_c-t_1t_2\wt v_a)]\\
&\times\prod_{\substack{d=1\\d\neq a,b,c}}^N(s_d^*-t_1^{-1}s_a^*)(s_d^*-t_2^{-1}s_a^*)(s_d^*-t_1t_2s_a^*)\,.
\eea
Consider the first term $\frac{1}{2}\partial_{s_b}\partial_{s_c}B_a(s^*,0)\Delta^{(1)}_\times s_b\Delta^{(1)}_\times s_c$ in \eqref{2nd order eqn}. Note that $S_{L,\text{ind}}$ does not exterior corners, so $\Delta^{(1)}_\times s_a= 0$ and the first two lines of \eqref{dBdsds contracted} vanish. If we next look at lines 3 to 6, they are zero if $a$ is the top-left box. If not, $a$ has at least $2$ neighbours in the positions drawn in \eqref{Bu0 neighbours}, and the product in line 6 will always include a contribution from when $c\neq a,b$ is one such neighbour. Since the neighbours drawn in \eqref{Bu0 neighbours} precisely satisfy $s_c^*-t_1^{-1}s_a^*=0$, $s_c^*-t_2^{-1}s_a^*=0$, or $s_c^*-t_1t_2s_a^*=0$, this product must be zero. Lastly, we consider lines 7 to 10. Since $\Delta^{(1)}_\times s_a=0$, the summand is proportional to $\Delta^{(1)}_\times s_b\Delta^{(1)}_\times s_c$, which is only nonzero if $b$ and $c$ are exterior corners. By construction, these boxes $b$, $c$ cannot be neighbours of $a$ in the positions drawn in \eqref{Bu0 neighbours}. Therefore the (at least) 2 neighbours of $a$ in positions \eqref{Bu0 neighbours} must be included in the product in the last line, which is then zero. We conclude that
\be
\frac{1}{2}\partial_{s_b}\partial_{s_c}B_a(s^*,0)\Delta^{(1)}_\times s_b\Delta^{(1)}_\times s_c=0\,.
\ee
Due to almost identical arguments from analysing \eqref{dBdsds contracted}, which we will omit, one can also show that
\be
\partial_{s_b}\partial_{s_c}B_a(s^*,0)\Delta^{(1)}_\times s_b\Delta^{(1)}_{0} s_c=0\,.
\ee
Next, we want to show that
\be
\partial_v\partial_{s_b}B_a(s^*,0)\Delta^{(1)}_\times s_b=0\,,
\ee
for which we need to analyze
\bea\label{dBduds on sol}
&\partial_v\partial_{s_b}B_a(s^*,0)v_b = \zeta t_1^\frac{1}{2}t_2^\frac{1}{2}v_a\prod_{\substack{b=1\\b\neq a}}^N(s_b^*-t_1s_a^*)(s_b^*-t_2s_a^*)(s_b^*-t_1^{-1}t_2^{-1}s_a^*)\\
&+\zeta t_1^\frac{1}{2}t_2^\frac{1}{2}(s_a^*-t_1^{-\frac{1}{2}}t_2^{-\frac{1}{2}})\sum_{\substack{b=1\\b\neq a}}^N[(v_b-t_1v_a)(s_b^*-t_2s_a^*)(s_b^*-t_1^{-1}t_2^{-1}s_a^*)+(s_b^*-t_1s_a^*)(v_b-t_2v_a)(s_b^*-t_1^{-1}t_2^{-1}s_a^*)\\
&+(s_b^*-t_1s_a^*)(s_b^*-t_2s_a^*)(v_b-t_1^{-1}t_2^{-1}v_a)]\prod_{\substack{c=1\\c\neq a,b}}^N(s_c^*-t_1s_a^*)(s_c^*-t_2s_a^*)(s_c^*-t_1^{-1}t_2^{-1}s_a^*)\,.
\eea
The first term above is directly zero since $\Delta^{(1)}_\times s_a=0$. Note that for the boxes $a\in S_{L,\text{ind}}$ with $i_a = 1$ or $j_a = 1$, $(i_a+1,j_a+1)$ being in the partition also implies that the two neighbours of $a$, $(i_a,j_a+1)$ and $(i_a+1,j_a)$ are in the partition. One of these neighbours must be part of the product in the third line above, where either $s_c^*-t_1s_a^*=0$ or $s_c^*-t_2s_a^*=0$, making the term in the second and third lines zero. In the leftover case where $a$ is an interior box with $i_a > 1$ and $j_a > 1$, the neighbour $(i_a-1,j_a-1)$ to the top-left is guaranteed to exist in the partition. Since $\Delta^{(1)}_\times s_a=0$, the summand in square brackets above is proportional to $\Delta^{(1)}_\times s_b$. For this to be nonzero, $b$ must be an exterior corner, and $(i_a-1,j_a-1)\neq b$. Therefore $c=(i_a-1,j_a-1)$ must be included in the product in the third line. Since $s_c^*-t_1^{-1}t_2^{-1}s_a^*=0$, the whole term is zero.

At this point, \eqref{2nd order eqn} is simplified to
\be\label{2nd order eqn nonzero canceled}
\frac{1}{2}\partial_{s_b}\partial_{s_c}B_a(s^*,0)\Delta^{(1)}_{0} s_b\Delta^{(1)}_{0} s_c+\partial_v\partial_{s_b}B_a(s^*,0)\Delta v\Delta^{(1)}_{0} s_b=0\qquad \forall a\in S_{L,\text{ind}}\,.
\ee
Next, we express $\Delta^{(1)}_0s$ in terms of the basis found in \eqref{eq: right null span} as
\be
\Delta^{(1)}_0s=\sum_{S\in\cC}x_Sv_S + \sum_{b\in S_{R,\text{ind}}}x_bv_b\,.
\ee
Substituting into \eqref{2nd order eqn nonzero canceled} gives
\bea
&\frac{1}{2}\sum_{S,\,S'\in\cC}\partial_{s_b}\partial_{s_c}B_a(s^*,0)(v_S)_b(v_{S'})_c\, x_S\,x_{S'}+\sum_{S\in\,\cC}\sum_{c\,\in S_{R,\text{ind}}}\partial_{s_b}\partial_{s_c}B_a(s^*,0)(v_S)_b\, x_S\,x_c\\
&+\frac{1}{2}\sum_{b,\,c\,\in S_{R,\text{ind}}}\partial_{s_b}\partial_{s_c}B_a(s^*,0)\,x_b\,x_c+\sum_{S\in\,\cC}\Delta v\,\partial_v\partial_{s_b}B_a(s^*,0)(v_S)_b\, x_S+\sum_{b\,\in S_{R,\text{ind}}}\Delta v\,\partial_v\partial_{s_b}B_a(s^*,0)x_b=0\,.
\eea
Since rescaling a subset $S\in\cC$ of the gauge fugacities is an \emph{exact} solution of $B_a(s,0)=0$, we must have
\be
\frac{1}{2}\partial_{s_b}\partial_{s_c}B_a(s^*,0)(v_S)_b(v_{S'})_c\,= 0\qquad \forall S, S'\in\cC\,.
\ee
In fact, any higher derivatives terms must also be zero when contracted with $v_S$. This leaves us with
\bea\label{2nd order eqn final}
&\sum_{S\in\,\cC}\sum_{c\,\in S_{R,\text{ind}}}\partial_{s_b}\partial_{s_c}B_a(s^*,0)(v_S)_b\, x_S\,x_c+\frac{1}{2}\sum_{b,\,c\,\in S_{R,\text{ind}}}\partial_{s_b}\partial_{s_c}B_a(s^*,0)\,x_b\,x_c\\
&+\sum_{S\in\,\cC}\Delta v\,\partial_v\partial_{s_b}B_a(s^*,0)(v_S)_b\, x_S+\sum_{b\,\in S_{R,\text{ind}}}\Delta v\,\partial_v\partial_{s_b}B_a(s^*,0)x_b=0\,,\quad\forall a\in S_{L,\text{ind}}\,.
\eea

In the case when $s^*$ is associated with a $1$-wheel partition whose Young diagram is shown in Figure \ref{fig:1-wheel partition}, the right null space of $J$ is $2$-dimensional. The cover $\cC$ only has one element which is the whole partition itself, and we denote the corresponding coefficient as $x$. The corresponding null vector is $s^*$ itself. $S_{R,\text{ind}}$ consists of the sole interior corner $a$ which is coloured blue in Figure \ref{fig:1-wheel partition}, and we denote the corresponding coefficient as $y$. For the left null space, $S_{L,\text{ind}}$ contains the top-left box with coordinates $(1,1)$, which we will denote as $b$, as well as the interior corner $a$. \eqref{2nd order eqn final} then consists of the $2$ equations  

\begin{equation}
    \partial_{s_c}\partial_{s_a}B_b(s^*,0)s_c^*\; xy = 0\,,\quad y\left(\frac{1}{2}\partial_{s_a}^2B_a(s^*,0)\,y + \Delta v\partial_v\partial_{s_a}B_a(s^*,0)\right)=0\,. 
\end{equation}
The solution $y = 0$ is unphysical since $s^* + \Delta^{\scriptscriptstyle{(1)}}_\times s + x s^*$ still contains fugacities related by \eqref{eq: wheel combi} for arbitrary $x$. We are left with the other solution
\begin{equation}
    x = 0\,,\qquad y = -\frac{2 \Delta v\partial_v\partial_{s_a}B_a(s^*,0)}{\partial_{s_a}^2B_a(s^*,0)}\,,
\end{equation}
which, when added to \eqref{eq: particular sol LO}, is precisely \eqref{eq: 1-wheel predictor}.

\bibliographystyle{JHEP}
\bibliography{bae}

\providecommand{\href}[2]{#2}\begingroup\raggedright\begin{thebibliography}{10}

\bibitem{bullimore2016vortices}
M.~Bullimore, T.~Dimofte, D.~Gaiotto, J.~Hilburn and H.-C.~Kim, \emph{Vortices and vermas}, {\emph{arXiv preprint arXiv:1609.04406} (2016) }.

\bibitem{bullimore2019twisted}
M.~Bullimore, A.~Ferrari and H.~Kim, \emph{Twisted indices of 3d $\mathcal{N}=4$ gauge theories and enumerative geometry of quasi-maps}, {\emph{Journal of High Energy Physics} {\bfseries 2019} (2019) 1}.

\bibitem{dimofte2011vortex}
T.~Dimofte, S.~Gukov and L.~Hollands, \emph{Vortex counting and lagrangian 3-manifolds}, {\emph{Letters in Mathematical Physics} {\bfseries 98} (2011) 225}.

\bibitem{Inglese:2023tyc}
M.~Inglese, D.~Martelli and A.~Pittelli, \emph{{Supersymmetry and Localization on Three-Dimensional Orbifolds}},  \href{https://arxiv.org/abs/2312.17086}{{\ttfamily 2312.17086}}.

\bibitem{Haouzi:2023doo}
N.~Haouzi, \emph{{A new realization of quantum algebras in gauge theory and Ramification in the Langlands program}},  \href{https://arxiv.org/abs/2311.04367}{{\ttfamily 2311.04367}}.

\bibitem{Kimura:2024xpr}
T.~Kimura and G.~Noshita, \emph{{Gauge origami and quiver W-algebras II: Vertex function and beyond quantum $q$-Langlands correspondence}},  \href{https://arxiv.org/abs/2404.17061}{{\ttfamily 2404.17061}}.

\bibitem{bullimore2016boundaries}
M.~Bullimore, T.~Dimofte, D.~Gaiotto and J.~Hilburn, \emph{Boundaries, mirror symmetry, and symplectic duality in 3d $\mathcal{N}=4$ gauge theory}, {\emph{Journal of High Energy Physics} {\bfseries 2016} (2016) 1}.

\bibitem{cheng20243}
M.C.~Cheng, S.~Chun, B.~Feigin, F.~Ferrari, S.~Gukov, S.M.~Harrison et~al., \emph{3-manifolds and voa characters}, {\emph{Communications in Mathematical Physics} {\bfseries 405} (2024) 44}.

\bibitem{dimofte2014gauge}
T.~Dimofte, D.~Gaiotto and S.~Gukov, \emph{Gauge theories labelled by three-manifolds}, {\emph{Communications in Mathematical Physics} {\bfseries 325} (2014) 367}.

\bibitem{dimofte20133}
T.~Dimofte, D.~Gaiotto and S.~Gukov, \emph{3-manifolds and 3d indices}, .

\bibitem{Hayashi:2024jof}
H.~Hayashi, T.~Nosaka and T.~Okazaki, \emph{{ADHM Wilson line defect indices}},  \href{https://arxiv.org/abs/2406.00413}{{\ttfamily 2406.00413}}.

\bibitem{Bullimore:2021rnr}
M.~Bullimore and D.~Zhang, \emph{{3d $\mathcal{N}=4$ Gauge Theories on an Elliptic Curve}}, \href{https://doi.org/10.21468/SciPostPhys.13.1.005}{\emph{SciPost Phys.} {\bfseries 13} (2022) 005} [\href{https://arxiv.org/abs/2109.10907}{{\ttfamily 2109.10907}}].

\bibitem{Dedushenko:2021mds}
M.~Dedushenko and N.~Nekrasov, \emph{{Interfaces and quantum algebras, I: Stable envelopes}}, \href{https://doi.org/10.1016/j.geomphys.2023.104991}{\emph{J. Geom. Phys.} {\bfseries 194} (2023) 104991} [\href{https://arxiv.org/abs/2109.10941}{{\ttfamily 2109.10941}}].

\bibitem{Dedushenko:2023qjq}
M.~Dedushenko and N.~Nekrasov, \emph{{Interfaces and Quantum Algebras, II: Cigar Partition Function}},  \href{https://arxiv.org/abs/2306.16434}{{\ttfamily 2306.16434}}.

\bibitem{Zhang:2022wwy}
D.~Zhang, \emph{{Boundaries, states and cohomology in three-dimensional $N = 4$ theories}}, Ph.D. thesis, Cambridge U., DAMTP, 2022.
\newblock 10.17863/CAM.96582.

\bibitem{Panerai:2020boq}
R.~Panerai, A.~Pittelli and K.~Polydorou, \emph{{Topological Correlators and Surface Defects from Equivariant Cohomology}}, \href{https://doi.org/10.1007/JHEP09(2020)185}{\emph{JHEP} {\bfseries 09} (2020) 185} [\href{https://arxiv.org/abs/2006.06692}{{\ttfamily 2006.06692}}].

\bibitem{benini2016black}
F.~Benini, K.~Hristov and A.~Zaffaroni, \emph{{Black hole microstates in AdS$_{4}$ from supersymmetric localization}}, \href{https://doi.org/10.1007/JHEP05(2016)054}{\emph{JHEP} {\bfseries 05} (2016) 054} [\href{https://arxiv.org/abs/1511.04085}{{\ttfamily 1511.04085}}].

\bibitem{Hosseini:2016tor}
S.M.~Hosseini and A.~Zaffaroni, \emph{{Large $N$ matrix models for 3d $\mathcal{N}{=}2$ theories: twisted index, free energy and black holes}}, \href{https://doi.org/10.1007/JHEP08(2016)064}{\emph{JHEP} {\bfseries 08} (2016) 064} [\href{https://arxiv.org/abs/1604.03122}{{\ttfamily 1604.03122}}].

\bibitem{Hosseini:2016ume}
S.M.~Hosseini and N.~Mekareeya, \emph{{Large $N$ topologically twisted index: necklace quivers, dualities, and Sasaki-Einstein spaces}}, \href{https://doi.org/10.1007/JHEP08(2016)089}{\emph{JHEP} {\bfseries 08} (2016) 089} [\href{https://arxiv.org/abs/1604.03397}{{\ttfamily 1604.03397}}].

\bibitem{Choi:2019dfu}
S.~Choi and C.~Hwang, \emph{{Universal 3d Cardy Block and Black Hole Entropy}}, \href{https://doi.org/10.1007/JHEP03(2020)068}{\emph{JHEP} {\bfseries 03} (2020) 068} [\href{https://arxiv.org/abs/1911.01448}{{\ttfamily 1911.01448}}].

\bibitem{Choi:2019zpz}
S.~Choi, C.~Hwang and S.~Kim, \emph{{Quantum vortices, M2-branes and black holes}},  \href{https://arxiv.org/abs/1908.02470}{{\ttfamily 1908.02470}}.

\bibitem{Aharony:2008ug}
O.~Aharony, O.~Bergman, D.L.~Jafferis and J.~Maldacena, \emph{{N=6 superconformal Chern-Simons-matter theories, M2-branes and their gravity duals}}, \href{https://doi.org/10.1088/1126-6708/2008/10/091}{\emph{JHEP} {\bfseries 10} (2008) 091} [\href{https://arxiv.org/abs/0806.1218}{{\ttfamily 0806.1218}}].

\bibitem{Kapustin:2010xq}
A.~Kapustin, B.~Willett and I.~Yaakov, \emph{{Nonperturbative Tests of Three-Dimensional Dualities}}, \href{https://doi.org/10.1007/JHEP10(2010)013}{\emph{JHEP} {\bfseries 10} (2010) 013} [\href{https://arxiv.org/abs/1003.5694}{{\ttfamily 1003.5694}}].

\bibitem{smirnov2020characters}
A.~Smirnov and H.~Dinkins, \emph{Characters of tangent spaces at torus fixed points and 3 d-mirror symmetry}, {\emph{Letters in Mathematical Physics} {\bfseries 110} (2020) 2337}.

\bibitem{Crew:2020psc}
S.~Crew, N.~Dorey and D.~Zhang, \emph{{Blocks and Vortices in the 3d ADHM Quiver Gauge Theory}}, \href{https://doi.org/10.1007/JHEP03(2021)234}{\emph{JHEP} {\bfseries 03} (2021) 234} [\href{https://arxiv.org/abs/2010.09732}{{\ttfamily 2010.09732}}].

\bibitem{feigin2013representations}
B.~Feigin, M.~Jimbo, T.~Miwa and E.~Mukhin, \emph{Representations of quantum toroidal gln}, {\emph{Journal of Algebra} {\bfseries 380} (2013) 78}.

\bibitem{hernandez2009quantum}
D.~Hernandez, \emph{Quantum toroidal algebras and their representations}, {\emph{Selecta Mathematica} {\bfseries 14} (2009) 701}.

\bibitem{hosseini2019gluing}
S.M.~Hosseini, K.~Hristov and A.~Zaffaroni, \emph{Gluing gravitational blocks for ads black holes}, {\emph{Journal of High Energy Physics} {\bfseries 2019} (2019) 1}.

\bibitem{smirnov2016rationality}
A.~Smirnov, \emph{Rationality of capped descendent vertex in $ k $-theory}, {\emph{arXiv preprint arXiv:1612.01048} (2016) }.

\bibitem{nekrasov2009supersymmetric}
N.A.~Nekrasov and S.L.~Shatashvili, \emph{Supersymmetric vacua and bethe ansatz}, {\emph{arXiv preprint arXiv:0901.4744} (2009) }.

\bibitem{nekrasov2009quantum}
N.~Nekrasov and S.~Shatashvili, \emph{Quantum integrability and supersymmetric vacua}, {\emph{Progress of Theoretical Physics Supplement} {\bfseries 177} (2009) 105}.

\bibitem{okounkov2015lectures}
A.~Okounkov, \emph{Lectures on k-theoretic computations in enumerative geometry}, {\emph{arXiv preprint arXiv:1512.07363} (2015) }.

\bibitem{Jockers:2018sfl}
H.~Jockers and P.~Mayr, \emph{{A 3d Gauge Theory/Quantum K-Theory Correspondence}}, \href{https://doi.org/10.4310/ATMP.2020.v24.n2.a4}{\emph{Adv. Theor. Math. Phys.} {\bfseries 24} (2020) 327} [\href{https://arxiv.org/abs/1808.02040}{{\ttfamily 1808.02040}}].

\bibitem{Bullimore:2020jdq}
M.~Bullimore, S.~Crew and D.~Zhang, \emph{{Boundaries, Vermas, and Factorisation}}, \href{https://doi.org/10.1007/JHEP04(2021)263}{\emph{JHEP} {\bfseries 04} (2021) 263} [\href{https://arxiv.org/abs/2010.09741}{{\ttfamily 2010.09741}}].

\bibitem{aganagic2017quasimap}
M.~Aganagic and A.~Okounkov, \emph{Quasimap counts and bethe eigenfunctions}, {\emph{arXiv preprint arXiv:1704.08746} (2017) }.

\bibitem{Aganagic:2016jmx}
M.~Aganagic and A.~Okounkov, \emph{{Elliptic stable envelopes}}, \href{https://doi.org/10.1090/jams/954}{\emph{J. Am. Math. Soc.} {\bfseries 34} (2021) 79} [\href{https://arxiv.org/abs/1604.00423}{{\ttfamily 1604.00423}}].

\bibitem{Ferrari:2023hza}
A.E.V.~Ferrari and D.~Zhang, \emph{{Berry Connections for 2d $(2,2)$ Theories, Monopole Spectral Data \& (Generalised) Cohomology Theories}},  \href{https://arxiv.org/abs/2311.08454}{{\ttfamily 2311.08454}}.

\bibitem{Colombo:2024mts}
E.~Colombo, S.M.~Hosseini, D.~Martelli, A.~Pittelli and A.~Zaffaroni, \emph{{Microstates of accelerating and supersymmetric AdS$_4$ black holes from the spindle index}},  \href{https://arxiv.org/abs/2404.07173}{{\ttfamily 2404.07173}}.

\bibitem{koroteev2021quantum}
P.~Koroteev, P.P.~Pushkar, A.V.~Smirnov and A.M.~Zeitlin, \emph{Quantum k-theory of quiver varieties and many-body systems}, {\emph{Selecta Mathematica} {\bfseries 27} (2021) 87}.

\bibitem{closset2019three}
C.~Closset and H.~Kim, \emph{Three-dimensional $\mathcal{N} = 2$ supersymmetric gauge theories and partition functions on seifert manifolds: A review}, {\emph{International Journal of Modern Physics A} {\bfseries 34} (2019) 1930011}.

\bibitem{nakajima1999lectures}
H.~Nakajima, \emph{Lectures on Hilbert schemes of points on surfaces}, no.~18, American Mathematical Soc. (1999).

\bibitem{mcgerty2018kirwan}
K.~McGerty and T.~Nevins, \emph{Kirwan surjectivity for quiver varieties}, {\emph{Inventiones mathematicae} {\bfseries 212} (2018) 161}.

\bibitem{feigin2015quantum}
B.~{Feigin}, M.~{Jimbo}, T.~{Miwa} and E.~{Mukhin}, \emph{{Quantum toroidal gl(1) and Bethe ansatz}}, \href{https://doi.org/10.48550/arXiv.1502.07194}{\emph{arXiv e-prints} (2015) arXiv:1502.07194} [\href{https://arxiv.org/abs/1502.07194}{{\ttfamily 1502.07194}}].

\bibitem{feigin2017integrals}
B.~Feigin, M.~Jimbo and E.~Mukhin, \emph{Integrals of motion from quantum toroidal algebras}, {\emph{Journal of Physics A: Mathematical and Theoretical} {\bfseries 50} (2017) 464001}.

\bibitem{maulik2012quantum}
D.~Maulik and A.~Okounkov, \emph{Quantum groups and quantum cohomology}, {\emph{arXiv preprint arXiv:1211.1287} (2012) }.

\bibitem{negut2012moduli}
A.~Negut, \emph{Moduli of flags of sheaves and their k-theory}, {\emph{arXiv preprint arXiv:1209.4242} (2012) }.

\bibitem{Vafa:1991uz}
C.~Vafa, \emph{{Topological mirrors and quantum rings}}, {\emph{AMS/IP Stud. Adv. Math.} {\bfseries 9} (1998) 97} [\href{https://arxiv.org/abs/hep-th/9111017}{{\ttfamily hep-th/9111017}}].

\bibitem{Benini:2013yva}
F.~Benini and W.~Peelaers, \emph{{Higgs branch localization in three dimensions}}, \href{https://doi.org/10.1007/JHEP05(2014)030}{\emph{JHEP} {\bfseries 05} (2014) 030} [\href{https://arxiv.org/abs/1312.6078}{{\ttfamily 1312.6078}}].

\bibitem{Crew:2023tky}
S.~Crew, D.~Zhang and B.~Zhao, \emph{{Boundaries \& localisation with a topological twist}}, \href{https://doi.org/10.1007/JHEP10(2023)093}{\emph{JHEP} {\bfseries 10} (2023) 093} [\href{https://arxiv.org/abs/2306.16448}{{\ttfamily 2306.16448}}].

\bibitem{Tanaka:2014oda}
A.~Tanaka, H.~Mori and T.~Morita, \emph{{Superconformal index on $\mathbb{RP}^2 \times \mathbb{S}^1$ and mirror symmetry}}, \href{https://doi.org/10.1103/PhysRevD.91.105023}{\emph{Phys. Rev. D} {\bfseries 91} (2015) 105023} [\href{https://arxiv.org/abs/1408.3371}{{\ttfamily 1408.3371}}].

\bibitem{dinkins2022exotic}
H.~Dinkins, \emph{Exotic quantum difference equations and integral solutions}, Ph.D. thesis, The University of North Carolina at Chapel Hill, 2022.

\bibitem{benini2015topologically}
F.~Benini and A.~Zaffaroni, \emph{A topologically twisted index for three-dimensional supersymmetric theories}, {\emph{Journal of High Energy Physics} {\bfseries 2015} (2015) 1}.

\bibitem{Ferrari:2024ksu}
A.E.V.~Ferrari and D.~Zhang, \emph{{On Spectral Data for $(2,2)$ Berry Connections, Difference Equations \& Equivariant Quantum Cohomology}},  6, 2024 [\href{https://arxiv.org/abs/2406.15448}{{\ttfamily 2406.15448}}].

\bibitem{closset2016comments}
C.~Closset and H.~Kim, \emph{Comments on twisted indices in 3d supersymmetric gauge theories}, {\emph{Journal of High Energy Physics} {\bfseries 2016} (2016) 1}.

\bibitem{Cacciatori:2009iz}
S.L.~Cacciatori and D.~Klemm, \emph{{Supersymmetric AdS(4) black holes and attractors}}, \href{https://doi.org/10.1007/JHEP01(2010)085}{\emph{JHEP} {\bfseries 01} (2010) 085} [\href{https://arxiv.org/abs/0911.4926}{{\ttfamily 0911.4926}}].

\bibitem{DallAgata:2010ejj}
G.~Dall'Agata and A.~Gnecchi, \emph{{Flow equations and attractors for black holes in N = 2 U(1) gauged supergravity}}, \href{https://doi.org/10.1007/JHEP03(2011)037}{\emph{JHEP} {\bfseries 03} (2011) 037} [\href{https://arxiv.org/abs/1012.3756}{{\ttfamily 1012.3756}}].

\bibitem{Hristov:2010ri}
K.~Hristov and S.~Vandoren, \emph{{Static supersymmetric black holes in AdS$_{4}$ with spherical symmetry}}, \href{https://doi.org/10.1007/JHEP04(2011)047}{\emph{JHEP} {\bfseries 04} (2011) 047} [\href{https://arxiv.org/abs/1012.4314}{{\ttfamily 1012.4314}}].

\bibitem{Katmadas:2014faa}
S.~Katmadas, \emph{{Static BPS black holes in U(1) gauged supergravity}}, \href{https://doi.org/10.1007/JHEP09(2014)027}{\emph{JHEP} {\bfseries 09} (2014) 027} [\href{https://arxiv.org/abs/1405.4901}{{\ttfamily 1405.4901}}].

\bibitem{Halmagyi:2014qza}
N.~Halmagyi, \emph{{Static BPS black holes in AdS$_{4}$ with general dyonic charges}}, \href{https://doi.org/10.1007/JHEP03(2015)032}{\emph{JHEP} {\bfseries 03} (2015) 032} [\href{https://arxiv.org/abs/1408.2831}{{\ttfamily 1408.2831}}].

\bibitem{Benini:2016rke}
F.~Benini, K.~Hristov and A.~Zaffaroni, \emph{{Exact microstate counting for dyonic black holes in AdS4}}, \href{https://doi.org/10.1016/j.physletb.2017.05.076}{\emph{Phys. Lett. B} {\bfseries 771} (2017) 462} [\href{https://arxiv.org/abs/1608.07294}{{\ttfamily 1608.07294}}].

\bibitem{Cvetic:1999xp}
M.~Cvetic, M.J.~Duff, P.~Hoxha, J.T.~Liu, H.~Lu, J.X.~Lu et~al., \emph{{Embedding AdS black holes in ten-dimensions and eleven-dimensions}}, \href{https://doi.org/10.1016/S0550-3213(99)00419-8}{\emph{Nucl. Phys. B} {\bfseries 558} (1999) 96} [\href{https://arxiv.org/abs/hep-th/9903214}{{\ttfamily hep-th/9903214}}].

\bibitem{Jafferis:2011zi}
D.L.~Jafferis, I.R.~Klebanov, S.S.~Pufu and B.R.~Safdi, \emph{{Towards the F-Theorem: N=2 Field Theories on the Three-Sphere}}, \href{https://doi.org/10.1007/JHEP06(2011)102}{\emph{JHEP} {\bfseries 06} (2011) 102} [\href{https://arxiv.org/abs/1103.1181}{{\ttfamily 1103.1181}}].

\bibitem{Herzog:2010hf}
C.P.~Herzog, I.R.~Klebanov, S.S.~Pufu and T.~Tesileanu, \emph{{Multi-Matrix Models and Tri-Sasaki Einstein Spaces}}, \href{https://doi.org/10.1103/PhysRevD.83.046001}{\emph{Phys. Rev.} {\bfseries D83} (2011) 046001} [\href{https://arxiv.org/abs/1011.5487}{{\ttfamily 1011.5487}}].

\bibitem{Benini:2015eyy}
F.~Benini, K.~Hristov and A.~Zaffaroni, \emph{{Black hole microstates in AdS$_{4}$ from supersymmetric localization}}, \href{https://doi.org/10.1007/JHEP05(2016)054}{\emph{JHEP} {\bfseries 05} (2016) 054} [\href{https://arxiv.org/abs/1511.04085}{{\ttfamily 1511.04085}}].

\bibitem{Hao:2013jqa}
W.~Hao, R.I.~Nepomechie and A.J.~Sommese, \emph{{Completeness of solutions of Bethe's equations}}, \href{https://doi.org/10.1103/PhysRevE.88.052113}{\emph{Phys. Rev. E} {\bfseries 88} (2013) 052113} [\href{https://arxiv.org/abs/1308.4645}{{\ttfamily 1308.4645}}].

\bibitem{doi:10.1142/5763}
A.J.~Sommese and C.W.~Wampler, \emph{The Numerical Solution of Systems of Polynomials Arising in Engineering and Science}, World Scientific (2005), \href{https://doi.org/10.1142/5763}{10.1142/5763}.

\bibitem{Jiang:2017phk}
Y.~Jiang and Y.~Zhang, \emph{{Algebraic geometry and Bethe ansatz. Part I. The quotient ring for BAE}}, \href{https://doi.org/10.1007/JHEP03(2018)087}{\emph{JHEP} {\bfseries 03} (2018) 087} [\href{https://arxiv.org/abs/1710.04693}{{\ttfamily 1710.04693}}].

\bibitem{compAlgGeom}
D.~Cox, J.~Little and D.~O'Shea, \emph{Ideals, Varieties, and Algorithms}, Springer New York, NY, 3~ed. (2007), \href{https://doi.org/10.1007/978-0-387-35651-8}{10.1007/978-0-387-35651-8}.

\end{thebibliography}\endgroup

\end{document}